\documentclass[mnsc,nonblindrev]{informs3_no_remark} 

\OneAndAHalfSpacedXI 

\usepackage{natbib}
 \bibpunct[, ]{(}{)}{,}{a}{}{,}%
 \def\bibfont{\small}%

\usepackage{subfigure}
\usepackage{lscape,booktabs,longtable}
\usepackage{amsmath}
\usepackage[pdftex,colorlinks=true,urlcolor=blue,citecolor=blue,pdfstartview=FitH]{hyperref}
\usepackage{comment}

\usepackage{caption}
\usepackage[ruled,vlined]{algorithm2e}

\newenvironment{lyxlist}[1]
 {\begin{list}{}
  {\settowidth{\labelwidth}{#1}
   \setlength{\leftmargin}{\labelwidth}
   \addtolength{\leftmargin}{\labelsep}
   }}
 {\end{list}}

\TheoremsNumberedThrough     
\ECRepeatTheorems

\EquationsNumberedThrough    

\MANUSCRIPTNO{MS-SMS-19-03009.R1}

\begin{document}



\RUNTITLE{Silent Abandonment in Contact Centers}

\TITLE{
Silent Abandonment in Contact Centers: Estimating Customer Patience from Uncertain Data}
\ARTICLEAUTHORS{%
\AUTHOR{Antonio Castellanos, Galit B. Yom-Tov, Yair Goldberg}
\AFF{\EMAIL{antonio.cas@campus.technion.ac.il,  gality@technion.ac.il, yairgo@technion.ac.il}} 
\AFF{Technion---Israel Institute of Technology} 
} 


\ABSTRACT{
In the quest to improve services, companies offer customers the opportunity to interact with their agents using texting. This has become a favorite channel of communication. However, text-based contact centers face operational challenges, since the measurement of common proxies for customer experience, such as  customers abandonment and willingness to wait (patience), are subject to information uncertainty.
We focus on the impact of a main source of such uncertainty: \emph{silent abandonment} by customers. 
These customers leave the system while waiting for a reply to their inquiry but give no indication of doing so. 
As a result, the system is unaware the customers left and wastes agents' time until this is realized. A sample of 33 companies shows that, on average, 69.5\% of abandoning customers do so silently. A more detailed analysis of one company with 12.6\% silent abandonment shows that such behavior reduces agent efficiency by 3.2\% and system capacity by 15.3\%. We develop methodologies to identify these customers in two types of contact centers: ones that allow or do not allow customers to write inquiries during waiting. Different tools are required since this ability changes the nature of uncertainty in the data. We use text analysis and classification models to obtain abandonment level. We then use a parametric approach and develop an expectation-maximization algorithm to accurately estimate patience. We show how accounting for silent abandonment in a queueing model dramatically improves the estimation accuracy of key performance measures. Finally, we suggest strategies to operationally cope with the phenomenon of silent abandonment.

}

\maketitle


%


\section{Introduction}
\label{sec:introduction}
The field of service engineering relies on measuring proxies for customer experience in a service system. Two of the most common operational measures used as such proxies are customer waiting and abandonment from the queue. Both are crucial performance measures for understanding customers' willingness to wait for service, which in turn is crucial for making operational decisions \citep{Mandelbaum2013Data,Garnett2002}. 
Waiting happens when a customer enters the service system, but the system does not have an available service agent. Abandonment naturally occurs when such waiting is too long and exceeds the customer's willingness to wait (henceforth, patience). Different streams of literature study various aspects of customer patience, such as its distribution (e.g., \citealt{Gans2003TelephoneProspects}), its connection to service utility (e.g., \citealt{Aksin2013}), its manipulation (e.g., \citealt{Armony2008,Aksin2017}), and more.

The literature on estimating customer patience and its implications for optimizing operational decisions (e.g., staffing and routing) assumes accurate and complete knowledge of customer abandonment.
However, in some service environments, such as text-based contact centers (henceforth ``contact centers'' for short), we face the problem of not always being able to know whether a customer abandoned or received service, as we will explain shortly. This uncertainty creates a situation where the company is unsure of the service quality they provide to their customers and how efficiently they use their resources, which in turn may lead to problematic operational decisions. The main goal of this paper is to overcome such uncertainties and allow companies to estimate customer patience and abandonment proportions correctly.

We concentrate on a specific type of uncertainty relating to a specific type of customer behavior in contact centers: \emph{silent abandonment} (Sab).
A Sab customer is one that leaves the system while waiting in the queue but gives no real-time indication of doing so (i.e., they do not close the chat window or application when abandoning). Therefore, when an agent becomes available, the (abandoning) customer is assigned to that agent. Only after the agent's inquiries go unanswered for some time does the agent (or system) realize that the customer has abandoned the queue without notifying the system, and the agent (or system) closes the conversation. We find that this situation creates two problems of {\bf information uncertainty}: (a) \textit{missing data}: the system may not be aware (even in retrospect) whether a customer silently abandoned the queue or was served. To the best of our knowledge, all companies assume the latter, thereby biasing quality measurements (for a detailed definition of the concept of missing data, see \citealt{little2002statistical});  and (b) \textit{censored data}: the system may be aware that the customer silently abandoned the queue but it does not know exactly when, thereby censoring the data on customer patience  (for a discussion of censored data, see \citealt{smith_2002}). In addition, Sab customers create two operational problems of {\bf agent efficiency}: (a) \textit{idleness}: the agent waits for inquiries from a customer that is no longer there; and (b) \textit{wasted work}: the agent tries to solve problems that have already been solved by the customer themself or by another agent (e.g., when the customer writes an inquiry while waiting in the queue and then abandons the queue and uses a different channel of communication such as a phone call), thereby creating confusion, frustration, and wasted effort. 
We note that silent abandonment is more likely to happen when the system is overloaded with customers and waits are long. During such periods, a significant number of agents are likely to be either idle  or ``busy" with abandoning customers, wasting critically needed capacity.  Moreover, the Sab customers are taking the places of customers that want service and are actually waiting in the queue. 
Finally, we note that silent abandonment results in inaccurate measurements of queue length. Therefore, any algorithm that uses that information (e.g., for  delay announcements; see \citealt{ibrahim2018sharing}) would need to be adjusted to allow for silent abandonment.

The abovementioned missing data problem is connected to the company's ability to retrospectively distinguish between Sab and served customers. In some contact centers, Sab is easy to identify in retrospect. Those are mostly contact centers that do not allow customers to write text while waiting for an agent to be assigned (we refer to these as \emph{no-write-in-queue} 
contact centers). In such contact centers, Sab instances are conversations where the customer initiated a service request, could not communicate anything regarding their problem during waiting, and did not reply to any agent inquiries (see chat examples in Appendix \ref{app:chat_examples}). In other companies, Sab customers are not easily identifiable from some of the served customers. Specifically, these companies allow customers to communicate their problems while waiting in the queue (we refer to these as \emph{write-in-queue} 
contact centers). In these centers, there is uncertainty regarding customers who wrote while waiting but did not write anything after an agent was assigned to serve them. Without reading the conversation text, it is impossible to distinguish whether such customers have abandoned the queue silently or were served but acted impolitely regarding that service (i.e.,  did not acknowledge being served even by a thank-you message).
We refer to a customer from this class as an \textit{uncertain silent abandonment} (uSab) customer, and to the classical abandoning customer, who closes the communication window while waiting for the agent, as a \emph{known abandonment} (Kab) customer. One of our goals is to shed light on the pros and cons of allowing customers to write during waiting and to investigate the impact of that choice on operations (\S\ref{sec:patience_coef}). Allowing writing provides a more natural communication environment (such as in social media), saves agents' time, and keeps customers busy, which in turn decreases perceived wait and increases patience  \citep{Maister1984TheLines}. In Section \ref{sec:patience_coef}, we show that customers write 22.1\% of their text while waiting in write-in-queue systems and that customers who write during waiting exhibit higher patience. 

The problem is the emergence of the missing data phenomena.  It is important to note that write-in-queue systems are becoming more common in the contact-center industry. In fact, our industry partner reports that by 2020 66\% of their clients, which include 35\% of \textit{Fortune}'s most valuable companies, have adopted systems that allow such operation, up from 5\% in 2016 \citep{LivePerson2020}. 

\begin{figure}[htb]
\centering
\subfigure[Proportion of Known  and (Uncertain) Silent Abandonment]{
\includegraphics[width=0.47\textwidth]{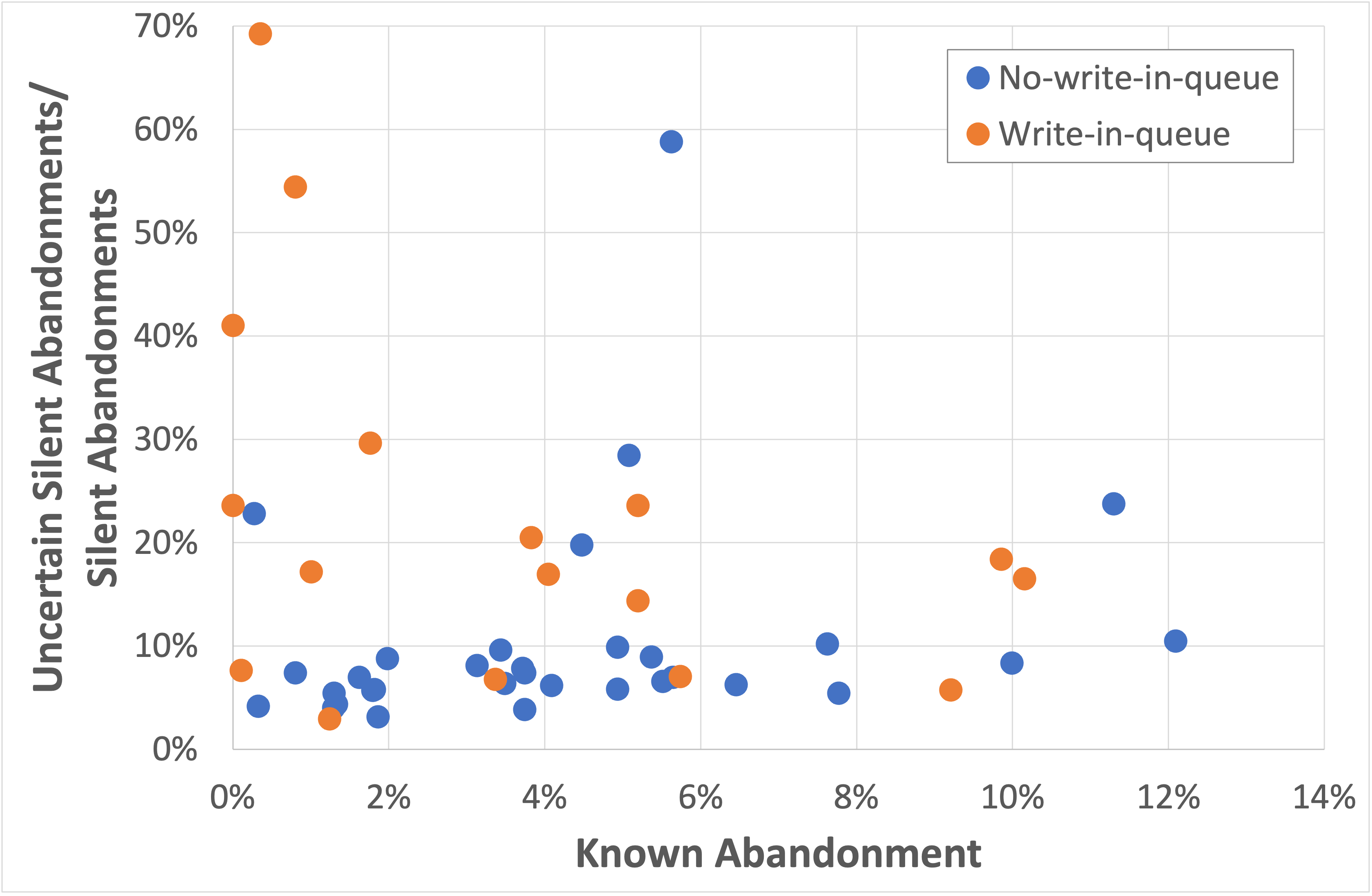} \label{fig:KabvsUsab}
} 
\subfigure[Proportion of Silent Abandonment from All Abandonments (No-write-in-queue Companies)]{
\includegraphics[width=0.47\textwidth]{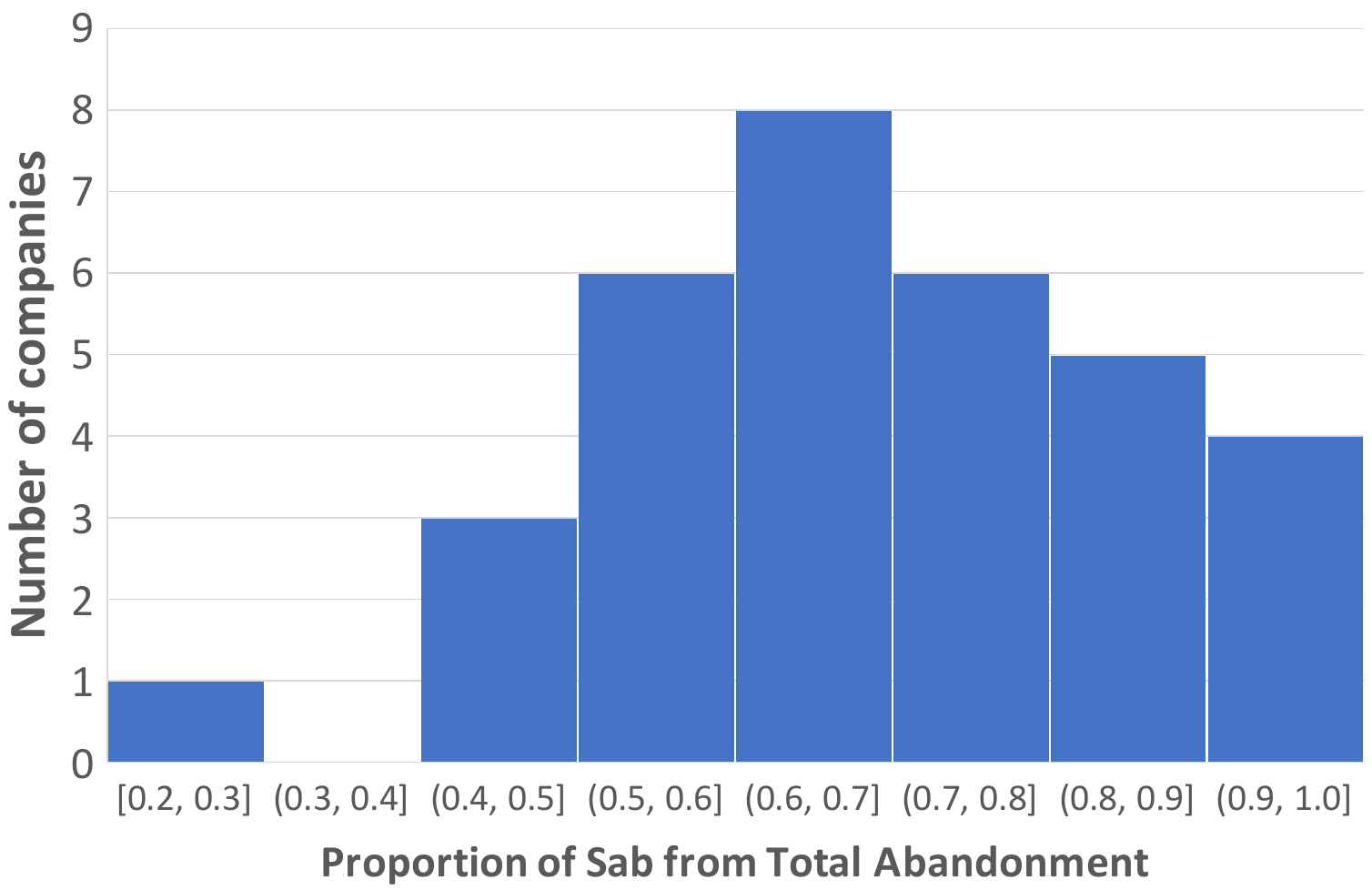}
\label{fig:SabHist}
}
\caption{Silent Abandonment in Different Contact Centers}
\label{fig:Sabfig}
\end{figure}

The phenomenon of Sab is very common in contact centers, as demonstrated by Figure \ref{fig:KabvsUsab}, which is based on data from fifty western companies. These companies are of different domains and sizes, from small contact centers with 9000 conversations per month to those of Fortune 500 companies that handle more than 100,000 conversations per month on average. We observe that the proportion of Sab in the no-write-in-queue companies is between 3\% and 59\%, much larger than the proportion of Kab (1\%--12\%). In these companies, the Sab customers represent, on average, 67.5\% of the total abandoning customers (see proportion distribution in Figure \ref{fig:SabHist}). Therefore, ignoring Sab customers profoundly biases performance measures. The proportion of uSab customers in write-in-queue companies has similar magnitude (3\%--70\%) but serves only as an upper bound to the Sab proportion since some uSab customers were served. uSab customers represent, on average, 22\% of the conversations; hence, inferring how many of them indeed abandoned is important. In Section \ref{sec:prob_abnd}, we will analyze in detail two contact centers, one of a telecommunications company that allows customers to write-in-queue and the other of a transportation company that does not (i.e., a no-write-in-queue contact center). These contact centers are typical in the scope of their silent abandonment. The Sab percentage in the no-write-in-queue contact center is 5.2\% in total and 27.4\% of the abandoning customers (\S\ref{sec:prob_abnd_NWQ}). The uSab percentage in the write-in-queue contact center is 24.4\%; using our models, we show that 51.8\% of these uSab conversations are indeed silent abandonment; hence, Sab represents 71.5\% of the total abandonment (\S\ref{sec:prob_abnd_WQ}). With all the above, we see that silent abandonment is a phenomenon that exists in all contact centers, and that we require solutions to understand and cope with it.

The context of this research is contact centers, which are an important part of the service industry's digital revolution. Service companies branch into more accessible and easy-to-use service channels such as mobile applications. Technology allows modern-day companies to replace traditional service encounters (face-to-face, telephone) with technology-mediated ones \citep{Massad2006CustomerEncounters,vanDolen2002ModeratedEncounter}, which allow customers and service employees to be in different locations and connect via a digital interface \citep{Schumann2012TechnologyAcademics,Froehle2004NewExperience}. 
%
Nowadays, employees and customers can interact through social media (e.g., Twitter or Facebook), corporate websites (e.g., chats), or messaging applications (e.g., WhatsApp and WeChat). This enables customers to interact with agents through platforms similar to those used to contact their family and friends. 
Therefore, it should come as no surprise that text-based contact centers are slowly replacing call centers as the preferred way for customers to communicate with companies. Indeed, as early as 2012, a survey conducted by a contact-center solution provider found that 78\% of customers preferred to text with a company rather than call their call center \citep{ringCentral_2012}. 

It is worth noting that the digital revolution provides the service industry not only new opportunities to improve services \citep{Rafaeli2017TheFuture,Altman2021} but also new operational challenges. Operating text-based contact centers is substantially different from operating call centers. For example, in textual service systems, unlike in call centers or face-to-face services, agents can provide service to multiple customers concurrently \citep{Tezcan2014RoutingCustomers,Goes2017WhenMultitasking,daw2021coproduction}. We claim that the information uncertainty that results from the phenomenon of silent abandonment creates a need to redefine the basic methods of measuring quality and efficiency, as well as to develop methodologies to estimate customer patience. This is the focus of this paper. 
We also show that models that account for the silent-abandonment phenomenon and incorporate the methodologies we develop here fit the data of contact centers much more accurately than a regular Erlang-A model \citep{Mandelbaum2007} (\S\ref{sec:model_fitting}). Finally, we measure and discuss the implications of silent abandonment on system performance and managerial practices.

We note that Sab is not exclusive to contact centers and appears in other environments such as ticket queues and emergency departments.
In ticket queues, an arriving customer is given a ticket with their queue number, sometimes with additional information regarding the estimated wait for service \citep{Xu2007, Kerner2017}. The customer then decides whether to join the queue or silently abandon. Even if the customer decides to join the queue, they might silently abandon sometime during the wait. In some systems with physical queues, the Sab is obvious, but in many cases---analogous to contact centers---the Sab is not realized until the customer is called for service and does not show up. Agents may waste time waiting for a Sab customer to appear. This wasted time is an example of the idleness that results from Sab. 
Silent abandonment in ticketing queues is influenced by delay announcements provided to  customers \citep{Kerner2017}. 
Our data does not include any delay announcements. Yet, in this paper, we deepen the understanding of how other aspects of the waiting experience impact patience. Specifically, we develop methods to estimate the average customer patience as well as the coefficients of variables that impact that patience from uncertain data (that include censored and missing information). We show that customers that engage in communicating their problem during waiting exhibit longer patience (\S\ref{sec:patience_coef}). This offers a new way for companies to impact their customers' patience.  
We further contribute to the literature on ticketing queues by developing  estimation methods for customers' patience in these systems (see Appendix \ref{sec:appendixPatienceTQ}). These methods are very similar to the one needed for contact centers.

Silent abandonments appear also in healthcare systems. For example, in emergency departments (EDs), a patient may abandon the queue before seeing a medical practitioner without telling anyone---a phenomenon called ``left without being seen (LWBS).'' According to \cite{Medicare}, the national average of LWBS patients in US EDs was 2\% during 2021. ED abandonment increases the risk of a patient suffering an adverse outcome, increases the probability of the patient returning \citep{Baker1991}, and impacts hospital revenue \citep{Batt2015}. 

In both of the above examples (ticketing queues and LWBS), the data of customer patience is censored in the same way as in contact-center data. \cite{Yefenof2018}---an important predecessor of this paper---developed maximum likelihood methods to overcome data censoring when estimating customer patience in EDs. While LWBS customers indeed abandon the ED silently, there is no uncertainty about them doing so, since physicians report on such events in the electronic medical record (EMR) system. Hence, uncertainty is not an issue there as it is in contact-center data. 

Silent abandonment is also related to the phenomena of no-show and service failure. 
A no-show customer does not arrive to a scheduled appointment and fails to notify the system in advance.
This creates censored data similar to silent abandonment, but not missing data, since complete information is observed, in hindsight, regarding patient service (or lack thereof). The scope of no-show customers can be as high as 23\% to 34\% \citep{Liu2016}. Healthcare providers can reduce this phenomenon by calling customers the day before or sending an automated reminder \citep{Geraghty2008}, but cannot eliminate it. \cite{HoLauNoshows1992} showed that no-shows strongly affect system performance because of capacity loss and forced idleness of physicians. We claim this also happens in contact centers. Several methodologies have been suggested to cope with no-show customers, such as overbooking \citep{Vissers1979} and reminders \citep{Geraghty2008}. However, in contact centers, arrivals are not known in advance, thus other coping mechanisms are needed. Another difference between no-show and Sab customers relates to our points on idleness and wasted work regarding agent efficiency, presented above. In medical appointments, it can be observed whether a patient shows up or not for their appointment, and this information is realized as soon as their service is supposed to start (without delay). Therefore, in the no-show case, the agent can immediately start serving the next patient instead of waiting for the one who did not show up (assuming there is a next patient at the clinic). But in contact centers, the customer is not physically present in front of the agent, so there is no indication that the customer has abandoned the queue; this information is realized only after a few minutes of wasted agent effort. Therefore, overbooking can mitigate the efficiency loss of no-shows, whereas Sab requires other solutions, as we  suggest in Sections \ref{sec:managerial} and \ref{sec:patience_coef}.
Another closely related phenomenon is service failure. For example, silent abandonments appear in virtual assistant systems such as Alexa and Siri, and regular interactive voice response (IVR) systems. In many of them, it is hard to distinguish whether the customer finished service or silently abandoned. \cite{Carmeli2019IVR} analyzed the impact of service failure (they called it abandonment during service) on the design of IVR systems and websites, where a customer may or may not successfully complete a self-service. They show the impact of estimating the proportion of customers that had an unsuccessful service (17\%) on system design. Here, we only consider abandonment from queue and make a similar claim that Sab has an impact on system design.  
Finally, our research can also be related to research on queue inference, where queue statistics are deduced from limited information. For example, \cite{Larson1990} deduced system load using service completion data, where no knowledge on arrival epochs or queue length is available. Our concern here is not about limited information but about missing data, where service outcome data is missing for \emph{some} customers but available for others. To the best of our knowledge, no work has addressed the problem of this type of missing data.

\subsection{Research Goals} \label{sec:goals}
The present  paper concentrates on the following goals: 

\textit{Estimate the scope of the silent-abandonment phenomenon}. We want to estimate how many customers silently abandon the queue in contact centers.  
This is similar to estimating the scope of no-shows in healthcare. Our goal is to be more precise than prior studies by analyzing silent abandonment at the individual customer level. Figure \ref{fig:Sabfig} shows that overall silent abandonment is a common phenomenon. It classified a Sab customer using the number and timing of customer messages, but for write-in-queue companies, such classification provides only an upper bound (see detailed explanation in Appendix \ref{app:chat_examples}). In Section \ref{sec:prob_abnd}, we construct classification models for write-in-queue contact centers' data that estimate the probability of silent abandonment by a specific customer. This model uses,  among other things, the text messages of the customer and agent. Implementation of these methods on two detailed datasets shows that around one-third of abandoning customers in the no-write-in-queue company dataset, and around two-thirds in the write-in-queue company dataset, are Sab customers. 

\textit{Create an algorithm to estimate customer patience in the presence of silent abandonment}. \cite{Gans2003TelephoneProspects} 
reviewed methods for estimating customer patience based on call-center applications. As we mentioned, customer behavior in contact centers differs from that in call centers. To our knowledge, no paper has attempted to estimate customer patience in contact centers, although finite patience has been considered in optimization models of contact centers  \citep{Tezcan2014RoutingCustomers}.  To estimate call-center patience, \cite{Mandelbaum2013Data} assumed that customer-patience time, $\tau$, and virtual wait time, $W$, are exponentially distributed with rates $\theta$ and $\gamma$, respectively.  
Specifically, they developed a maximum likelihood estimator for estimating customer patience from right-censored data. Inspired by the LWBS phenomenon in EDs, \cite{Yefenof2018} extended their estimator to left-censored patience data, which is created by patients who do not announce their abandonment time. (To that end, they developed both parametric and nonparametric methods.) Their estimators are suitable for estimating customer patience in no-write-in-queue systems, where the only type of information uncertainty is censored data. However, in write-in-queue systems, system design and silent abandonment create both of the abovementioned types of uncertainty in the data (i.e., censored data and missing data). Therefore, we develop a new method for estimating customer patience that also addresses the additional problem of missing data. This method uses an expectation-maximization algorithm (see Section \ref{sec:patience_estimate}).
It is important to estimate system parameters as accurately as possible, since performance measures of queueing systems are sensitive to inaccuracies in such estimations \citep{Whitt2006}. We show in Section \ref{sec:managerial} that, indeed, a more accurate estimation of customer patience, one that takes silent abandonment into account, significantly improves the fit of the queueing model to the data. (In Appendix \ref{subsection:EM_covaritates}, we extend this method to account for different coefficients that may impact customer patience.)

\textit{Analyze the operational implications of silent abandonment}. In Section \ref{sec:prob_abnd}, we estimate the amount of time companies waste due to the phenomenon of silent abandonment. In Section \ref{sec:managerial}, we develop a queueing model that captures the dynamics of contact centers in the presence of silent abandonment. 
We then analyze the implications of Sab wasted time on system performance and staffing. We discuss how a bot or a classification model for identifying silent abandonment in real time may be used to reduce the impact of Sab customers on the system.
Finally, in Section \ref{sec:patience_coef}, we show the advantage of write-in-queue systems in prolonging customer patience. 

\section{Data and Research Setting} \label{sec:dataAndReseatch}

For the purposes of our research, we have acquired and analyzed data from both of the aforementioned contact centers, namely, one write-in-queue and one no-write-in-queue. The data was provided by LivePerson Inc., which builds computational infrastructures for the contact-center industry.  
As mentioned above, differences in the way these contact centers operate and in the way people use them have an impact on information uncertainty. 
For clarity, we show example conversations  in Appendix \ref{app:chat_examples} for each scenario of conversation dynamics---served, Kab, and Sab---in both types of systems, and denote graphically the timing of customer and agent message dynamics.

The full interaction contains several agent and customer messages the two parties send one another. Due to concurrency, agent response time may also include short waits (if the agent is busy answering another customer). Conversations are closed either by the customer, the agent, or the system. The latter happens automatically when the customer has been inactive for some predetermined time. 


\subsection{No-write-in-queue Contact-Center Data}
\label{subsec:Chat}
The data is extracted from 18,479 service interactions conducted in February 2017 (out of which 780 conversations (4.2\%) were excluded during the data cleaning process since they contain unrealistic or no information). It includes general information on each conversation as well as on each line written. Each conversation is identified by conversation ID, employee ID, date, the amount of time the customer waited in the queue before the chat started, whether the customer abandoned the queue by closing the conversation window and at what time, the time an agent was assigned to that conversation, the time the conversation ended, the device used for the communication, type of service (e.g., sales or support), and more. Each conversation message in the data contains the following information: a time stamp of when the message was sent, who wrote that message (customer, agent, or system), and the number of words in the message.  
The data also includes information on the work status of each service agent (online, offline, on break, or idle) during the workday. Each agent's load is estimated by analyzing the agent's activities with customers when the agent is online.

The contact center is open 7 days a week from 8:00 to 22:00. The average number of arrivals per hour is 51.58 customers, and each agent serves up to 3 customers concurrently. The arrival rate varies with the hours of the day. The pattern of the hourly arrival rate is typical of service systems (see Table \ref{tbl:ParametersHour}). The mean number of agents working per hour is 12.45, out of which 8.34 are actively serving customers. 
Table \ref{tbl:general_stat} provides general statistics for conversations in this data (such as the number of messages) both for all conversations and by conversation type (served, Kab, or Sab). Specifically, the average total time a customer is in the system, from entering the system until conversation closure, is 12.59 minutes ($SD=16.90$) (the average includes Sab customers). This measure includes three main parts: wait time, service time, and closure time. The average wait time in queue is 1.41 minutes ($SD=5.56$). The average service time, from assignment to agent until the last message was written in that conversation, is 10.26 minutes ($SD=15.86$), and the average time from the last message until conversation closure is 0.93 minutes ($SD=1.73$). The company automatically closes conversations that were inactive for 2 minutes. 

\subsection{Write-in-queue Contact-Center Data} \label{subsec:MessagingData}

From a write-in-queue contact center, we acquired data on 332,978 service interactions conducted during May 2017 (out of which 1,391 conversations (0.4\%) were excluded during the data cleaning process since they contain unrealistic or no information). It includes detailed information on all the conversations (like the no-write-in-queue data) as well as the text written (for uSab conversations). 
This contact center operates 24/7. The number of arrivals, which typically varies along the day, is 594.79 per hour on average.
The mean number of online agents per hour is 134.69. The mean concurrency level of agents is 5.4 customers per agent ($SD=4$).
Average total time a customer is in the system (from entering the system until conversation closure) is 120.57 minutes ($SD=95.80$) (the average includes Sab customers). 
The average wait in the queue is 8.30 minutes ($SD=18.28$), the average service duration is 46.34 minutes ($SD=63.56$), and average closure time is 65.93 minutes ($SD=75.08$). The company automatically closes conversations that were inactive for 2 hours.

\section{Estimating the Scope of Silent Abandonment as a Source of Information Uncertainty}
\label{sec:prob_abnd}
In this section, we aim to build models to define which conversations can be classified as silent abandonment with high probability. This will enable us to estimate the percentage of Sab customers. In addition, such information also enables us to estimate the time it takes for a service agent to realize that a Sab customer has abandoned the queue, and to estimate the agent's wasted work on a Sab customer.
We conduct separate analyses of the two types of contact centers due to the difference in estimation method each requires.

\subsection{Estimating the Scope of Silent Abandonment in No-write-in-queue Systems}
\label{sec:prob_abnd_NWQ}
The company that provided us with the no-write-in-queue system dataset erroneously estimates the percentage of abandoning customers by counting only customers that left the system by closing the interaction window, thus providing a clear indicator that they abandoned the system---the Kab customers. 
The proportion of Kab customers in the no-write-in-queue data is 13.9\%. We claim that this is an underestimation of the proportion of abandoning customers since it ignores silent abandonment. That is, the no-write-in-queue company does not account for customers that arrived to the system and were assigned to an agent but did not communicate with that agent at all---but instead clearly abandoned the system during wait time. Since these customers gave no indication they were leaving, the system was unaware of their abandonment and assigned them to an agent. 
Therefore, we can identify the conversations in which customers silently abandoned by indicating whether the conversations include system and agent messages but do not include any customer messages.  
Using this method, we find that Sab customers constitute 5.2\% of all customers arriving to the no-write-in-queue contact center. Therefore, the correct estimation of the probability of abandonment is 19.1\%, emphasizing our claim that the company is unaware of the actual service level it provides. Moreover, out of all abandoning customers, 27.4\% abandon the queue silently.


We use the silent-abandonment classification to estimate the time it takes for an agent to realize that the customer actually (silently) abandoned the queue: 2.32 minutes on average ($SD=3.8$) from agent assignment to last agent line. This is the time in which the agent keeps trying to communicate with the customer (the agent might be serving other customers at the same time), and gets no reply. 
If we subtract the silent-abandonment conversations from the conversation data, we see that the average \emph{served} customer service duration (from agent assignment to last agent line) is 12.54 minutes ($SD=16.82$). 

To provide an estimation of agents' wasted effort, we compute the number of messages and words agents spend on Sab customers. We find that, on average, for Sab customer an agent writes 1.39 messages ($SD=0.53$) with an average of 18.02 words ($SD=12$) in each message. From the agent's perspective, 6.1\% of the customers handled during the day are silent-abandonment customers 
and 93.9\% are served. Hence, on average 1.5\% (0.8\%) of messages (words) written by agent are wasted on Sab customers. 
To provide an additional estimation of wasted effort on Sab customers from the system perspective, we compute the percentage of agents' concurrency that is wasted on these Sab customers.
We compute the percentage of concurrency time (service time plus closure time) agents spent on silent-abandonment customers by dividing the concurrency time spent on Sab conversations by total concurrency time. Average closure time is 1.27 ($SD=1.28$) for Sab customers and 1.06 ($SD=1.85$) minutes for served customers, respectively. Hence, 
\[
\text{Time effort} = \frac{0.061\cdot(2.32+1.27)}{0.061\cdot(2.32+1.27)+0.939\cdot(12.54+1.06)} = 0.017;
\]
thus, the system wastes 1.7\% of their agent's concurrency time engaging in Sab conversations. During those time intervals, there may be other customers in the queue, blocked from entering service due to the concurrency limit of three customers. Hence, this wasted time effectively reduces agent capacity by 1.7\%. We will show the impact of these efforts in Section \ref{sec:managerial}.

\subsection{Estimating the Scope of Silent Abandonment in Write-in-queue Systems} 
\label{sec:prob_abnd_WQ}

In the case of the write-in-queue system, the company also underestimates the proportion of abandoning customers by taking into account only the known abandonments. The proportion of Kab customers in the write-in-queue dataset is 5.1\% of the customer population. As mentioned in Section  \ref{sec:introduction}, here the identification of silent abandonment is much more problematic due to the problem of missing data resulting from the customers' ability to write while waiting. 
USab conversations account for 24.4\% of the conversations in the write-in-queue dataset. We suggest that in order to classify which of these uSab customers silently abandon the queue and which were served, we need to take a closer look at the conversation text. (See an example of such conversation texts in Figures EC.\ref{fig:ShortServConvo}--EC.\ref{fig:SabConvo}.) In this section, in order to distinguish between served customers and uSab customers who will be classified as served, we call the latter group \emph{served-in-one-exchange} (Sr1) customers (this emphasizes that the service of these customers includes only one exchange of information: the customer writing while waiting and the agent writing after assignment).

We built an automated classification model to distinguish the conversations of customers who silently abandoned the queue from the Sr1 conversations.
We manually tagged a random sample of 650 uSab conversations into the two groups---Sr1 and Sab---by reading the text of the whole service interaction. The sample included 342 Sr1 conversations and 308 Sab conversations. 
We then extracted textual features from the conversation transcript for all conversations as well as their metadata features described in Section \ref{subsec:MessagingData}. To obtain the textual features, we used natural language processing (NLP) techniques. We built a sparse matrix by tokenizing the transcript of the conversations and filtering words that do not convey information (like ``a'' and ``the'') using the English stop words dictionary in scikit-learn \citep{scikit-learn}. For the model-development stage below, we used a random subset of 550 conversations (and left the other 100 conversations for a final out-of-sample test). For each word in the sparse matrix, we computed its mutual information \citep{kraskov2004estimating}, which measures dependencies between attributes, to assess how much information each word contributes to the silent-abandonment tag. (This results in a mutual information matrix.)
To reduce the number of words used in our models, we took the top 50 agent words and top 50 customer words that provide the highest dependency with the Sab tag according to the words' mutual information matrix. Then, each word is represented by a variable stating the number of times it was used in the conversation. 
The 550-conversation set was then randomly separated into training and test sets containing 75\% and 25\% of the conversations, respectively. 
We denote by $\pi_{i}$ the probability that customer $i$ silently abandoned the queue, given that this conversation is part of the uSab group. Formally, $\pi_{i} \triangleq Pr\left\{ Sab_{i}\mid uSab_i\right\}$.

\begin{table}[!htb]
   \begin{minipage}{.5\textwidth}
    \centering
    \includegraphics[width=0.75\textwidth]{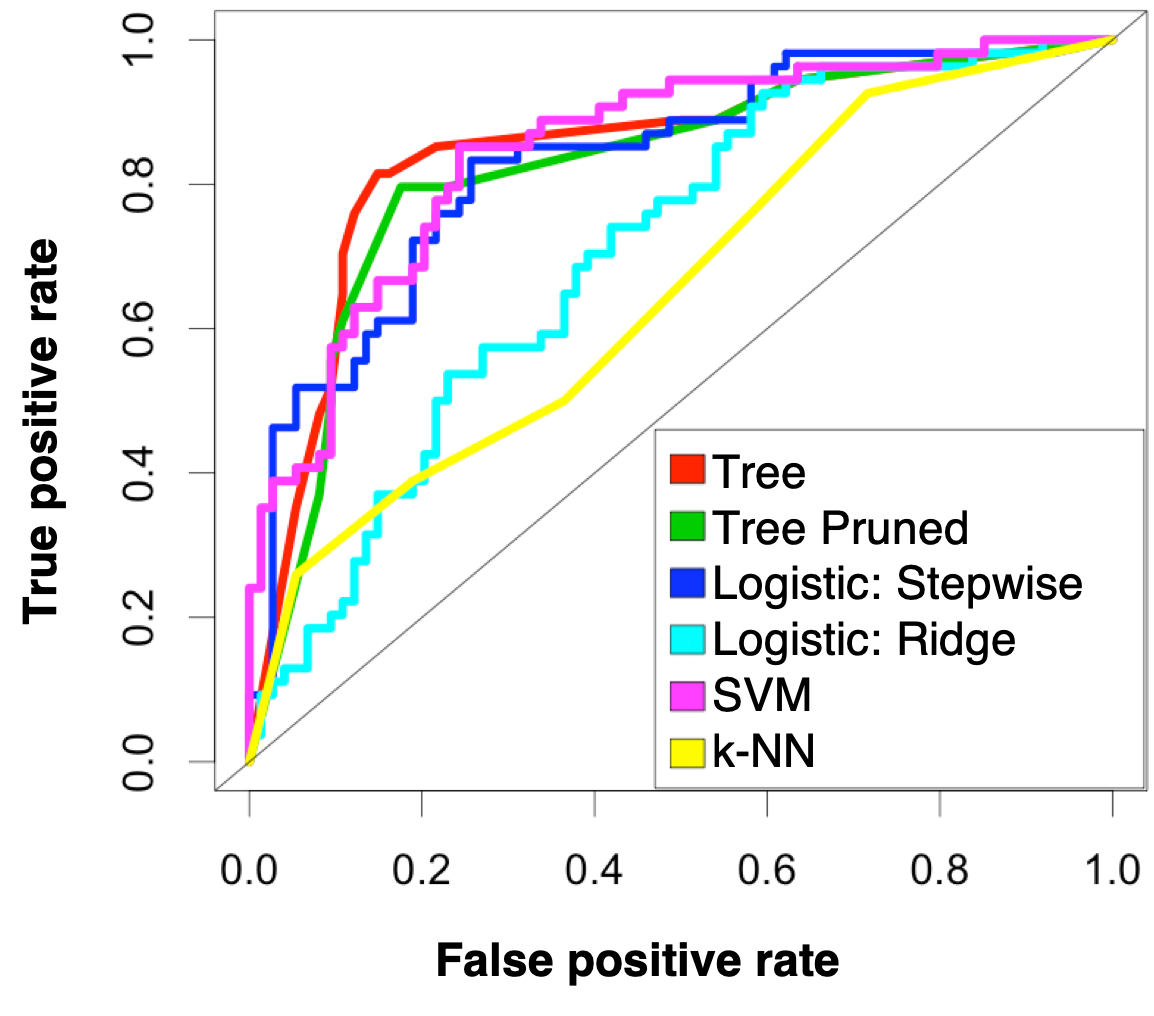}%
    \captionof{figure}{ROC Curve on Test Dataset}
    \label{fig:rocC}
  \end{minipage}
  \begin{minipage}{0.3\textwidth}
    \caption{Area under the ROC}  
    \label{tbl:logridge}
    \centering
    \begin{small}
    \centering
    \begin{tabular}{lc} 
    \toprule
    Model & AUC\\ \midrule 
    SVM & 0.85\\
    Tree & 0.85\\
    Logistic Regression: Stepwise & 0.83\\
    Tree Pruned & 0.82\\
    Logistic Regression: Ridge & 0.71\\
    k-NN & 0.65\\     
    \bottomrule 
    \end{tabular}
    \end{small}
  \end{minipage}\hfill
\end{table} 

We examined the performance of several machine learning classification methods: logistic regression (stepwise backward and with a ridge penalty), support vector machines (SVM), k-nearest neighbors (k-NN), and classification tree (additionally, we pruned the tree). We trained each model on the training subset of 412 conversations defined above, and then estimated $\hat{\pi}_i$ for each conversation $i$ in the test subset (138 conversations).
In Figure \ref{fig:rocC}, we compare the accuracy of these classification models using the receiver operating characteristic (ROC) curve. The ROC plots the true positive rate (TPR) against the false positive rate (FPR) for varying threshold levels. The ROC curve is a recognized visual method for comparing performance of different classification methods and for selecting the best threshold to work with \citep{Fawcett2006ROC}. A standard characteristic in that regard is the area under the ROC curve (AUC), presented in Table \ref{tbl:logridge}.   Using this criterion, we conclude that the best classification methods for our problem, with $\text{AUC} = 0.85$, are the SVM model and the classification tree. Models with an AUC above 0.80 are considered ``excellent" classification models \citep {Hosmer2000}. More details about feature selection procedures used for the SVM and the classification tree models can be found in Appendix \ref{sec:ClassMod}. The SVM model includes the following features, among others: specific words written in the conversation, customer experience (e.g., amount of time the customer waited in the queue), and agent's work time (e.g., amount of time the agent engaged with the customer). The SVM model includes words showing whether the customer is asking for the agent's immediate attention along with specific words related to the service the company provides. Due to privacy and legal reasons, we cannot share the exact expressions; however, we can summarize that the customers that were in fact Sab customers were ones that expressed some urgency. 

Our chosen model is the SVM. To select a specific threshold level for the model, we searched for the threshold that maximizes the sensitivity (TPR) and specificity (1-FPR) proportions; that is, it maximizes the proportion of Sab and Sr1 conversations that are correctly identified. To that end, we use the Youden index, defined as $\max_{c}\{\mathrm{sensitivity\left(c\right)+specificity\left(c\right)-1}\}$, where $c$ is the threshold \citep{berrar_2019}. Note that in our setting, this optimization problem is not weighted, since we care as much about maximizing the proportion of Sab correctly identified as we do the proportion of Sr1 customers. 
We find that the optimal threshold ($c$) is 0.47 with a sensitivity proportion (TPR) of 85\%, a specificity proportion (1-FPR) of 76\%, and an error rate\footnote{Error rate is computed using the false positive rate and the false negative rate as follows: $\left(\frac{FPR+FNR}{N}\right)$, where $N$ is the sample size.} of 20\%. An additional way to choose the best threshold is by maximizing the F1-score (a harmonic mean of precision and TPR) \citep{Chinchor1992}, that is, by maximizing the sum of precision and recall together. The highest F1 score in our case is 0.78 obtained with the same threshold $c=0.47$.
We then performed a validation test on an out-of-sample set defined above (including 100 random uSab conversations separate from the 550 conversations used for the model and threshold selection). We find that 91\% of the out-of-sample conversations are correctly classified by the SVM with that specific threshold (0.47). 

To obtain results for the full write-in-queue dataset, we processed the transcript of every uSab conversation and indicated whether it contained the SVM's words. We classify each conversation as Sab if the $\hat{\pi_{i}}$ estimated by the SVM is larger than 0.47. 
We find that out of the group of uSab conversations, which constitute 24.4\% of all the conversations in the data, 51.8\% are classified as Sab conversations (have a SVM score larger than 0.47) and 48.2\% are Sr1 conversations. This means that the actual proportion of abandoning customers in this dataset is
17.7\%, far above the estimation of 5.1\% abandonment that the company currently has. Moreover, out of all abandoning customers, we find that 77.8\% are Sab customers (12.6\% of all arriving customers). This again highlights the importance of taking Sab customers into account in order to correctly evaluate performance levels in contact centers. 

To estimate the service time of customers that (silently) abandoned the queue, we calculate the average service duration of the Sab conversations classified above (using the $\hat{\pi_{i}}> 0.47$ threshold). We find that it takes, on average, 20.06 minutes ($SD=40.65$) for an agent to identify a Sab conversation in that contact center (from assignment until last message). According to this company policy, an agent must reach out to an unresponsive customer at least twice (after the initial message) with at least 10 minutes between each reach (see Table \ref{tbl:general_stat} for statistics regarding within conversation dynamics). While the agent might be serving other customers concurrently during that period, they need to be attentive to the passage of time to apply this policy and write the required messages. Note that during these 20.06 minutes, concurrency slot capacity is wasted while the agent tries to communicate with a departed customer. Given the uSab conversation classification, we estimate that the average service duration of Sr1 conversations is 53.67 minutes ($SD=71.36$). This finding reveals that on average Sr1 customers have a longer service time than Sab customers and a similar service time to the Sr customers, which is not easily observed in the distribution of the LOS of uSab.\footnote{We note that the dynamics of Sr1 and Sr conversations  are different, which results from the abovementioned policy of how to handle unresponsive customers. Specifically, the data shows that the agent sends an Sr1 customer a mean of 4.31 $(SD=2.45)$ messages before ``giving-up" on them and either closes the conversation or lets the system close it automatically 2 hours later. On the other hand, during Sr conversations, when the customer is responsive, the agent sends  a mean of 11.8 $(SD=9.36)$ messages but does not need to wait the entire 10-minute interval, and therefore the time between messages is much shorter---2.99 minutes on average throughout the conversation (see Table \ref{tbl:general_stat}).} Overall, we find no reason for the agents to be actively reaching for the Sr1 customers for a longer period of time than for the Sab customers. Moreover, we note that Sab customers have a longer average closure time, 113.64 minutes ($SD=65.88$), than Sr1 customers, 94.98 minutes ($SD=71.03$). 

We find that 13.3\% of the inquiries the agent answers are from Sab customers, 12.4\% from Sr1 customers, and 74.3\% from (regular) served customers. Measuring agent effort in treating Sab customers,
we find that agents waste on average 2.29 messages ($SD=1.29$) and 26.2 words per message ($SD=20.29$) per Sab customer. Taking an overall perspective, this comprises 3.2\% (3.6\%) of messages (words) written by the agents in that contact center.
The effort wasted on unresponsive customers in the write-in-queue system is twice that of the no-write-in-queue system and is connected to this company's investment in reaching unresponsive customers, both in more reaches towards such customers (3.2 vs.\ 1.5 messages on average) and in the level of unresponsiveness it allows for its customers (automatic closure after 2 hours vs.\ 2 minutes).

To provide an additional estimation of wasted effort on Sab customers from the system perspective, we compute the amount of agents' concurrency time wasted on those customers.  Dividing the concurrency time spent on Sab conversations by the total concurrency time reveals that the system spends 15.3\% of its agents' concurrency capacity dealing with Sab conversations:
\begin{equation}
\label{eq:Effort_msg}
    \text{Time effort} = \frac{0.133\cdot(20.06+113.64)}{0.133\cdot(20.06+113.64)+0.124\cdot(53.67+94.98)+0.743\cdot(48.57+59.18)} = 0.153.
\end{equation}
This finding highlights the importance of identifying silent abandonment as soon as possible to improve system efficiency. 

\section{Estimating Customer Patience with Silent Abandonment}
\label{sec:patience_estimate}
Our next goal is to estimate customer patience in contact centers, where customer data is uncertain due to the presence of silent abandonment. 
As mentioned in Section \ref{sec:introduction}, contact-center data on customer patience is censored.  When the customer abandons the queue and provides an indication of doing so---a known abandonment---they provide exact information regarding their patience. Indeed, their patience equals the (observed) wait time. However, how long the customer would be required to wait if they were to stay in the queue, the \emph{virtual wait time}, is unknown. Therefore, patience acts as a lower bound for virtual wait time. When the customer is served, the wait time is actually a lower bound for their true patience. Therefore, the data is right-censored by the virtual wait time (itself uncensored). This type of right-censoring 
was studied by \cite{Mandelbaum2013Data} using call-center data; we refer to their estimator as \emph{Method 1}\footnote{Detailed formulas of Methods 1 and 2 are provided in Appendix EC.\ref{app:mandelbaum_yefenot_formulas}}. In contrast to call centers, the contact centers' data regarding patience is also left-censored due to silent abandonment. Indeed, when a customer abandons the queue without indicating they have done so---a silent abandonment---the wait time equals the virtual wait time (itself uncensored); thus, the wait time is an upper bound for the customer's real patience, which was clearly less than the wait time. \cite{Yefenof2018} addressed this situation, motivated by LWBS in EDs; we refer to their estimator as \emph{Method 2}.
As we mentioned, we have complete data in no-write-in-queue systems; therefore, Method 2 can be used to estimate customer patience since similar conditions exist. We apply Method 2 to no-write-in-queue data in Section \ref{sec:managerial}. But write-in-queue systems require a new methodology for patience estimation because of the added complexity missing data brings to customer classification. This different approach is the focus of this section. In Section \ref{subsec:EM_model}, we develop our expectation-maximization (EM) algorithm for estimating customer patience in write-in-queue systems, and in Section \ref{subsection:EM_validation}, we validate its accuracy, sensitivity, and robustness. Later, in Section \ref{subsection:EM_covaritates}, we extend this EM algorithm to a more general model that can estimate how different variables, like the amount of customer writing while waiting, impact customer's patience. 

\subsection{The EM Algorithm: Model Assumption and Formulation}
\label{subsec:EM_model}
The problem of missing information on uSab customers stems from not knowing whether they received one-exchange service (Sr1), in which case their patience would be right-censored, or they silently abandoned, in which case their patience would be left-censored. Nonetheless, we know that the time these customers waited in the queue, and hence their virtual wait time, is uncensored. 

Following the formulation of \cite{Yefenof2018}, let $\tau$ be customer-patience time (failure time) and assume it has a cumulative distribution function (cdf) $F$ and a probability distribution function (pdf) $f$. Assume that $\tau\sim exp(\theta)$. This assumption follows  \cite{Brown2005}, who showed using call-center data (with no delay announcements) that patience distribution has an exponential tail. We show, in Section \ref{sec:managerial}, that queueing models with exponentially distributed patience fit no-write-in-queue contact-center data better than do queueing models with generally distributed patience, providing further support for our assumption.
Let $W$ be the virtual wait time (censoring time)---the time the customer is required to wait by the system---and assume it has a cdf $G$ and a pdf $g$.
We know from queueing theory that in overloaded systems, like the contact centers we are investigating, wait time is close to exponentially distributed \citep{Kingman1962}. In addition, \cite{Brown2005} showed that in call centers with no delay announcements, virtual wait time is close to exponentially distributed. In our dataset, we also have no delay announcements that may change customer patience while waiting \citep{ibrahim2018sharing}; hence, we can make a realistic assumption that the virtual waiting times are exponentially distributed. 
Formally, assume that $W\sim exp(\gamma)$.
Let $\Delta$  be an indicator for the case where the customer lost patience before the agent replied: $\Delta \triangleq 1_{\{\tau\leq W\}}$. Conversations in which information regarding $\Delta_i$ is missing are assigned a null value. 
Let $Y$ be a random variable indicating whether the customer will inform the system when abandoning. We assume that $Y\sim Bernoulli(q)$, where $q$ is the probability that the customer will inform the system when abandoning; formally, $q \triangleq Pr\left\{ Indicate\;abandonment\right\}$.

Assume that $W$ and $\tau$ are independent, as is frequently done in right-censoring survival analysis (e.g., \citealp{smith_2002,Mandelbaum2013Data,Yefenof2018}). Moreover, this is a natural assumption in contact centers since patience is decided by the individual customer while the virtual wait time is decided by the company. This is indeed the case in our contact centers where no delay information, such as their place in queue, is provided to the customer.
Additionally, we assume that $Y$ and $W$ are independent. That is, the decision of a customer to indicate whether they abandoned the queue is independent of their wait time. For example, a customer might tend to leave windows open in the computer even if they are not using them; therefore, this tendency would be independent of the wait time. The independence assumption of $Y$ and $W$ is for tractability reasons; currently, we do not have evidence to support this assumption and suggest that it be relaxed in future research. Finally, let $U$ be the system's observed time. For each arriving customer $i$, we observe the vector of data $(U_{i},Y_{i},\Delta_{i})$, $i=1,...,n$. We also assume that our sample of customers is homogeneous: all customers have the same  patience parameter $\theta$ and abandonment indicator $q$.

Summarizing, our model rests on the following assumption:
\begin{assumption} \label{assumption:1}
The EM algorithm developed in Section \ref{subsec:EM_model} assumes:
(a) Patience time is $\tau\sim exp(\theta)$.
(b) Virtual wait time is $W\sim exp(\gamma)$.
(c) Customer abandonment indicator is $Y\sim Bernoulli(q)$.
(d) $W$ and $\tau$ are independent.
(e) $Y$ and $W$ are independent.
(f) Customers are homogeneous.
\end{assumption}

\subsubsection{Customer Classes with Complete Data.}
\label{subsec:ClassComplete}
In Table \ref{tbl:CompeteDtaNotation}, we follow \cite{Yefenof2018} in formally defining three customer classes under the assumption of complete data regarding which customers abandoned. The table identifies each customer class by type, notation indicator, formal definition of that indicator (based on values $\Delta$ and $Y$), what variable is observed in $U$, and what variable is censored by $U$ and in what direction. 

\begin{table}[!htb]
  \centering
  \caption{Classes of Customers: Complete Data}
  \begin{small}
  \begin{tabular}{lcccccl} 
  \toprule
  Class Type & Notation Indicator & Formal Definition & $\Delta$ & $Y$ & Observed - $U$ & Censored (direction)\tabularnewline
  \midrule 
  Service &  $C_{1}=1$ & $1-\Delta$ & 0 & 0 & $W$ & $\tau$ (right-censored)\tabularnewline
  Kab & $C_{2}=1$ & $Y\Delta$ & 1 & 1 & $\tau$ & $W$ (right-censored) \tabularnewline
  Sab & $C_{3}=1$ & $(1-Y)\Delta$ & 1 & 0 & $W$ &$\tau$ (left-censored) \tabularnewline
  \bottomrule

  \end{tabular}
  \end{small}

  \label{tbl:CompeteDtaNotation}
\end{table}

Remark: Note that the data of no-write-in-queue systems is complete. Therefore, we can categorize the conversations into the above three classes with complete certainty.

\subsubsection{Customer Classes with Missing Data.} \label{subsec:ClassesMissingDa}
Due to the problem of missing data on the uSab conversations in the write-in-queue systems, we are not able to categorize all the conversations into just one of the classes we defined in Section \ref{subsec:ClassComplete}. Therefore, we need to formulate additional class indicators. Let $M$ denote the customer classes in a system in which there is missing data on which individual customers abandoned. These classes are defined in Table \ref{tbl:MissingDtaNotation}.
\begin{table}[!htb]
  \centering
  \caption{Classes of Customers: Missing Data}
  \begin{small}
  \begin{tabular}{lcccccl} 
  \toprule
  Class Type & Notation Indicators & Formal Definition & $\Delta$ & $Y$ & Observed - $U$ & Censored (direction) \tabularnewline
  \midrule 
  Service & $C_1=1; M=1$ & $1-\Delta$ & 0 & 0 & $W$ & $\tau$ (right-censored)\tabularnewline
  Kab & $C_2=1; M=2$ & $Y\Delta$ & 1 & 1 & $\tau$  & $W$ (right-censored) \tabularnewline
  uSab: &  &  &  &  &  &  \tabularnewline
  ~~~uSab is Sab & $C_3=1;M=0$ & $(1-Y)\Delta$ & null & 0 & $W$  & $\tau$ (left-censored) \tabularnewline
  ~~~uSab is Sr1 & $C_1=1;M=0$ & $1-\Delta$ & null & 0 & $W$  & $\tau$ (right-censored) \tabularnewline
  \bottomrule

  \end{tabular}
  \end{small}

  \label{tbl:MissingDtaNotation}
\end{table}
%

\subsubsection{The EM Algorithm Formulation.}
\label{subsection:EM_algorithm}
The EM algorithm estimates the following parameters simultaneously: the rate at which customers lose patience, $\theta$; the probability of informing the system when abandoning, $q$; and the rate of the virtual wait time distribution, $\gamma$. The optimization problem is defined to maximize the likelihood function, which measures the probability that the observations are given from the assumed distributions given the parameters ($\theta, q, \gamma$). We write the likelihood of the observed data $D\triangleq \{(U_{i},Y_{i},\Delta_{i}),$  $i=1,...,n\}$ as follows:
\begin{equation}
\begin{aligned} \label{eq:Likelihood}
L(D;\theta,q,\gamma)= & \prod_{i=1}^{n}\left\{ e^{-\theta U_{i}}\gamma e^{-\gamma U_{i}}\right\} ^{C_{1}^{i}}\left\{ q\theta e^{-\theta U_{i}}e^{-\gamma U_{i}}\right\} ^{C_{2}^{i}}\left\{ (1-q)(1-e^{-\theta U_{i}})\gamma e^{-\gamma U_{i}}\right\} ^{C_{3}^{i}}\\
= & \prod_{i=1}^{n}\left\{ e^{-\theta U_{i}}\gamma e^{-\gamma U_{i}}\right\} ^{1-\Delta_{i}}\left\{ q\theta e^{-\theta U_{i}}e^{-\gamma U_{i}}\right\} ^{\Delta_{i}Y_{i}}\left\{ (1-q)(1-e^{-\theta U_{i}})\gamma e^{-\gamma U_{i}}\right\} ^{(1-Y_{i})\Delta_{i}.}
\end{aligned}
\end{equation}
The function is formulated following \cite{Yefenof2018}: the first part is for the served customer ($C_{1}^{i}=1$), where we multiply the survival function of the customer patience ($1-F_{\tau}\left(u\right)$) by the pdf of the customer's wait time. The second part is for the Kab customer ($C_{2}^{i}=1$), where we multiply the probability of informing when abandoning by the pdf of the customer patience and the survival function of the customer's wait time ($1-G_{W}\left(u\right)$). Finally, the third part is for the Sab customer ($C_{3}^{i}=1$), where we multiply the probability of not informing when abandoning by the cdf of the customer patience and the pdf of the customer's wait time.  

However, this likelihood function depends on knowing the complete data. Recall that some of the observations belong to the class $M=0$ since they have missing data in $\Delta$. Therefore, we cannot find the parameters by simply solving the maximization problem. 
Instead, we need to formulate an EM algorithm (see Algorithm \ref{EM}), a well-known computing strategy for dealing with problems of missing data and censoring
\citep{little2002statistical}. The algorithm estimates the parameters ($\theta, q$, $\gamma$) 
using Theorems \ref{Theorem1} and \ref{Theorem2}. Specifically, it estimates starting parameter values and subsequently iterates between the expectation step (E-step)---using Theorem \ref{Theorem1}---and the maximization step (M-step)---using Theorem \ref{Theorem2}---and updates these estimators until convergence.  
In the $t$th iteration, the E-step consists of finding a surrogate function (given in Equation~\eqref{eq:loglikelihoodMisg}) that is a lower bound on the log-likelihood function (given in Equation~\eqref{eq:loglikelihood}) and is tangent to the log-likelihood at $(\widehat{\theta^{(t)}},\widehat{q^{(t)}},\widehat{\gamma^{(t)}})$. In practice, it is enough to compute the expectation of the log-likelihood given the information of the previous iteration, which is presented in Equation \eqref{eq:ci} of Theorem \ref{Theorem1}.

\begin{equation}
\begin{aligned}\label{eq:loglikelihoodMisg}
l(D,\theta,q,\gamma) & =\sum_{i=1}^{n}\left\{ \left(\widehat{C_{1,t}^{i}}\right)\left(\log\gamma-\gamma U_{i}-\theta U_{i}\right)\right\} \\
 & +\sum_{i=1}^{n}\left\{ \left(\widehat{C_{2,t}^{i}}\right)\left[\log\theta-\theta U_{i}-\gamma U_{i}+\log q\right]\right\} \\
 & +\sum_{i=1}^{n}\left\{ \left(\widehat{C_{3,t}^{i}})\right)\left[\log\left(1-q\right)+\log(1-e^{-\theta U_{i}})+\log\gamma-\gamma U_{i}\right]\right\} .
\end{aligned}
\end{equation}

\begin{algorithm}
\SetAlgoLined
\KwResult{$\widehat{\theta^{(t+1)}}$, $\widehat{q^{(t+1)}}$ and $\widehat{\gamma^{(t+1)}}$.}

\vspace{3pt}
Initialization: For every customer $i$, use Equation \eqref{eq:ci} to calculate $\widehat{C_{1,0}^{i}}$ and $\widehat{C_{2,0}^{i}}$ and $\widehat{C_{3,0}^{i}}=\hat{\pi}_{i} 1_{\{M^{i}=0\}}$, where $\hat{\pi_{i}}\in\left[0,1\right]$ is chosen randomly. 
To obtain the starting
parameters, $(\widehat{\theta^{(1)}},\widehat{q^{(1)}},\widehat{\gamma^{(1)}})$, solve Equations \eqref{eq:PartialDTheta} and \eqref{eq:PartialDGammaq}, respectively. Set $t=0$.\\
\vspace{6pt}
 \While{ $ \mid\widehat{\theta^{(t)}}-\widehat{\theta^{(t+1)}}\mid+\mid\widehat{q^{(t)}}-\widehat{q^{(t+1)}}\mid+\mid\widehat{\gamma^{(t)}}-\widehat{\gamma^{(t+1)}}\mid>\epsilon$}{
  E-step: Compute given the observed data
  $D=\{(U_{i},Y_{i},\Delta_{i})$ $i=1,...,n\}$ and the current estimations of the parameters $(\widehat{\theta^{(t)}},\widehat{q^{(t)}},\widehat{\gamma^{(t)}})$, $\widehat{C_{j,t}^{i}}$, $j=1,2,3$ $\forall i=1,...,n$ using Eq.\ \eqref{eq:ci}. \\
\vspace{3pt}
  M-step: Maximize to obtain $(\widehat{\theta^{(t+1)}},\widehat{q^{(t+1)}},\widehat{\gamma^{(t+1)}})$.\
  That is, update the estimations of the parameters using Equations \eqref{eq:PartialDTheta} and \eqref{eq:PartialDGammaq}, respectively.\\
  \vspace{3pt}
  Update $t\leftarrow t+1$.
 }
 \caption{The EM Algorithm}
 \label{EM}
\end{algorithm}


\begin{theorem} \label{Theorem1}
Under Assumption \ref{assumption:1}, $\widehat{C_{1,t}^{i}}$, $\widehat{C_{2,t}^{i}}$, and $\widehat{C_{3,t}^{i}}$ are given by
\begin{equation} 
\begin{split}
\label{eq:ci}
&\widehat{{C_{1,t}^{i}}}=(1-\widehat{C_{3,j}^{i}})1_{\{M^{i}=0\}}+1_{\{M^{i}=1\}};  \\
&\widehat{C_{2,t}^{i}}=1_{\{M^{i}=2\}}; \\
&\widehat{C_{3,t}^{i}}=1_{\{M^{i}=0\}}\left(1-e^{-\widehat{\theta^{(t)}}U_{i}}\right).
\end{split}
\end{equation}
\end{theorem}
The proof is given in Appendix \ref{sec:appendixEproof}.

The notations $\widehat{C_{1,t}^{i}}$, $\widehat{C_{2,t}^{i}}$, and $\widehat{C_{3,t}^{i}}$ represent the probabilities (weights) for the $i$th customer to belong to class $C_1, C_2,$ or $C_3$, respectively, given the parameters from iteration $t-1$, $(\widehat{\theta^{(t)}},\widehat{q^{(t)}},\widehat{\gamma^{(t)}})$, and the observed data. Note that the EM algorithm's update of the weights with missing data in the $t-1$ iteration, $\widehat{C_{j,t-1}^{i}}$
$j=1,3$, is different for each observation $i$ in the data class $M^{i}=0$. 
That is, $\widehat{C_{3,t-1}^{i}}$ need not equal $\widehat{C_{3,t-1}^{k}}$, given that $M^{i}=M^{k}=0.$
In the M-step of the $t$th iteration, $(\widehat{\theta^{(t+1)}},\widehat{q^{(t+1)}},\widehat{\gamma^{(t+1)}})$ are found (in Equations \eqref{eq:PartialDTheta} and \eqref{eq:PartialDGammaq}, respectively) to be the maximizers of the surrogate function in Equation \eqref{eq:loglikelihoodMisg}.

\begin{theorem} \label{Theorem2}
Under Assumption \ref{assumption:1}, the parameters $\widehat{q^{(t+1)}}$,  $\widehat{\gamma^{(t+1)}}$ are given by
\begin{equation}
\begin{split} \label{eq:PartialDGammaq}
&\widehat{q^{(t+1)}}=\left\{ \sum_{i=1}^{n}\widehat{C_{2,t}^{i}}\right\} \left\{ \sum_{i=1}^{n}\left(1-\widehat{C_{1,t}^{i}}\right)\right\} ^{-1},  \\
&\widehat{\gamma^{(t+1)}}=\left\{ \sum_{i=1}^{n}\left(1-\widehat{C_{2,t}^{i}}\right)\right\} \left\{ \sum_{i=1}^{n}U_{i}\right\} ^{-1}, 
\end{split}
\end{equation}
and the parameter $\widehat{\theta^{(t+1)}}$ is given as a solution to the following equation:
\begin{equation}\label{eq:PartialDTheta}
\widehat{\theta^{(t+1)}}\left\{ \sum_{i=1}^{n}\left(\widehat{C_{3,t}^{i}}-1\right)U_{i}\right\} +\sum_{i=1}^{n}\widehat{C_{2,t}^{i}}+\widehat{\theta^{(t+1)}}\left\{ \sum_{i=1}^{n}\widehat{C_{3,t}^{i}}\frac{U_{i}e^{-\widehat{\theta^{(t+1)}}U_{i}}}{1-e^{-\widehat{\theta^{(t+1)}}U_{i}}}\right\} =0.
\end{equation}

\end{theorem}
The proof is given in Appendix \ref{sec:appendixMproof}.

We repeat the E-step and the M-step until convergence for some predetermined $\epsilon>0$. 
The procedure ends when we find a maximum of  the likelihood function that yields estimators for the parameters $(\widehat{\theta^{(t+1)}},\widehat{q^{(t+1)}},\widehat{\gamma^{(t+1)}})$.

Regarding convergence of our EM algorithm, by Theorem 8.1 of \cite{little2002statistical}, each iteration of the EM algorithm increases the likelihood function. Then, because the smoothness conditions hold for the exponential distribution the sequence of estimators converges to a stationary point by Theorem 8.2.

%
More details on the EM algorithm and proofs are provided in Appendix \ref{sec:proofE}.


\subsection{Validation of the EM Algorithm}
\label{subsection:EM_validation}
We perform several performance evaluations to validate the use of our EM algorithm in practice. Due to space constraints, most of the validations appear in Appendix \ref{app:validation}. In Appendixes \ref{subsubsec:EMaccuracy} and \ref{sec:ApendixEMqueueing}, we compare the accuracy of the EM algorithm to previous methods of estimating customer patience, \cite{Mandelbaum2013Data} (Method 1) and \cite{Yefenof2018} (Method 2). These basic comparisons use simulated data and demonstrate that our EM algorithm provides the most accurate estimation of all parameters ($\theta, \gamma$ and $q$), regardless of the level of load in the system. (The simulations in Appendix \ref{sec:ApendixEMqueueing} use simulated data from a queueing model that relaxes some of the conditions under Assumption \ref{assumption:1}. We find that the differences with the other methods are more pronounced with that simulated data.) 
Here, in Section \ref{subsubsec:EMValidation_data}, we validate the accuracy of the EM estimators using \emph{real data}, concluding that the EM algorithm is the only method that provides an accurate estimation of customer patience in reality. 
In Appendix \ref{subsubsec:Sensitivity}, we examine the algorithm's sensitivity under the initial conditions. We show that the parameter estimations are stable and do not change when different initial values are inserted in the EM algorithm. This suggests that one does not need to use the output of the classification model we developed in Section \ref{sec:prob_abnd}  (or any model with similar sensitivity and specificity proportions) as starting probabilities in the EM algorithm. Finally, in Appendix \ref{sec:ApendixEMqueueing}, we provide additional robustness tests.
(In all tests throughout this paper we set $\epsilon =10^{-6}$.)

As mentioned, the EM algorithm can cope with the missing data, but Methods 1 and 2 cannot.
To use them for these comparisons, we must make certain assumptions on how they cope with uSab conversations ($M=0$). To apply  \cite{Yefenof2018}, we have two options of how to classify uSab conversations: either as served (Sr1) customers ($C_{1}=1$) or as Sab customers ($C_{3}=1$). 
To apply the method of \cite{Mandelbaum2013Data}, we can classify uSab conversations as either served customers ($C_{1}=1$) or Kab ($C_{2}=1$), since this method cannot deal with left-censored conversations. These options result in all possible (four) baseline methods used for our accuracy comparisons.

\subsubsection{Estimating Patience in Write-in-queue System Data: Accuracy and Robustness Tests.}
\label{subsubsec:EMValidation_data}
As mentioned, most of the tests provided in the appendixes use simulated data that clearly adhere to our model assumptions. In this section, we perform tests that rely on real data that may not adhere to those assumptions. This will provide us with greater confidence in applying the method we developed here in practice. Using the write-in-queue contact-center dataset described in Section \ref{sec:dataAndReseatch}, we compare the EM algorithm to the other four methods for estimating customer patience.

The results are presented in Table  \ref{tbl:AvPatience}. The differences between the patience estimations (rows 2--6) are huge (13--188 minutes). Note that the estimations are consistent with other accuracy tests, where Methods 1 and 2 overestimate and underestimate customer patience depending on the variation of the method.

The main challenge we are confronted with is the lack of ground truth---we do not know the true value of customer patience. We overcome this challenge by using the manually tagged data described in Section  \ref{sec:prob_abnd_WQ}.  
Since this data is tagged, it has complete information on which customers abandoned, allowing us to apply the method of \cite{Yefenof2018}. The resulting estimation of customer patience, based on that labeled data, is 81.9 minutes (row 1 of Table \ref{tbl:AvPatience}). This is very similar to the EM algorithm's estimation of customer patience that is based on the monthly data: 81.11 minutes (row 6 of Table \ref{tbl:AvPatience}). On the other hand, it is very far from the estimations done using the other methods. Therefore, we can conclude that the EM algorithm (Algorithm \ref{EM}) is the only one able to cope with the missing data and obtains an accurate estimation of customer patience.

Going back to Table \ref{tbl:AvPatience}, we notice the large bias that missing data generates in the estimations. When we ignore both silent abandonment and missing data by regarding all uSab ($M=0$) as service ($C_{1}$=1) and by estimating customer patience using either Method 1 or 2,  we overestimate patience by twice or more (rows 2 and 3 of Table \ref{tbl:AvPatience}). Note that this is many companies' current practice. They use  Method 1 (row 2) while ignoring the concept of silent abandonment that creates left-censoring and missing data.
A more advanced company may have a better understanding of its system and an awareness of silent abandonment. However, if it is still unaware of the existence of missing data, it will consider all conversations in class $M=0$ to be Sab conversations ($C_{3}=1$) and apply either Method 1 (ignoring left-censoring) or Method 2 (not ignoring the left-censoring). In both cases, it underestimates customer willingness to wait (rows 4 and 5).

One might comment on our finding that the customers in write-in-queue systems are willing to wait for more than 1 hour (row 1 of Table \ref{tbl:AvPatience}). We think such enduring patience is reasonable for four reasons: (a) When reading the content of the conversations, we see that in this particular contact center customers receive an automatic message instructing them to ``go on with their daily activities" (while waiting for a reply) and to address the service as if ``talking to a friend." These customers therefore expect longer waits and adjust their patience accordingly. (b) Write-in-queue systems are used to support the ongoing relationship between customers and companies. As a result, they have a high proportion of returning customers that are expected to have realistic expectations of the virtual wait time, which was found to be 8.77 minutes. The fact that customer patience outlasts the virtual wait time is consistent with similar results from call centers \citep{Brown2005}. (c) \cite{Mandelbaum2013Data} showed that customers are willing to wait around two (or more) times longer than their service requirement. Recall that here service time is 46.34 minutes, which fits our findings well. (d) The ability to write while waiting influences patience, as we will show and explain later in Section \ref{sec:patience_coef}. 

\begin{table}[!htb]
  \centering
  \caption{Comparison of Estimations of Average Customer Patience: Write-in-queue System Dataset (May 2017)}
  \begin{scriptsize}
  \begin{tabular}{llcc} 
  \toprule
  Row & Method & Avg.\ Patience (Minutes) & Data used\tabularnewline
  \midrule
  1 & Method 2---Using sample of labeled conversations & ~81.90 & Sample$^*$\tabularnewline
  2 & Method 1---Uncertain silent abandonment is service & 166.42 & Full data\tabularnewline
  3 & Method 2---Uncertain silent abandonment is service & 188.07& Full data\tabularnewline
  4 & Method 1---Uncertain silent abandonment is abandonment  & ~28.27& Full data\tabularnewline
  5 & Method 2---Uncertain silent abandonment is silent abandonment & ~13.17& Full data\tabularnewline
  6 & {EM} & {~81.11}& Full data\tabularnewline
\bottomrule
\multicolumn{4}{l}{$^*$The sample includes 2500 conversations, where all uSab conversations are manually labeled to Sab or Sr1 classes.}
  \end{tabular}
  \end{scriptsize}

  \label{tbl:AvPatience}
\end{table}


\section{Incorporating Silent Abandonment into a Queueing Model and Managerial Implications} \label{sec:managerial}

In this section, we analyze how the phenomenon of silent abandonment affects system efficiency and what decision-makers can do about it. As explained, silent abandonment affects system efficiency in two ways: (a) the Sab customer holds a service slot within the concurrency system, preventing other customers from entering service while idling the agent who is waiting for the customer's response. (b) The agent may waste time on solving the no longer relevant problem of the Sab customer.  Both forms of system inefficiency reduce the system's capacity in high-load moments, when available capacity is most crucial. In our no-write-in-queue and write-in-queue system datasets, system capacity is reduced  by 1.7\% and 15.3\%, respectively. According to queueing theory, such a reduction in agent availability should have a large impact on system performance in overloaded systems \citep{Koole2002}. The aims of the present section are to, first, introduce a queueing model that takes the Sab phenomenon into account and show that such a model is able to predict contact-center performance measures better than those that neglect to account for silent abandonment  (\S\ref{sec:model_fitting}). 
Then, we use the model to analyze how much the loss of capacity due to Sab harms system performance, and we evaluate the staffing changes needed when Sab occurs (\S\ref{sec:staffing}). Finally, we discuss several ways one might avoid such a problem (\S\ref{sec:avoid_Sab}).

\subsection{Service System  with Customer Silent Abandonment: Evaluation of Queueing Model Fitting to Data}
\label{sec:model_fitting}
\begin{figure}[!htb]
\centering
\includegraphics[width=0.40\textwidth]{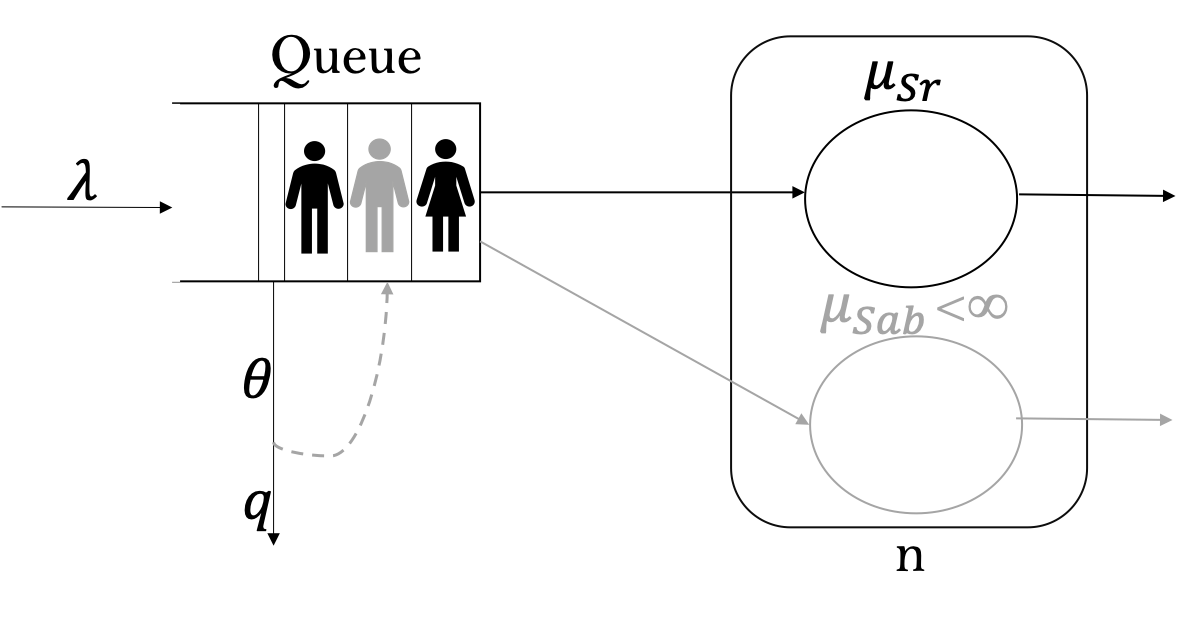}
\caption{A Queueing Model with Silent Abandonment}
\label{fig:QmSab}
\end{figure}

We propose the queueing model presented in Figure \ref{fig:QmSab} to capture the phenomenon of Sab. We assume that the arrival rate is according to a Poisson process with rate $\lambda$. Customers entering the queue have finite patience that is exponentially distributed at rate $\theta$. The probability that an abandoning customer will indicate their abandonment is denoted by $q$.  Customers who don't provide that indication stay in the queue and are assigned to a service agent (when one becomes available). Queueing policy is first-come, first-served (FCFS). The company can provide service to $n$ customers in parallel; that is, there are $n$ service slots of statistically identical agents. Service time is exponentially distributed with rate $\mu_{Sr}$ for served customers (those who belong to class $C_1$) and rate $\mu_{Sab}$ for Sab customers (those who belong to class $C_3$).  This model is very similar to the Erlang-A (M/M/N+M) model, with the important difference that a customer that abandons the queue but does not notify the system of their abandonment is assumed to be in the queue (e.g., the gray customer in Figure \ref{fig:QmSab}) and, when assigned to an agent, receives some service time, albeit at a different service rate. This enables us to capture the loss of capacity resulting from Sab.

To verify that this queueing model is of merit---it fits real data well and much better than classical models that ignore the Sab phenomena---we fit the model to the no-write-in-queue system dataset described in Section \ref{subsec:Chat}.
The Erlang-A (M/M/N+M) model \citep{Mandelbaum2007} is used as a baseline (see Model (1) below). 
We compare the following four variants of fitting a queueing model to the no-write-in-queue data:

\begin{lyxlist}{00.00.0000}
\item[\emph{Model (1):}]  A classic Erlang-A queueing model that ignores Sab by considering Sab customers as served. Hence, Sab customers are considered as served in the parametric estimations of customer patience and service rate, and in calculating performance measures.  Labeled as \textit{``(1) Ignoring Sab."} 
\item[\emph{Models (2) and (3):}] A queueing model with Sab,  no loss of capacity due to Sab, and that considers Sab in the estimation of customer patience, that is, the model in Figure \ref{fig:QmSab} with $\mu_{Sab}=\infty$. We check two versions of this model: one with a nonparametric estimation of customer patience (Model (2)) and the other with a parametric estimation of customer patience (Model (3)). Labeled as \textit{``(2) Sab as left-censored, nonparametric"} and  \textit{``(3) Sab as left-censored, parametric,"} respectively.

\item [\emph{Model (4):}] A queueing model with Sab, loss of capacity due to Sab, and that considers Sab in the parametric estimation of customer patience, that is, the model in Figure \ref{fig:QmSab}. Labeled as \textit{``(4) Considering Sab as time-consuming."}

\end{lyxlist}

In Model (1), we estimate customer patience based on \cite{Mandelbaum2013Data} (Method~1), and the service rate is calculated by averaging the service time of all the customers that were assigned to an agent, regardless of whether they were served or silently abandoned the queue. These estimations resulted in average patience of 33.9 minutes and average service rate, $\mu=\mu_{All}$, that changes over time and is provided in Table \ref{tbl:ParametersHour}. 
In Models (2) and (3), we estimate customer patience based on \cite{Yefenof2018} (Method 2). The  average patience is estimated to be 7.8 or 2 minutes for the nonparametric and parametric estimation methods, respectively. In both models, service time is calculated only for served customers: $\mu=\mu_{Sr}$ (see Table \ref{tbl:ParametersHour}); Sab customers' service time is 0. 
Finally, in Model (4), we estimate customer patience based on the parametric version of \cite{Yefenof2018} (Method 2) (i.e., 2 minutes). Service time is calculated separately for served customers, $\mu=\mu_{Sr}$, and Sab customers, $\mu=\mu_{Sab}$ (see Table \ref{tbl:ParametersHour}). 
(Note that using our EM algorithm in the case of Models (3) and (4) gives the same customer-patience estimation, since the no-write-in-queue system has complete information.)
The simulation parameters $\lambda_t$, $\mu_t$, and $n_t$ were estimated for each hour over the month (see Appendix \ref{sec:apexMOPS}), while the parameters of customer-patience distribution as well as $q=0.7$ were kept constant over time. 
Note that $n_t$ is the number of available slots: the number of online agents times a fixed concurrency level of 3 customers per agent. (As mentioned in Section \ref{subsec:Chat}, the concurrency level is the maximal number of customers that can be served in parallel, such that if all the slots are occupied and an additional customer enters the system, they will need to wait in the queue.) 

\begin{table}[!htb]
 \centering
  \caption{RMSE between Queueing Models and the No-write-in-queue System Dataset (Weekdays, February 2017)}
  \begin{scriptsize}
  \begin{tabular}{lccccc} \toprule
  Performance & Avg. [SD] & (1) Ignoring & (2) Sab as left-censored, &(3) Sab as left-censored, &(4) Considering Sab \\
  Measure && Sab & nonparametric & parametric & as time-consuming\\
  \midrule
  $P\{\text{Wait}>0\}$ &0.59 [0.03]& \textbf{0.27} & 0.28 &0.31 & \textbf{0.27} \tabularnewline
  $P\{\text{Ab}\}$ &0.23 [0.04] & 0.12$^*$ & 0.09 & 0.08 & \textbf{0.07}  \tabularnewline
  E[Queue] & 1.72 [0.46] & 3.27 & 1.18 &0.96 & \textbf{0.87} \tabularnewline
  E[Wait] (min)& 2.41 [0.72]& 2.83 & 1.26 &1.17 & \textbf{1.05} \tabularnewline
  E[Wait$|$Served] (min)& 2.04 [0.58]& 3.31$^*$ & 1.39 & 1.06 &\textbf{1.04}  \tabularnewline
  \bottomrule
  \multicolumn{6}{l}{$^*$To provide a fair comparison, RMSE of this performance measure was calculated with respect to Kab only.}
  \end{tabular}
  \end{scriptsize}
 \label{tbl:RMSEQ}
\end{table}
We ran simulations for four weeks duration with 10 repetitions and calculated the performance measures over each hour. We compared the differences between the simulated performance measures of these four queueing models and the real performance measures calculated from the dataset (shown in Appendix \ref{sec:apexMOPS}). Table \ref{tbl:RMSEQ} presents the differences using the root mean square error (RMSE) score.
Model (1) was designed to provide a baseline of the fit between the model and the data when the phenomenon of silent abandonment is ignored altogether. We see that the fit of the queueing model to the data in this case is the worst among all the compared models. 
In Models (2) and (3),  the company understands that silent abandonment occurs and that the data is left-censored but ignores the impact of Sab on the available capacity. By comparing the two versions, we note that the fit of the parametric model to the data is much better than that of the nonparametric model. This gives us higher confidence that the assumptions we made in Assumption \ref{assumption:1} are actually very reasonable for the contact-center environment. 
Finally, we observe that Model (4), which considers Sab both in terms of patience estimation and in terms of efficiency loss, is the best fit. 

\begin{figure}[!htb]
\centering
\subfigure[As a Function of Time (Models (3) and (4) and Real)]{
\includegraphics[width=0.5\textwidth]{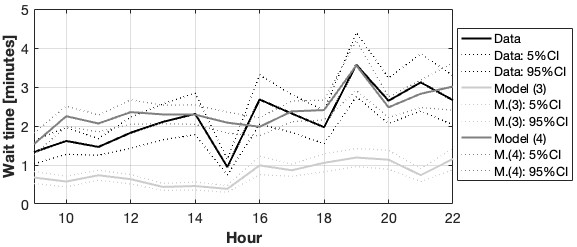} \label{fig:E_W_Models45}
} 
\subfigure[As a Function of $1/\mu_{Sab}$ (Model (4))]{
\includegraphics[width=0.46\textwidth]{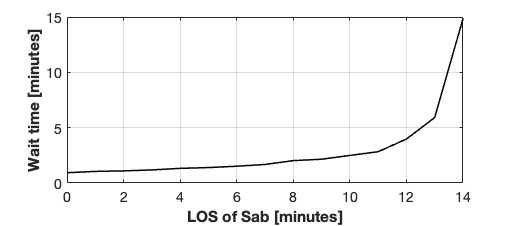} \label{fig:E_WAIT_musab}
} 
\caption{Estimations of E[Wait]}
\label{fig:comp_Models45}
\end{figure}

Figure  \ref{fig:E_W_Models45} compares the estimation of E[Wait] for Models (3) and (4), and the real E[Wait] in the dataset of the no-write-in-queue system, as a function of the hour of the day (with 95\% confidence intervals). We clearly see that Model (3) underestimates customers' wait time relative to Model (4). Comparing Models (3) and (4) enables us to understand the impact that capacity loss caused by the Sab customers has on performance measures. If the company is able to eliminate all capacity loss (5\% in our case), the expected wait time of all customers would be reduced by 1.6 minutes (67\% in absolute percentage), the expected wait time of served customers would be reduced by 1.5 minutes (83\%), the probability of waiting by 0.03 (8\%), the probability of abandonment by 0.04 (16\%), and the expected number of people waiting in queue---E[Queue]---by 0.16 (21\%).

Figure \ref{fig:E_WAIT_musab} illustrates the implications of identification time on performance.
Here, we simulated Model (4) with various LOS of Sab customers ($1/\mu_{Sab}$); as $1/\mu_{Sab}$ increases, it takes longer to understand that the Sab customer indeed abandoned the queue. For this graph, we use the parameters on a typical Monday (13:00--14:00), where 
$\lambda=56$ customers per hour, LOS of served customers is $12.3$ minutes, $q=0.7$, and average patience is 2 minutes ($\theta=30$ customers per hour). We notice in this figure that the effect of $1/\mu_{Sab}$ is not linear: after 12 minutes, the impact on waiting time increases steeply. Expecting the company to detect Sab in 12 minutes is realistic; in our datasets, one company is well below this number, while the other one is above (\S\ref{sec:prob_abnd}).
With long waits, more customers will abandon silently, creating a vicious cycle where the company loses customers and money. 



\subsection{Impact of Silent Abandonment on Staffing}
\label{sec:staffing}

\begin{figure}[!htb]
\centering
\includegraphics[width=0.52\textwidth]{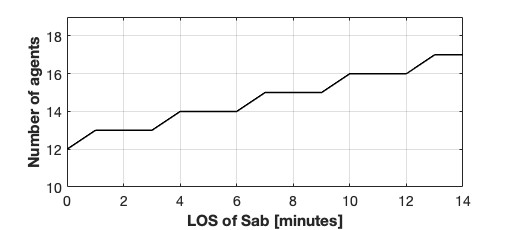}
\caption{Number of Agents Needed to Achieve Average Wait of 1 Minute as a Function of $1/\mu_{Sab}$. Simulation Study with Parameters of Hour 13:00 on a Typical Monday}
\label{fig:StaffingLOSsab}
\end{figure}

Figure \ref{fig:E_W_Models45} shows the fundamental effect Sab has on performance measures. A company wanting to reduce the long waits Sab customers create must hire more contact-center agents to compensate for the efficiency loss those customers cause. 
Indeed, the number of agents needed is a direct function of the time it takes to detect a Sab customer and the concurrency level each agent handles. Here, we assume a constant concurrency level of three  customers per agent. 
Figure \ref{fig:StaffingLOSsab} calculates the number of agents the company needs as a function of  $1/\mu_{Sab}$ (parameters as in \ref{fig:E_WAIT_musab}) so that the average wait time is below one minute. The longer it takes the company to detect their Sab customers, the higher the efficiency loss, and the more agents needed to handle the conversations. 
Specifically, according to Figure \ref{fig:StaffingLOSsab}, reducing two minutes from the time it takes the company to detect a Sab customer (e.g., from 4 minutes to 2 minutes), can save the company 7.1\% of its workforce. 

\begin{figure}[!htb]
\centering
\subfigure[$P\{Wait\}$ as a Function of $\beta$ ]{
\includegraphics[width=0.48\textwidth]{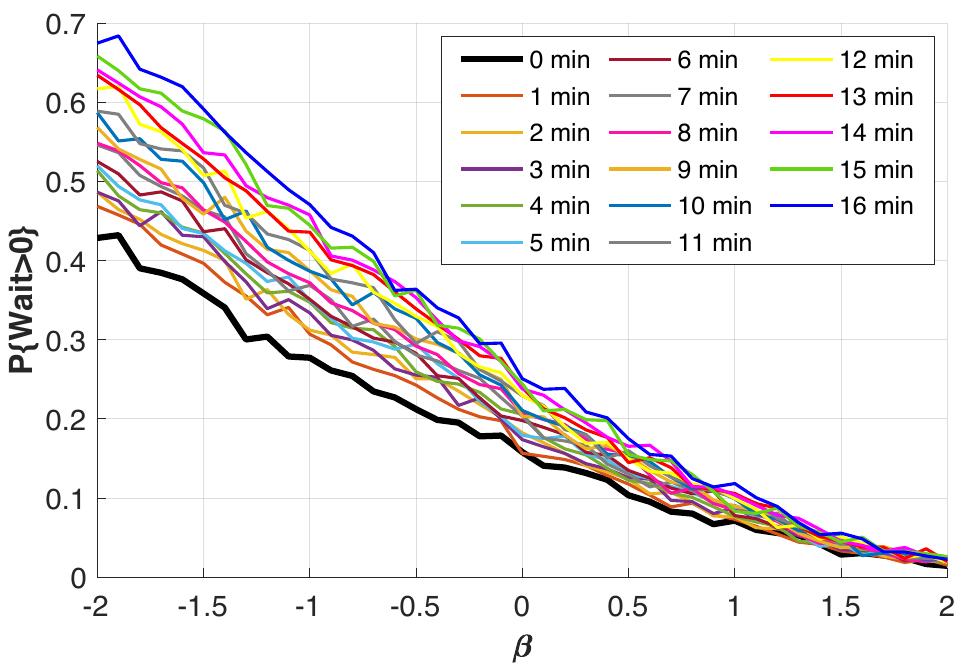} \label{fig:Garnett_Pwait}
} 
\subfigure[$P\{Ab\}$ as a Function of $\beta$ ]{
\includegraphics[width=0.48\textwidth]{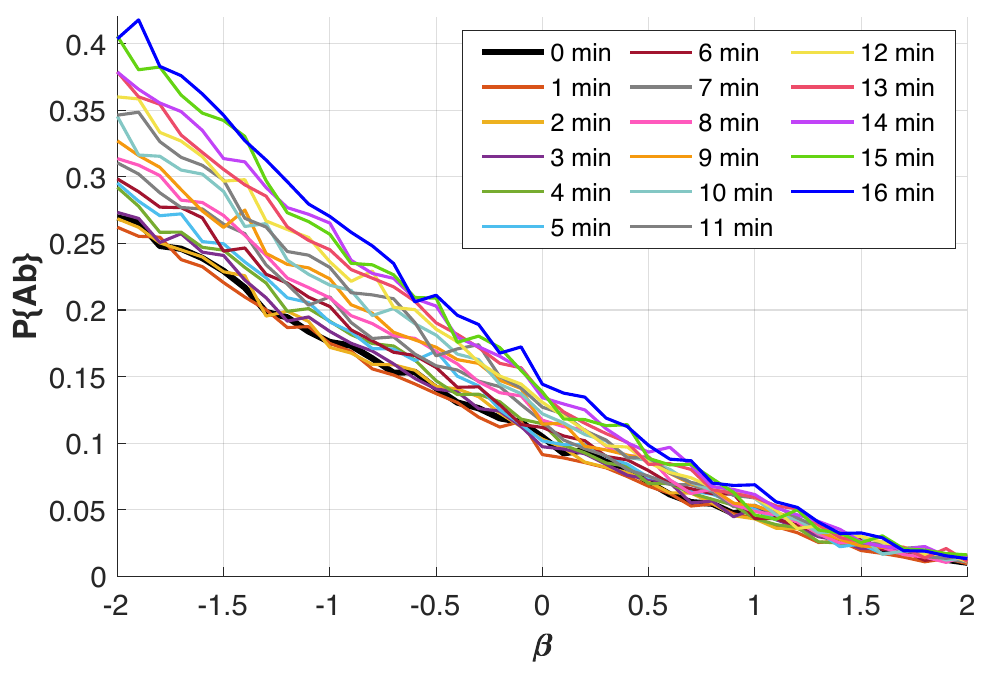} \label{fig:Garnett_Pab}
} 
\caption{Staffing as a Function of Service Time of Sab Customers. Simulation Study with Parameters of Hour 14:00 in Table \ref{tbl:ParametersHour}.}
\label{fig:staffing}
\end{figure}

Figure \ref{fig:staffing} presents the staffing level needed to fit specific performance measures as a function of the time it takes to handle Sab customers. Figure \ref{fig:Garnett_Pwait} for the probability of waiting, $P(Wait)$, and Figure \ref{fig:Garnett_Pab} for the abandonment probability, $P(Ab)$. This is a modification of the Garnett functions \citep{Garnett2002}, presented in the bold black lines (where $\mu_{Sab}=\infty$). The staffing level is represented by the parameter $\beta$ from the square root staffing rule $n=R+\beta\sqrt{R}$ \citep{Garnett2002}, where  $R$ is the offered load (the amount of work, $\lambda/\mu_{Sr}$, that enters the system). We vary the detection time for Sab customers from 0 to 16 minutes. We observe that when staffing is high---in the quality-driven regime where $\beta>1$---the detection time has almost no impact on performance.  But, in the QED regime, where $-1<\beta<1$, the detection time can increase abandonment from around 10\% to 15\% and waiting probability from 16\% to 25\%. And in the ED regime, it can increase abandonment an alarming 13 percentage points (from 27\% to 40\%). Note that operating in the quality-driven regime provides a very low abandonment but at a very high cost, while operating in the other regimes results in losses of abandoning customers. 

We summarize that detection time of Sab has high economic implications by increasing staffing requirements. Next, we discuss alternative ways to reduce that cost.

\subsection{Reducing the Operational Impact of Silent-Abandonment Customers}
\label{sec:avoid_Sab}
We propose three ways a company can reduce capacity loss by (a) using bots to handle \emph{suspected} Sab customers, (b) changing the concurrency algorithm, and (c) changing suspected Sab customer prioritization. 
These ideas rely on identifying suspected Sab customers. To that end, the company can design a prediction model in the spirit of the classification model we presented in Section \ref{sec:prob_abnd}, wherein information about customers' wait time, class, and initial messages is used to identify Sab customers.
As all classification models have some margin of error, even the best system will assign some Sab customers to agents and waste agent time, but hopefully to a lesser extent than before.

The first way we propose to reduce capacity loss is to design a bot that can identify silent abandonment without involving an agent. Such a bot can manage the beginning of the interaction  automatically and transfer the conversation to the agent only after the customer reacts. In no-writing-in-queue systems, such a bot can manage the initial stages of the conversation, namely, the introduction and an inquiry about the customer's problem. In write-in-queue systems, the bot can ask whether the customer's inquiry is still relevant. As some customers might find such a question annoying, the bot can be programmed to use that method only for suspected Sab customers. Using the simulation of Section \ref{sec:managerial}, we can estimate that if the company can automatically identify 50\% of its Sab customers, and hence reduce the average time of identifying Sab from 4 to 2 minutes, then the staffing required to achieve average wait of 1 minutes will be reduced by 7.1\%.

The second way we propose to reduce capacity loss is to change the concurrency algorithm by considering  suspected Sab customers as fractional customers (as opposed to full ones) until they write something. For example, as long as a customer writes nothing, they will be considered a suspected Sab customer and be assigned a value of 0.5, but as soon as they write something they will be assigned a value of 1. Therefore, an agent that has two suspected Sab customers and two responsive ones will be considered to be equivalent to an agent that handles three responsive customers. This will reduce the amount of blocking that Sab customers impose on the other customers in the queue (instead of reducing the time of customer assignment to the agent). 

A final possible solution is to handle queue priorities according to existing information on suspected Sab customers. For example, the system can send a suspected Sab customer to the end of the queue. Therefore, when the suspected Sab customer's turn for service arrives, the agent will have reached their idle period, which means that the effect of the Sab customer on system performance would be diminished. We think that this solution is appropriate mostly for customers who enter the queue when the contact center is closed (e.g., at night) and who would be loading the agent's capacity at the beginning of the workday without actually being there, thereby delaying new arrivals significantly. In such a scenario,  the ``cost" imposed by this unfair policy of requiring suspected Sab  customers to wait for one extra busy period may be worth it.

\section{The Influence of System Design on Customer Behavior and Silent Abandonment 
}
\label{sec:patience_coef}




Throughout this article, we have analyzed two contact-center designs---write-in-queue and no-write-in-queue systems. Figure \ref{fig:KabvsUsab} shows that companies choose both options.  The underlying conclusion that comes from our analysis of the Sab phenomena (\S\ref{sec:prob_abnd}) is that writing in queue creates the missing data problem and, hence, complicates the company's analysis of true performance measures. This is a large drawback and raises complementary questions: What are the benefits of allowing writing in queue? Should companies allow customers to write messages while waiting?

\begin{figure}[!htb]
\centering
\includegraphics[width=0.55\textwidth]{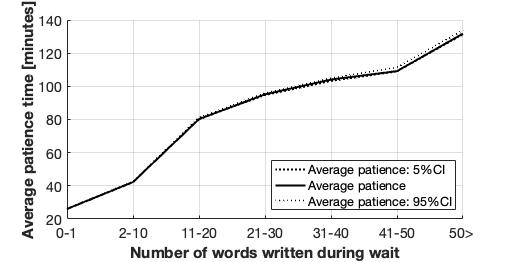}
\caption{Patience as a Function of the Number of Words Written during Waiting in Queue. Write-in-queue System. (Confidence Intervals are Computed Using Bootstrapping, Sampling with Replacement.)}
\label{fig:Patience&Words}
\end{figure}

The first benefit is to allow the customers a more natural communication platform that operates similar to  instant messaging applications (like WhatsApp) in which there are no communication constraints on people's message timing. A second benefit is operational: writing while waiting reduces agent service time, since the customer uses the waiting time to effectively start the service and the agent has most of the information needed when they are assigned. To provide some content to that claim, we measure customers writing while waiting in the write-in-queue dataset and find that customers write, on average 1.2 ($SD=0.51$) messages while waiting (before assignment to agent), which is 20.1\% of their  total messages and 27.8\% of their written text (see data in Table \ref{tbl:general_stat}).  A third benefit draws on behavioral operations literature; people that are kept busy are less likely to feel the burden of waiting, and the best way to reduce such burden is to start self-service while waiting \citep{Maister1984TheLines}. Therefore, allowing customers to write while waiting can be viewed as a way to reduce customer perception of wait time and increase patience. This may also be a case of sunk cost dynamics. \cite{Ulku2022SocialQueues} found a sunk cost effect of customer waiting by which  customer service time increased with customer waiting time. Hence, we can conjecture that customers who invest more effort in writing will be less likely to abandon.  
To provide evidence for the connection between writing in queue and patience, we analyze the impact of customer writing in the write-in-queue system, 
as shown in Figure \ref{fig:Patience&Words}. To create this figure, we split the conversations into bins based on the number of words the customer wrote during their waiting, and estimate the average patience in each bin.
We see that as customers write more words while waiting in queue, investing more effort,  their patience increases. Specifically, customers who write 21.8 words (the average in the data) while waiting exhibit patience that is more than three times larger than that of customers that write up to one word.
We note that companies can also use this observation to estimate individual customer patience while waiting, which can be used to determine prioritization. 


\section{Discussion} \label{sec:discussion}
In this article, we identified and defined the phenomenon of silent abandonment as an important source of uncertainty in contact centers. Our work exposed and analyzed how a small difference in the service design---allowing customers to write while waiting---changes the way we estimate performance and patience. Specifically, we showed that the timing of the submission of a customer's inquiry (i.e., before entering the queue or after being assigned to an agent) and the customer's management of their service window/application create uncertainty that affects a company's ability to know which customers have abandoned the queue and which have been served. While enabling or denying customer messages before entering the queue is a design decision, the fact that some people do not close their application (abandoning without indication) or are otherwise impolite (leaving without acknowledging service) is a behavioral phenomenon that the company cannot control but needs to deal with. 

We further analyzed the impact of silent abandonment on estimations of customer patience and abandonment proportions. We showed that in order to obtain accurate performance measures, silent abandonment needs to be considered as left-censored observations of customer patience and as time-consuming tasks. We suggested a queueing model that takes Sab customers into account, and showed that it captures system dynamics well, whereas queueing models that ignore Sab customers do not fit the data. Using our queueing model, we showed the impact that capacity loss caused by customer behavior has on performance measures. We then made several suggestions for operational changes in staffing, concurrency management, and prioritization to reduce that problem. 

When comparing customer patience in our two detailed datasets, we notice a huge difference between the two companies. The EM algorithm estimated customer patience in the write-in-queue system to be 81.1 minutes, and Method 2 estimated customer patience in the no-write-in-queue system to be much shorter, only 2 minutes. The service context is surely a major contributor
to such differences. Yet, in Section \ref{sec:patience_coef}, we showed that writing while waiting is associated with longer patience. Here, we provide additional behavioral explanations. First as we mentioned in Section \ref{subsubsec:EMValidation_data}, the higher patience in the write-in-queue system is consistent with previous literature that shows a connection between customer service time and willingness to wait \citep{Mandelbaum2013Data}, here, customer service time in the write-in-queue system is much longer than in the no-write-in-queue system: 46.34 and 10.26 minutes, respectively. Even so, patience in no-write-in-queue systems seems short. 
We claim that the difference in customer patience between the two contact centers is also related to the different platforms used for service in those systems. In the write-in-queue company, customer communication was usually through a smartphone (57.8\% of conversations), which are always with us, whereas, in the no-write-in-queue system, the communication was only through a desktop computer, which requires customers to remain stationary. When analyzing customer patience in the write-in-queue company dataset, we find big differences in customer patience between desktop and cellphone users (176.71 vs.\ 42.62 min, respectively). Another explanation relates to the systems' opening message when entering the queue. Specifically, like we mentioned in Section \ref{subsubsec:EMValidation_data}, in the write-in-queue contact center  customers are instructed to ``go on with their daily activities" and to regard the conversation as ``talking to a friend". These vague opening messages may have increased customer patience. 
Finally, customers may be aware of the company's automatic closure policy (see Section \ref{sec:dataAndReseatch}), which is vastly different between the two companies. As stated in Section 2, the write-in-queue contact center automatically closes communication after 2 hours of inactivity, while the no-write-in-queue company closes them after 2 inactive minutes. A customer that is aware of that policy may feel pressured in the no-write-in-queue contact center to stay alert. Pressure and anxiety are known to decrease patience \citep{Maister1984TheLines}.  

When analyzing the total percentage of abandoning customers in both environments, we see that it is almost the same, around 19\%. However, the percentage of Sab customers is higher in the write-in-queue system, where the wait is also longer (8.1 vs.\ 1.4 minutes, see Table \ref{tbl:general_stat}). This is somewhat similar to the increase in the no-show rate in medical offices as the wait time from appointment booking to physician visit increases \citep{Folkins1980,Galucci2005,Liu2010}. \cite{Folkins1980} claim that in the setting of a mental health center, it may be the case that no-shows happen since customers that wait longer solve their problems on their own. We conjecture that this may also be true for textual services. This raises the question of whether there is a connection between $q$ and wait time. We therefore think that future research on patience estimation can relax the assumption we made for the EM algorithm on the independence between $q$ and wait time.

Another interesting comparison can be made between silent abandonment and no-shows vis-\`{a}-vis the scope of these phenomena and their operational implications. 
Our findings suggest that 3\%--59\% of customers abandon no-write-in-queue contact-center queues without notification (see Figure \ref{fig:KabvsUsab}), compared to 23\%--34\% of no-shows in medical appointments. 
In terms of operational implications, \cite{Moore2001} found that in a family medical practice, no-shows are responsible for 25.4\% of scheduled wasted time. Here too, we showed that silent abandonment reduces system capacity, but at a lower magnitude of 1.7\%--15.3\%. However, in contact centers it translates to wasted tasks performed by the agent and occupied slots held by the silent-abandonment customers in the system.


From a different perspective, we note that agents may use the silent-abandonment phenomenon to their advantage. If a Sab customer is assigned to an agent, the agent seems to be busy while in practice they may rest a little. Therefore, agents may lack incentive to close suspected Sab conversations quickly. 
The company will want to prevent such strategic behavior by agents but should proceed carefully in order  to avoid situations where a long-waiting customer conversation is prematurely terminated. For example, it is possible that the customer did not notice that the agent finally answered. Hence, finding technological answers to handling capacity loss, like the ones we suggested in Section \ref{sec:managerial}, 
is important. 
Investigating the strategic behavior of agents may be interesting in its own right and a worthy topic of future research. 

To conclude, we believe that the phenomenon of Sab has an impact beyond the framework discussed in this paper, and therefore calls for further mathematical and behavioral modeling of textual-based services. 


%
%
%



\bibliographystyle{informs2014} 
\bibliography{Mendeley1.bib} 

\ECSwitch

\ECDisclaimer


\section{Examples of Customer Conversations for Each Class of Customer} \label{app:chat_examples}

\begin{figure}[tbh]
\centering
\subfigure[Served Customer]{
\includegraphics[width=0.25\textwidth]{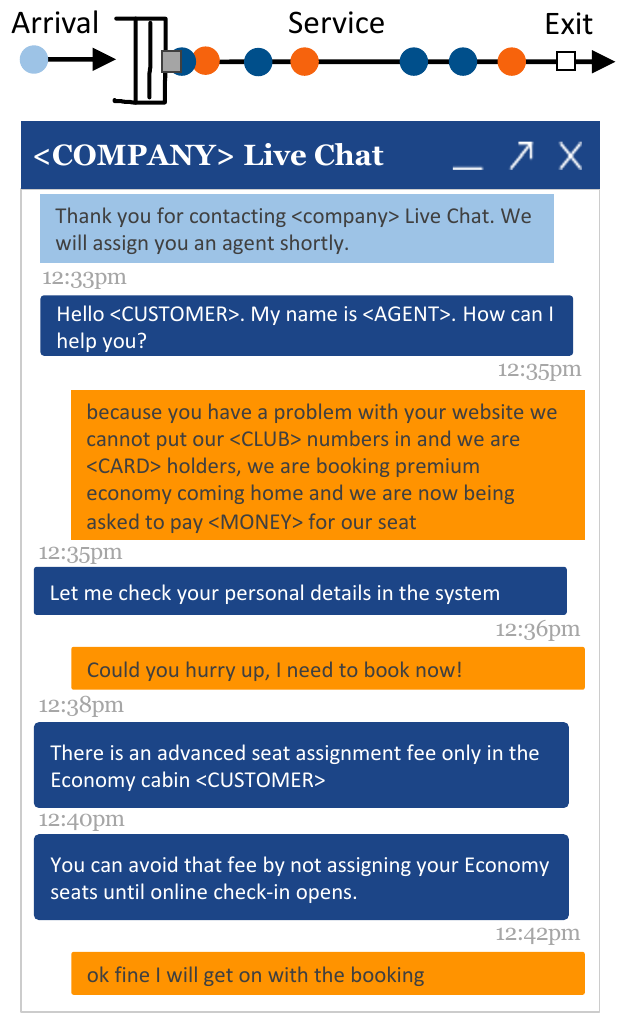} \label{fig:SRNW}
} 
\subfigure[Kab Customer ]{
\includegraphics[width=0.245\textwidth]{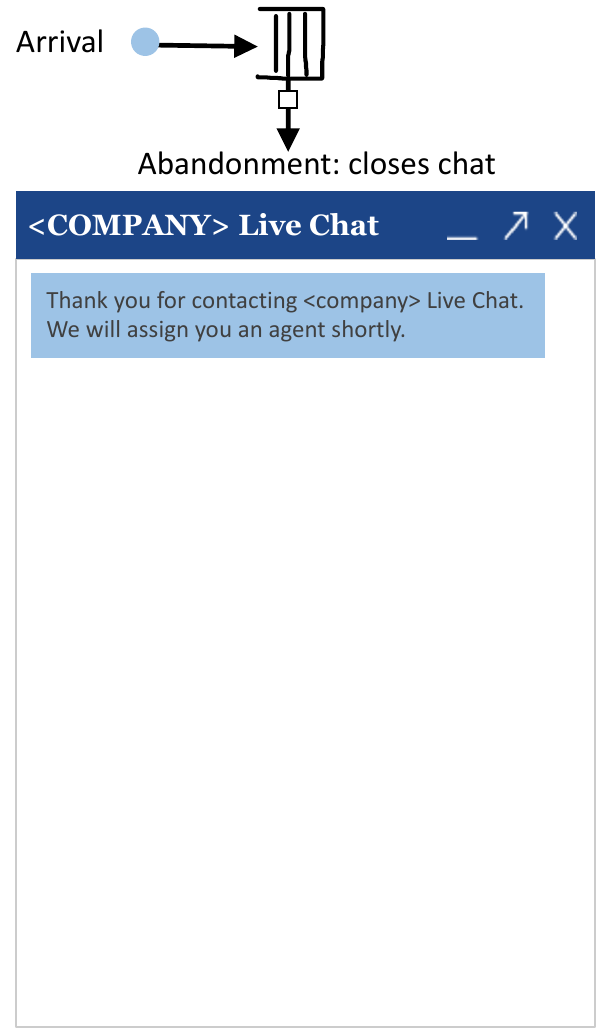} \label{fig:KABNW}
} 
\subfigure[Sab Customer]{
\includegraphics[width=0.24\textwidth]{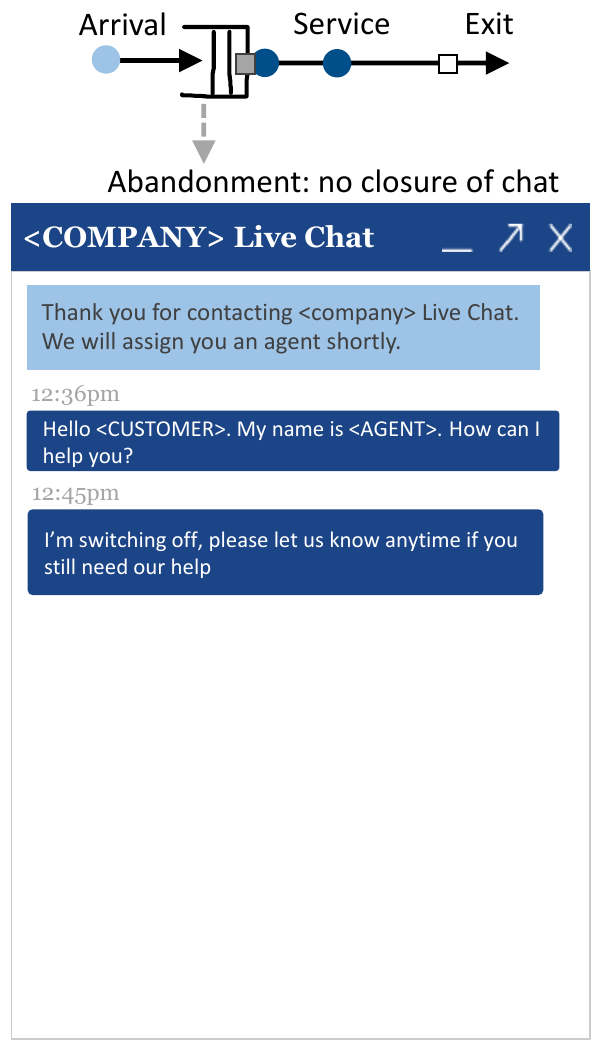} \label{fig:SABNW}
} 
\subfigure{
\includegraphics[width=0.16\textwidth]{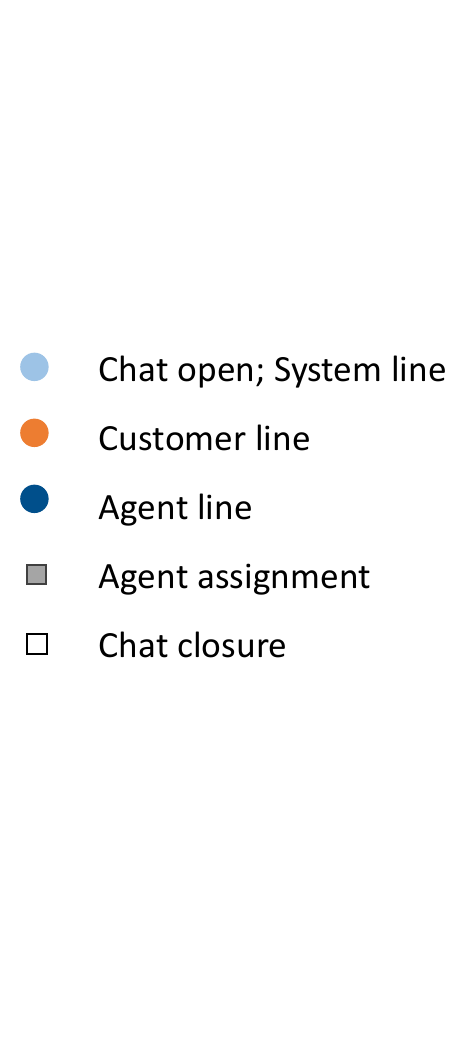} \label{fig:legend}
}
\caption{Redacted Conversation Example and Process Flow of Different Types of No-write-in-queue Customers}
\label{fig:cust_Process_flow}
\end{figure}

\begin{figure}[tbh]
\centering
\subfigure[Served Customer]{
\includegraphics[width=0.22\textwidth]{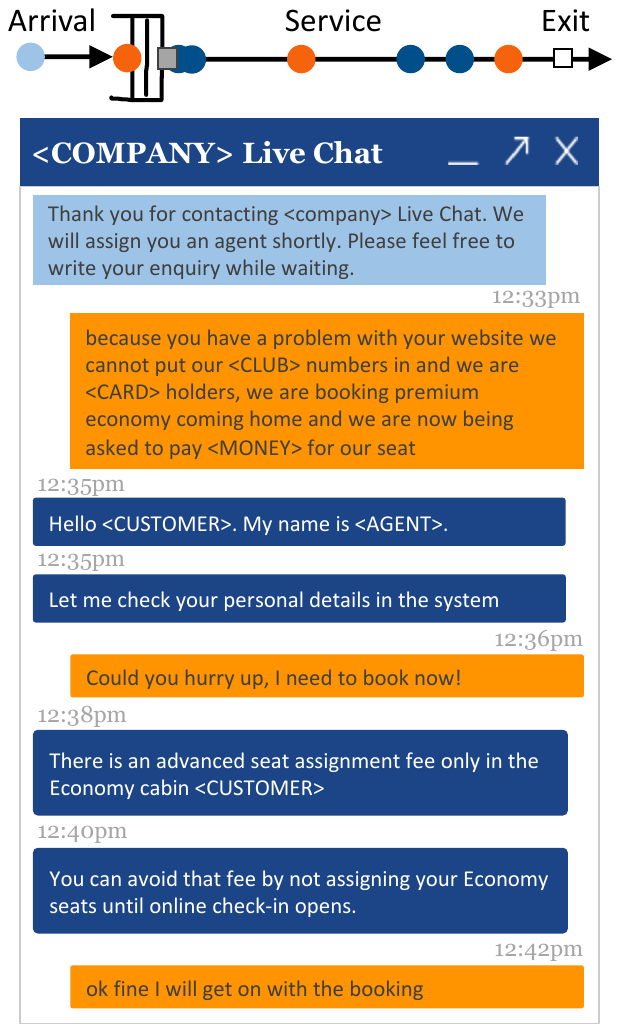} \label{fig:SRW}
} 
\subfigure[Kab Customer]{
\includegraphics[width=0.21\textwidth]{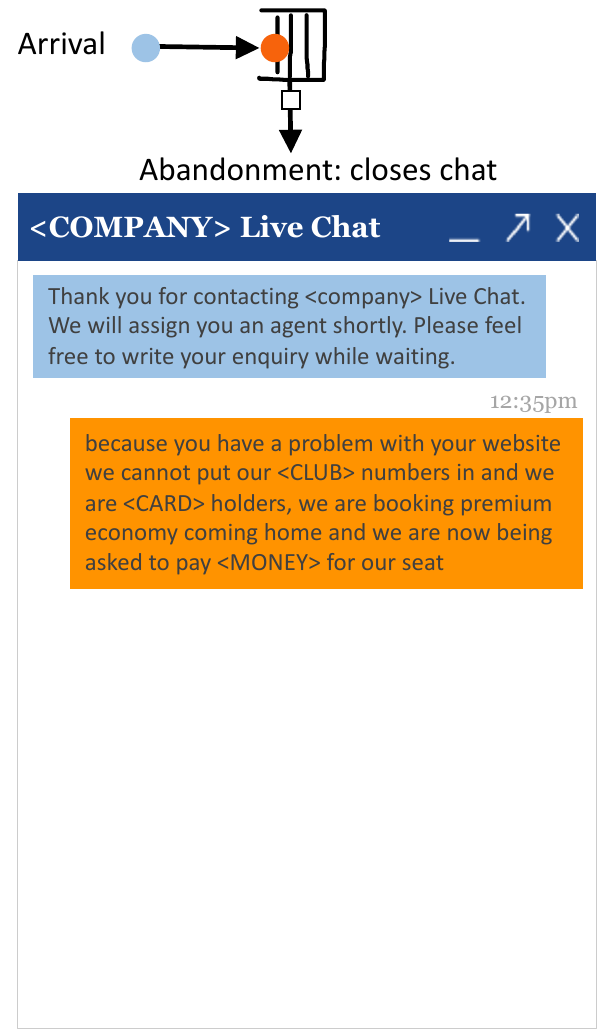} \label{fig:KABW}
} 
\subfigure[uSab: Sr1 Customer]{
\includegraphics[width=0.21\textwidth]{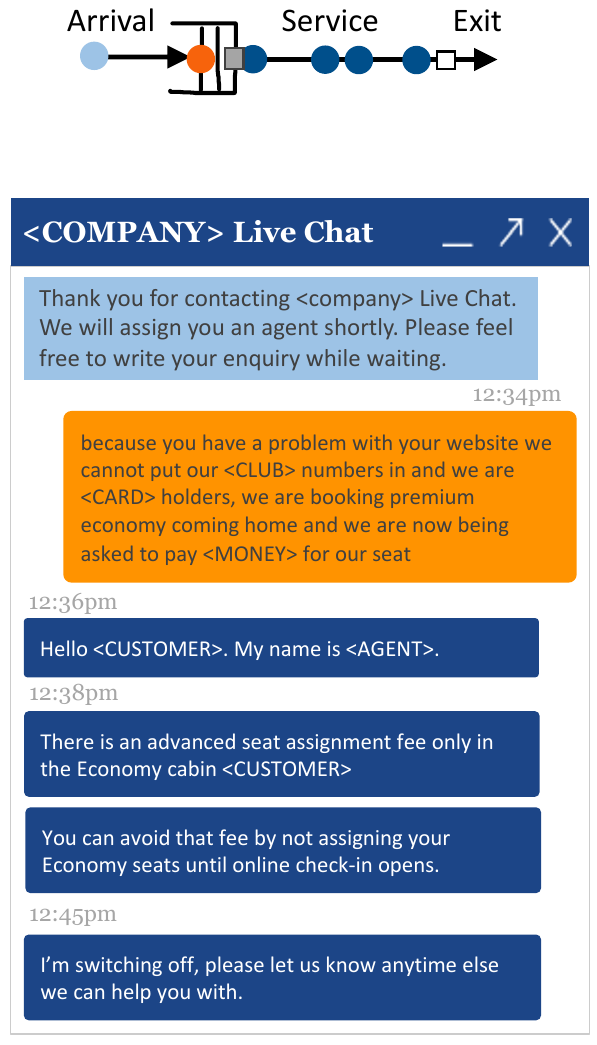} \label{fig:ShortServConvo}
} 
\subfigure[uSab: Sab Customer]{
\includegraphics[width=0.21\textwidth]{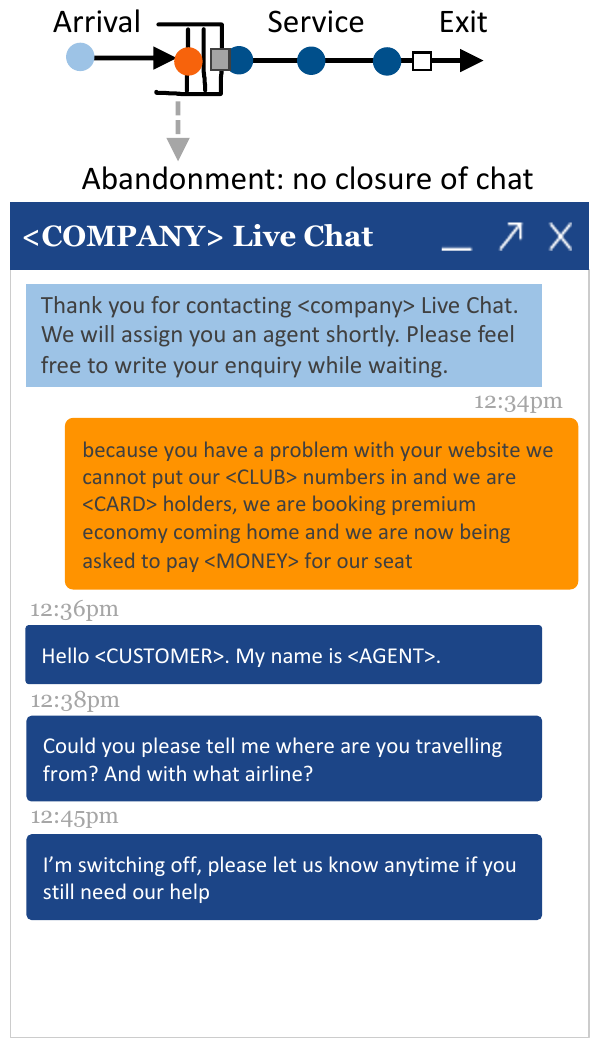}
\label{fig:SabConvo}
}
\caption{Redacted Conversation Example and Process Flow of Different Types of Write-in-queue Customers}
\label{fig:cust_Process_flow_WQ}
\end{figure}

We examine three different scenarios of conversation dynamics---served, Kab, and Sab---in both write-in-queue and no-write-in-queue systems. We demonstrate here a conversation detail for each scenario and denote graphically the timing of customer and agent messages dynamics. Customer messages are marked in orange (text/dot), agent messages in dark blue (text/dot), and automated system messages in light blue (text/dot). Figure EC.\ref{fig:SRW} shows that, in a write-in-queue system, a served (Sr) customer arrives and enters the queue. While waiting for an available agent in the queue, they can write an inquiry. Once an agent is assigned, indicated with a dark grey rectangle, the conversation begins---the agent introduces themself and replies to the customer inquiry, and then the customer and the agent communicate with each other until the inquiry is solved and the conversation ends and is closed by the customer or the agent manually or by the system after some grace time. Conversation closure is indicated with a white rectangle. The Sr customer in the no-write-in-queue system (Figure EC.\ref{fig:SRNW}) differs only in the fact that they write the inquiry only after the agent assignment. This difference does not impact the customer classification as a served one. Still, having information regarding the customer inquiry while they are in the queue may be exploited to change customer prioritization, influence customer routing, and potentially reduce wasted time at the beginning of service because the agent does not need to wait for the customer writing after assignment. A Known abandonment (Kab) customer abandons the system before they were assigned to an agent and indicates that they abandoned by closing the communication window. Figures EC.\ref{fig:KABW} and EC.\ref{fig:KABNW} show examples of such cases for write-in-queue and no-write-in-queue systems, respectively. A write-in-queue system has information regarding the customer's inquiry, but this fact does not affect the customer classification as Kab. Notice that a Kab causes no capacity loss because the customer is never assigned to an agent, and clearly the agent never writes anything. Finally, we have conversations in which customers did not write anything after the agent assignment. Here, the difference between the write-in-queue (Figure EC.\ref{fig:ShortServConvo}--EC.\ref{fig:SabConvo}) and the no-write-in-queue systems EC.\ref{fig:SABNW} is important. In the no-write-in-queue system, we know for sure, but only in retrospect, that such customers abandoned and classify them as Sab, but we don't know exactly when (which results in censored data). In the write-in-queue system, it could be that the customer abandoned but  did not indicate leaving the system, that is, the customer should be classified as a silent-abandonment customer, or 
it could be that the customer wrote a very simple inquiry that could be solved in a single exchange with the agent, and the agent solved it, but the customer was impolite and did not write, for example, ``thank you" at the end. (Examples of these two types of conversations are provided in Figures EC.\ref{fig:ShortServConvo}--EC.\ref{fig:SabConvo}.) Hence, just relying on the trace of events---the metadata---is not enough to determine if the customer abandoned. We have to examine the written text of the conversation to determine the actual classification of such customers to Sab or Sr. These are the uSab customers. In both cases, since the agents are uncertain whether or when such customers will appear, they give such uSab customers some grace time to reply and in many cases let the system close the conversation automatically after some predetermined time. Now that we have explained the types of customers in both systems, we can proceed to give more details of both systems and their statistics. 

From the above description, it is clear that conversation length is a combination of several elements: wait time of customer to agent assignment, customer response time (RT) (the time since an agent message was sent until a customer message is sent), agent response time (the time since a customer message and until an agent message is sent), and time between the last message sent and the conversation closure. Each of those is affected by the operational decisions and behavioural aspects. Wait time is determined by agent availability which is a function of the number of agents (staffing), the maximal concurrency level, arrival rate, service time, and behavioral aspects (e.g., customer patience). Agent response time is a function of operational decisions such as current concurrency level, but also customer behavior \citep{Altman2021,daw2021coproduction}. Customer response time is affected by customer expectations and agent response time \citep{Gallino2022Need,daw2021coproduction}. Statistics about each of those elements in the two contact centers are provided in Table \ref{tbl:general_stat}.

The no-write-in-queue dataset included 18,497 conversations.
We excluded 780 observations with unrealistic or no information (4.2\% of the conversations), for example, when their closure time was before the last message was sent, when no events of a conversation were registered, etc. This resulted in 17,717 conversations. 
The write-in-queue dataset included 332,978 conversations. We excluded 1,391 observations (0.4\% of the conversations) for the same technical reasons as for the no-write-in-queue data set. This resulted in 331,587 conversations.

\begin{table}[!htb]
 \centering
 \caption{General Statistics}
  \label{tbl:general_stat}

  \begin{scriptsize}
  \begin{tabular}{llcccccccc} 
  \toprule
Dataset&Cust.\ &\multicolumn{1}{c}{\% of}&\# agent msg.\ &\# agent words&\multicolumn{5}{c}{Agent RT in minutes [Mean (SD)]}\\  \cmidrule{6-10}
&type&\multicolumn{1}{c}{conv.\ }&\multicolumn{1}{c}{Mean (SD)}&\multicolumn{1}{c}{ per msg. }&\multicolumn{1}{c}{All}&\multicolumn{1}{c}{Turn 1}&\multicolumn{1}{c}{Turn 2}&\multicolumn{1}{c}{Turn 3}&\multicolumn{1}{c}{Turn 4}\\ \midrule
write-in-&All&&9.7 (8.98)&23.0 (22.74)&3.6 (15.16)&2.2 (5.3)&3.8 (15.1)&5.2 (18.4)&5.2 (18.9)\\
queue&Sr&70.6\%&11.8 (9.36)&22.2 (22.16)&3.0 (12.74)&1.9 (3.8)&2.1 (8.2)&3.0 (10.4)&3.5 (13.3)\\
&Kab&5.1\%&\multicolumn{1}{c}{--}&\multicolumn{1}{c}{--}&\multicolumn{1}{c}{--}&\multicolumn{1}{c}{--}&\multicolumn{1}{c}{--}&\multicolumn{1}{c}{--}&\multicolumn{1}{c}{--}\\
&uSab&24.4\%&3.3 (2.18)&31.4 (27.02)&10.8 (30.36)&3.1 (8.3)&10.2 (28.0)&16.5 (37.5)&18.2 (40.4)\\
&~-Sr1&48.2\%$^*$&4.3 (2.45)&34.5 (29.81)&12.3 (33.30)&3.4 (8.7)&10.6 (28.5)&17.3 (39.2)&18.4 (41.1)\\
&~-Sab&51.8\%$^*$&2.3 (1.29)&26.2 (20.29)&8.4 (24.68)&2.8 (5.3)&9.8 (27.8)&15.5 (35.4)&17.7 (37.7)\\ \midrule
no-write-&All&&5.7 (4.36)&32.4 (26.19)&1.1 (1.46)&0.8 (1.0)&1.1 (1.4)&1.1 (1.5)&1.2 (1.9)\\
in-queue&Sr&80.9\%&5.9 (4.36)&32.6 (26.26)&1.0 (1.46)&0.8 (0.9)&1.1 (1.4)&1.1 (1.5)&1.2 (1.9)\\
&Kab&13.9\%&\multicolumn{1}{c}{--}&\multicolumn{1}{c}{--}&\multicolumn{1}{c}{--}&\multicolumn{1}{c}{--}&\multicolumn{1}{c}{--}&\multicolumn{1}{c}{--}&\multicolumn{1}{c}{--}\\
&Sab&5.2\%&1.4 (0.53)&18.0 (12.00)&1.8 (1.56)&1.5 (1.5)&2.7 (1.5)&1.8 (0.9)&0.9 (0.6)\\ \bottomrule
\multicolumn{10}{l}{$^*$Percentage of uSab. Estimated using SVM with threshold of 0.47 (\S\ref{sec:prob_abnd_WQ}).}\\
\multicolumn{10}{l}{Note: Standard deviations are provided in parentheses.}\\
 \end{tabular}

\vspace*{5mm}

\begin{tabular}{llcccccccc} \toprule
Dataset&Customer&\multicolumn{2}{c}{\# of cust.\ messages}&&\multicolumn{2}{c}{Customer words per msg. }&&\multicolumn{2}{c}{Customer RT per msg.\ [min]}\\ \cmidrule{3-4} \cmidrule{6-7} \cmidrule{9-10}
&type&All&While waiting&&All&While waiting&&All&First turn$^1$\\ \midrule
write-in-queue&All&6.0 (6.45)&1.2 (0.51)&&13.1 (16.00)&18.2 (20.37)&&1.0 (3.58)&1.4 (5.23)\\
&Sr&7.9 (6.70)&1.2 (0.54)&&12.9 (15.65)&18.2 (20.25)&&1.0 (3.52)&1.4 (5.23)\\
&Kab&1.2 (0.54)&1.2 (0.54)&&16.4 (20.90)&16.4 (20.90)&&2.1 (5.30)&\multicolumn{1}{c}{--}\\
&uSab&1.1 (0.39)&1.1 (0.40)&&18.2 (20.54)&18.5 (20.64)&&1.1 (4.26)&\multicolumn{1}{c}{--}\\
&~-Sr1&1.1 (0.42)&1.1 (0.42)&&22.0 (22.55)&22.0 (22.55)&&1.0 (4.13)&\multicolumn{1}{c}{--}\\
&~-Sab&1.1 (0.40)&1.1 (0.40)&&15.3 (18.20)&15.3 (18.20)&&1.2 (4.74)&\multicolumn{1}{c}{--}\\ \midrule
no-write-in-queue&All&5.9 (4.69)&\multicolumn{1}{c}{--}&&14.1 (14.94)&\multicolumn{1}{c}{--}&&0.8 (0.87)&0.9 (0.88)\\
&Sr&5.9 (4.69)&\multicolumn{1}{c}{--}&&14.1 (14.94)&\multicolumn{1}{c}{--}&&0.8 (0.87)&0.9 (0.88)\\
&Kab&\multicolumn{1}{c}{--}&\multicolumn{1}{c}{--}&&\multicolumn{1}{c}{--}&\multicolumn{1}{c}{--}&&\multicolumn{1}{c}{--}&\multicolumn{1}{c}{--}\\
&Sab&\multicolumn{1}{c}{--}&\multicolumn{1}{c}{--}&&\multicolumn{1}{c}{--}&\multicolumn{1}{c}{--}&&\multicolumn{1}{c}{--}&\multicolumn{1}{c}{--}\\ \bottomrule
\multicolumn{9}{l}{$^1$First turn after agent assignment.}

\end{tabular}
\vspace*{5mm}

\begin{tabular}{llccccc} \toprule
Dataset&Customer&Wait time&Service time&Closure time&Concurency time&Total time\\ 
&type&\multicolumn{1}{c}{Mean (SD)}&\multicolumn{1}{c}{Mean (SD)}&\multicolumn{1}{c}{Mean (SD)}&\multicolumn{1}{c}{Mean (SD)}&\multicolumn{1}{c}{Mean (SD)}\\ \midrule
write-in-queue&All&8.3 (18.28)&46.3 (63.56)&65.9 (75.08)&112.3 (92.73)&120.6 (95.80)\\
&Sr&5.3 (11.59)&53.8 (65.36)&59.2 (73.51)&113.0 (95.61)&118.3 (97.04)\\
&Kab&7.3 (8.64)&\multicolumn{1}{c}{--}&\multicolumn{1}{c}{--}&\multicolumn{1}{c}{--}&7.3 (8.64)\\
&uSab&17.3 (29.29)&34.3 (58.76)&99.1 (73.11)&133.4 (75.85)&150.7 (82.36)\\
&~-Sr1&18.9 (30.26)&53.7 (71.36)&95.0 (74.99)&148.7 (77.00)&167.5 (81.86)\\
&~-Sab&19.4 (31.12)&20.1 (40.65)&113.6 (65.88)&133.7 (63.62)&153.1 (71.29)\\ \midrule
no-write-in-queue&All&1.4 (5.56)&10.3 (15.86)&0.9 (1.73)&11.2 (16.04)&12.6 (16.90)\\
&Sr&1.1 (2.72)&12.5 (16.82)&1.1 (1.85)&13.6 (16.91)&14.7 (17.21)\\
&Kab&2.0 (4.32)&\multicolumn{1}{c}{--}&\multicolumn{1}{c}{--}&\multicolumn{1}{c}{--}&2.0 (4.32)\\
&Sab&4.2 (20.43)&2.3 (3.80)&1.3 (1.28)&3.6 (4.09)&7.8 (20.99)\\ \bottomrule
\multicolumn{7}{l}{Note: Times are measured in minutes.}
 \end{tabular}

\vspace*{5mm}

\begin{tabular}{llcccccccc} \toprule
Dataset&Customer &\multicolumn{4}{c}{\% terminated by$^2$}&&\multicolumn{3}{c}{Platform (\%)}\\ \cmidrule{3-6} \cmidrule{8-10}
&type&System&Agent&Customer&Manager&&App&Web&Other\\ \midrule
write-in-queue&All&32.0\%&47.8\%&19.9\%&0.3\%&&56.5\%&42.7\%&0.9\%\\
&Sr&28.0\%&53.1\%&18.5\%&0.4\%&&57.8\%&42.0\%&0.2\%\\
&Kab&0.0\%&0.0\%&100.0\%&0.0\%&&28.7\%&71.3\%&0.0\%\\
&uSab&50.3\%&42.2\%&7.2\%&0.3\%&&58.4\%&38.8\%&2.8\%\\
&~-Sr1&47.4\%&50.6\%&1.8\%&0.3\%&&59.9\%&37.1\%&3.0\%\\
&~-Sab&58.2\%&37.7\%&4.0\%&0.2\%&&58.4\%&38.8\%&2.9\% \\ \midrule
no-write-in-queue&All&\multicolumn{1}{c}{--}&\multicolumn{1}{c}{--}&\multicolumn{1}{c}{--}&\multicolumn{1}{c}{--}&&0.0\%&100.0\%&0.0\%\\
&Sr&\multicolumn{1}{c}{--}&\multicolumn{1}{c}{--}&\multicolumn{1}{c}{--}&\multicolumn{1}{c}{--}&&0.0\%&100.0\%&0.0\%\\
&Kab&\multicolumn{1}{c}{--}&\multicolumn{1}{c}{--}&\multicolumn{1}{c}{--}&\multicolumn{1}{c}{--}&&0.0\%&100.0\%&0.0\%\\
&Sab&\multicolumn{1}{c}{--}&\multicolumn{1}{c}{--}&\multicolumn{1}{c}{--}&\multicolumn{1}{c}{--}&&0.0\%&100.0\%&0.0\% \\ \bottomrule
\multicolumn{9}{l}{$^2$Information recorded only in write-in-queue dataset.}
 \end{tabular}

  \end{scriptsize}

\end{table}

\section{Classification Models} \label{sec:ClassMod}
We provide here more details on the two best classification models described in Section \ref{sec:prob_abnd}: the SVM model and the classification tree. For both models, we selected the features for the final models using a wrapper method with recursive feature elimination \citep{guyon2002gene}.
\subsection{Support Vector Machines} \label{subsec:SVM}
The fitted model has 259 support vectors. The variables included in this model are:
\begin{enumerate}
    \item AgentChars: the number of characters written by the agent in the conversation.
    \item AgentDuration: the time it takes the agent to write their messages.
    \item QueueTime: the time the customer waits for agent assignment in the queue.
    \item TotalDuration: the time from agent assignment until manual closure of the conversation by the agent or automatic closure by the system.
    \item wordag46, wordag1, wordag20, wordag31: specific words written by the agent during the conversation. 
    \item wordcust1: a specific word written by the customer in the initial inquiry.
\end{enumerate}   
We provide only coded words due to  privacy concerns of the company providing the dataset.

\subsection{Classification Tree}

\begin{figure}[!htb] 
\centering

\includegraphics[width=.7\textwidth]{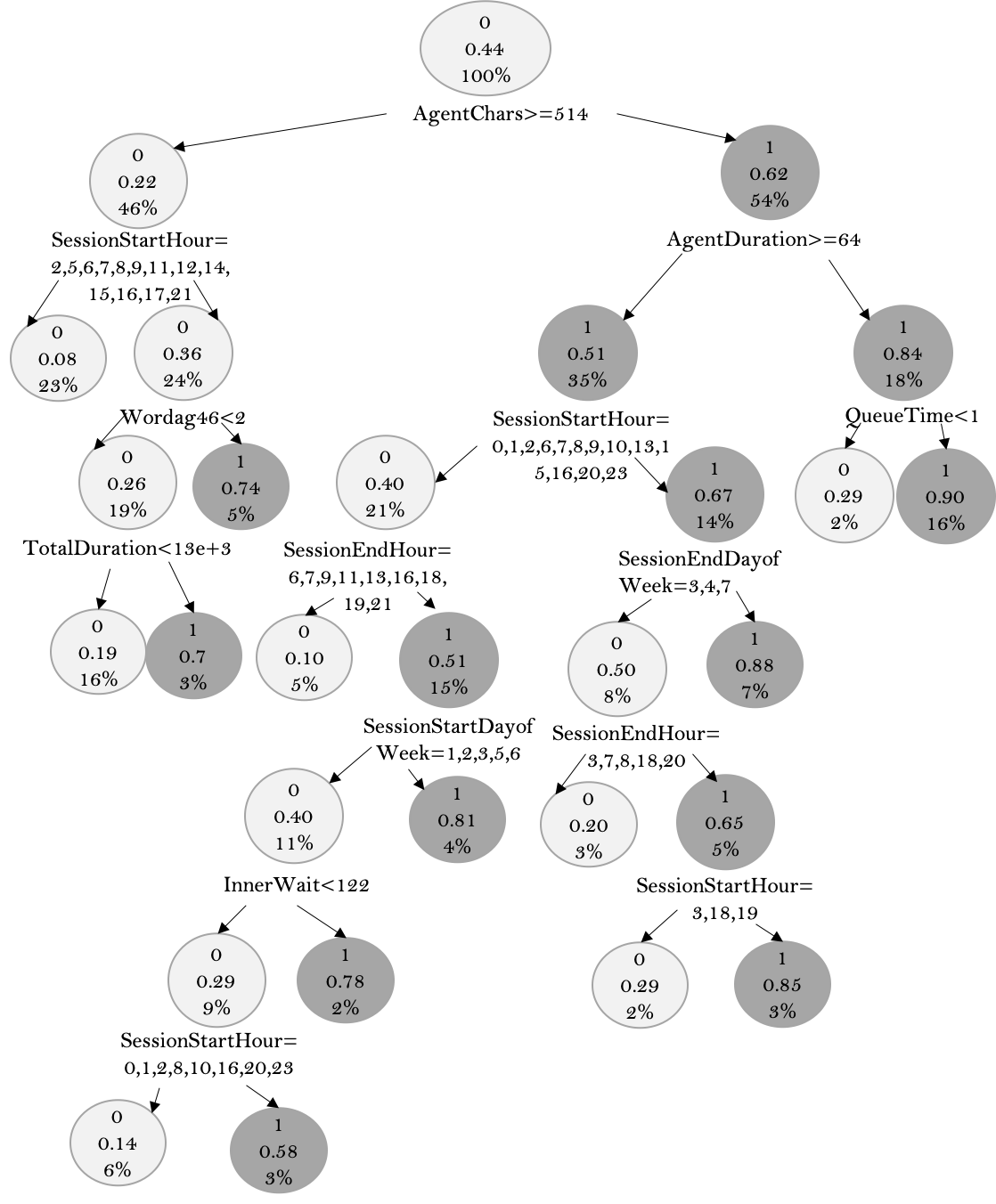}
\caption{Classification Tree for the Probability of a Conversation Being a Silent Abandonment. Splitting Variable on the Bottom of the Nodes. Nodes Show 1 for Silent Abandonment (Grey) and 0 for Served-in-one-exchange (Light Grey); Probability of Obtaining that Classification; Percentage of the Data that Falls into That Node. }
\label{fig:ClasfTree}
\end{figure}

The fitted classification tree is presented in Figure \ref{fig:ClasfTree}. The variables included in this model are:
\begin{enumerate}
\item AgentChars, AgentDuration, QueueTime, TotalDuration, wordag46: see \ref{subsec:SVM}.
\item SessionStartHour: the hour that the customer arrived to the system.
\item SessionStartDayofWeek: the day of the week on which the customer arrived to the system.
\item InnerWait: the time the customer waits for the agent's reply during service (i.e., after assignment to the agent).
\item SessionEndHour: the hour that the conversation was closed; the conversation may be closed manually by the agent (usually within a few hours of no customer reply) or automatically by the system (after a threshold time has passed).
\item SessionEndDayofWeek: the day of the week on which the conversation is closed.
\end{enumerate}   


\section{EM Algorithm: Proof and Explanation} \label{sec:proofE}
The log of the likelihood in Eq.\ \eqref{eq:Likelihood} is
\begin{equation} \label{eq:loglikelihood}
\begin{split}l(D,\theta,q,\gamma)=\sum_{i=1}^{n}\left\{ \left(1-\Delta_{i}\right)\left(\log\gamma-\gamma U_{i}-\theta U_{i}\right)\right\} +\sum_{i=1}^{n}\left\{ \left(\Delta_{i}Y_{i}\right)\left[\log\theta-\theta U_{i}-\gamma U_{i}+\log(q)\right]\right\} \\
+\sum_{i=1}^{n}\left\{ \left(\Delta_{i}(1-Y_{i})\right)\left[\log(1-q)+\log(1-e^{-\theta U_{i}})+\log\gamma-\gamma U_{i}\right]\right\},
\end{split}
\end{equation}
wherein, if the data is complete, the possible classes for  conversation $i$ are $C_{1}^{i}=1-\Delta_{i}$, $C_{2}^{i}  =Y_{i}\Delta_{i}$, and $C_{3}^{i}=(1-Y_{i})\Delta_{i}$. Therefore, the log-likelihood in Eq.\ \eqref{eq:loglikelihood} when the data is complete can be written as
\begin{equation} \label{eq:loglikelihoodComplete}
\begin{split}l(D,\theta,q,\gamma)=\sum_{i=1}^{n}\left\{ \left(C_{1}^{i}\right)\left(\log\gamma-\gamma U_{i}-\theta U_{i}\right)\right\} +\sum_{i=1}^{n}\left\{ \left(C_{2}^{i}\right)\left[\log\theta-\theta U_{i}-\gamma U_{i}+\log(q)\right]\right\} \\
+\sum_{i=1}^{n}\left\{ \left(C_{3}^{i})\right)\left[\log(1-q)+\log(1-e^{-\theta U_{i}})+\log\gamma-\gamma U_{i}\right]\right\}.
\end{split}
\end{equation}

In the case of missing data in $\Delta$, we cannot maximize the log-likelihood in Equation \eqref{eq:loglikelihoodComplete} because some of the observations might belong to class $C_{1}=1$ or $C_{3}=1$. To solve this problem, we use the expectation-maximization (EM) algorithm. It calculates starting parameters with random starting weights for conversation classes (see Algorithm \ref{EM}) and then iterates between two steps---the expectation and the maximization steps---until convergence. 
(The convergence criterion is given in Algorithm \ref{EM}.) 



\subsection{Expectation Step, Proof of Theorem \ref{Theorem1}} 
\label{sec:appendixEproof}
In the $t$th iteration, the expectation step consists of finding a surrogate function that is a lower bound on the log-likelihood in Eq.~\eqref{eq:loglikelihoodComplete} but is tangent to the log-likelihood at $(\widehat{\theta^{(t)}},\widehat{q^{(t)}},\widehat{\gamma^{(t)}})$, the vector of the parameters of the latest iteration, $t-1$. We achieve this by computing the expectation given what we know up to the $t$th iteration, that is, the $t-1$th estimations of the parameters $(\widehat{\theta^{(t)}},\widehat{q^{(t)}},\widehat{\gamma^{(t)}})$ and the data that is complete. Formally,
\begin{equation} \label{eq:expectLogLikelihood}
E\left[l(D,\theta,\gamma,q)\mid U_{i},M^{i},\widehat{\theta^{(t)}},\widehat{\gamma^{(t)}},\widehat{q^{(t)}}\right],
\end{equation}
where $M$ is defined as in Section \ref{subsec:ClassesMissingDa}.

When $M^i=2$, the data is complete, implying that for that conversation, $C^i_{2}=1$. Therefore, we can compute Eq.\ \eqref{eq:expectLogLikelihood} for such $i$th observations as follows:
\begin{align*}
E\left[C_{2}^{i}=1\mid U_{i},M^{i},\widehat{\theta^{(t)}},\widehat{q^{(t)}},\widehat{\gamma^{(t)}}\right]=E\left[\Delta_{i}Y_{i}\mid U_{i},M^{i},\widehat{\theta^{(t)}},\widehat{q^{(t)}},\widehat{\gamma^{(t)}}\right] & =1_{\{M^{i}=2\}}+\left(0\right)\left(1_{\{M^{i}\neq2\}}\right).
\end{align*}

We define $\widehat{C_{2,t}^{i}}$ as the probability that the $i$th conversation is a known abandonment in the $t$th iteration. This is exactly what the previous function represents. Therefore, 
\[
\widehat{C_{2,t}^{i}}=1_{\{M^{i}=2\}}.
\]

When $M^i=0$, there is missing data. This implies that for conversation $i$ either $C^i_{1}=1$ or $C^i_{3}=1$. We first compute Eq.\ \eqref{eq:expectLogLikelihood} by conditioning on knowing that the $i$th observation is a member of class $C_{3}=1$:
\begin{align*}
E\left[C_{3}^{i}=1\mid U_{i},M^i,\widehat{\theta^{(t)}},\widehat{q^{(t)}},\widehat{\gamma^{(t)}}\right] & =E\left[\Delta_{i}\left(1-Y_{i}\right)\mid U_{i},M^{i},\widehat{\theta^{(t)}},\widehat{q^{(t)}},\widehat{\gamma^{(t)}}\right]\\
 & =\left(1_{\{M_{i}=0\}}\right)Pr\left\{ \Delta_{i}=1\mid U_{i},M^{i}=0,\widehat{\theta^{(t)}},\widehat{q^{(t)}},\widehat{\gamma^{(t)}}\right\} +\left(0\right)\left(1_{\{M_{i}\neq0\}}\right)\\
 & =\left(1_{\{M_{i}=0\}}\right)Pr\left\{ \tau_{i}\leq W_{i}\mid U_i = W_i,\widehat{\theta^{(t)}},\widehat{\gamma^{(t)}}\right\} \\
 & =\left(1_{\{M_{i}=0\}}\right)Pr\left\{ \tau_{i}\leq U_{i}\mid U_{i},\widehat{\theta^{(t)}}\right\} \\
 & =1_{\{M_{i}=0\}}\left(1-e^{-\widehat{\theta^{(t)}}U_{i}}\right).
\end{align*}

The first equality follows since a customer $i$ for which
$M^i=0$ does not give an indicator when abandoning the queue, that is, $Y_i=0$, and by the independence of $\Delta_i$ and $Y_i$. The second equality follows since a customer $i$ that belongs to class $C_{3}=1$  must have $M^i=0$, and clearly, that customer abandoned. Hence, $\Delta_i=1$ by definition, which formally means that $\tau_i\leq W_i$. Additionally, when $ M^{i}=0$, the observed time $U_i$ is their wait time $W_i$; hence, the third equality follows.
The fourth equality is implied from the third.
The fifth equality follows since the fourth equality represents the cdf $F$
of patience time $\tau$ that has an exponential distribution by Assumption \ref{assumption:1}.

We define $\widehat{C_{3,t}^{i}}$ as the estimated probability that a customer $i$ is a silent abandonment in the $t$th iteration, which by the above calculation is
\[
\widehat{C_{3,t}^{i}}=1_{\{M^{i}=0\}}\left(1-e^{-\widehat{\theta^{(t)}}U_{i}}\right).
\]

Note that in the case where $M^{i}=0$ for conversation $i$, we need to consider also that the customer might belong to $C_{1}^{i}=1$. In addition, we have another group that belongs to class $C_{1}=1$, namely, the conversations where $M^{i}=1$. For them, the data is complete. So, the computation of Eq.\ \eqref{eq:expectLogLikelihood} for a customer $i$ that belongs to class $C_{1}^{i}=1$ is as follows:
\begin{align*}
E\left[C_{1}^{i}=1\mid U_{i},M^{i},\widehat{\theta^{(t)}},\widehat{q^{(t)}},\widehat{\gamma^{(t)}}\right] & =E\left[1-\Delta_{i}\mid U_{i},M^{i},\widehat{\theta^{(t)}},\widehat{q^{(t)}},\widehat{\gamma^{(t)}}\right]\\
 & =\left(1_{\{M^{i}=0\}}\right)Pr\left\{ \Delta_{i}=0\mid U_{i},M^{i}=0,\widehat{\theta^{(t)}},\widehat{q^{(t)}},\widehat{\gamma^{(t)}}\right\} +1_{\{M^{i}=1\}}\\
 & =\left(1-\widehat{C_{3,t}^{i}}\right)1_{\{M^{i}=0\}}+1_{\{M^{i}=1\}}.
\end{align*}
The second term on the right-hand side of the second equality ($1_{\{M^{i}=1\}}$) follows since for the customers that are classified as $M=1$, the data is complete. The first term follows since some of the customers in $M=0$ belong as well to $C_1=1$, and these customers are the ones who do not abandon, that is, $\Delta=0$.
The third equality follows since the probability of a  customer $i$ in $M^{i}=0$ being served (not abandoned) is exactly the complement of $\widehat{C_{3,t}^{i}}$.

We define $\widehat{C_{1,t}^{i}}$ as the probability that a customer $i$ is a served customer, which by the above computation is
\begin{align*}
\widehat{C_{1,t}^{i}} & =\left(1-\widehat{C_{3,t}^{i}}\right)1_{\{M^{i}=0\}}+1_{\{M^{i}=1\}}.
\end{align*}


Finally, we can rewrite our log-likelihood, Eq.~\eqref{eq:loglikelihood}, in the $t$th iteration with missing data as Eq.~\eqref{eq:loglikelihoodMisg}. This is exactly the surrogate function that is a lower bound on the log-likelihood: the E-step in Algorithm \ref{EM}. 

\begin{flushright}
$\blacksquare$
\par\end{flushright}

\subsection{Maximization Step, Proof of Theorem \ref{Theorem2}} \label{sec:appendixMproof}
In the $t$th iteration of the maximization step, the parameters $(\widehat{\theta^{(t+1)}},\widehat{q^{(t+1)}},\widehat{\gamma^{(t+1)}})$ are found to be the maximizers of the surrogate function defined in Eq.~\eqref{eq:loglikelihood}. We obtain the parameters $(\widehat{\theta^{(t+1)}},\widehat{q^{(t+1)}},\widehat{\gamma^{(t+1)}})$, where the partial derivatives of the surrogate function \eqref{eq:loglikelihoodMisg} are equal to zero. 

The partial derivative with respect to $q$ is
\[
\begin{split}\frac{\partial\ell}{\partial q} & =\end{split}
\left(\frac{1}{q}\right)\sum_{i=1}^{n}\widehat{C_{2,t}^{i}}-\left(\frac{1}{1-q}\right)\sum_{i=1}^{n}\widehat{C_{3,t}^{i}},
\]
which yields
\[
\widehat{q^{(t+1)}}=\left\{ \sum_{i=1}^{n}\widehat{C_{2,t}^{i}}\right\} \left\{ \sum_{i=1}^{n}\left(1-\widehat{C_{1,t}^{i}}\right)\right\} ^{-1}.
\]


The partial derivative with respect to $\gamma$ is
\[
\frac{\partial\ell}{\partial\gamma}=\frac{1}{\gamma}\sum_{i=1}^{n}\left(1-\widehat{C_{2,t}^{i}}\right)-\sum_{i=1}^{n}U_{i},
\]
which yields
\[
\widehat{\gamma^{(t+1)}}=\left\{ \sum_{i=1}^{n}\left(1-\widehat{C_{2,t}^{i}}\right)\right\} \left\{ \sum_{i=1}^{n}U_{i}\right\} ^{-1}.
\]


The partial derivative with respect to $\theta$ is

\begin{align*}
\frac{\partial\ell}{\partial\theta} & =\sum_{i=1}^{n}\left(\widehat{C_{1,t}^{i}}\right)\left(-U_{i}\right)+\sum_{i=1}^{n}\left(\widehat{C_{2,t}^{i}}\right)\left(-U_{i}+\frac{1}{\theta}\right)+\sum_{i=1}^{n}\widehat{C_{3,t}^{i}}\frac{U_{i}e^{-\theta U_{i}}}{1-e^{-\theta U_{i}}}\\
 & =\sum_{i=1}^{n}\left(U_{i}\right)\left(-\widehat{C_{1,t}^{i}}-\widehat{C_{2,t}^{i}}\right)+\sum_{i=1}^{n}\left(\frac{\widehat{C_{2,t}^{i}}}{\theta}\right)+\sum_{i=1}^{n}\widehat{C_{3,t}^{i}}\frac{U_{i}e^{-\theta U_{i}}}{1-e^{-\theta U_{i}}}\\
 & =\sum_{i=1}^{n}\left(U_{i}\right)\left(\widehat{C_{3,t}^{i}}-1\right)+\sum_{i=1}^{n}\left(\frac{\widehat{C_{2,t}^{i}}}{\theta}\right)+\sum_{i=1}^{n}\widehat{C_{3,t}^{i}}\frac{U_{i}e^{-\theta U_{i}}}{1-e^{-\theta U_{i}}}.
\end{align*}
The second equality follows from simplifying the terms. The third equality follows from the relation $-\widehat{C_{1,t}^{i}}-\widehat{C_{2,t}^{i}}-\widehat{C_{3,t}^{i}}=-1$ for the $i$th customer in the $t$th iteration. 


Finally, we set the derivative to zero and multiply by $\widehat{\theta^{(t+1)}}$: 
\[
\widehat{\theta^{(t+1)}}\left\{ \sum_{i=1}^{n}\left(\widehat{C_{3,t}^{i}}-1\right)U_{i}\right\} +\sum_{i=1}^{n}\widehat{C_{2,t}^{i}}+\widehat{\theta^{(t+1)}}\left\{ \sum_{i=1}^{n}\widehat{C_{3,t}^{i}}\frac{U_{i}e^{-\widehat{\theta^{(t+1)}}U_{i}}}{1-e^{-\widehat{\theta^{(t+1)}}U_{i}}}\right\} =0.
\]

To show that the solutions are indeed maximizers, we calculated the Hessian of Eq.\  \eqref{eq:loglikelihoodComplete} and showed that it is a negative-definite matrix, as follows:
\[
\begin{aligned}\frac{\partial^{2}\ell}{\partial q^{2}}= & -\frac{1}{q}\sum_{i=1}^{n}\widehat{C_{1,t}^{i}}-\frac{1}{\left(1-q\right)^{2}}\sum_{i=1}^{n}\widehat{C_{3,t}^{i}},\\
\frac{\partial^{2}\ell}{\partial\gamma^{2}}= & -\frac{1}{\gamma^{2}}\sum_{i=1}^{n}\left(1-\widehat{C_{2,t}^{i}}\right),\\
\frac{\partial^{2}\ell}{\partial\theta^{2}}= & -\frac{1}{\theta^{2}}\sum_{i=1}^{n}\left(\widehat{C_{2,t}^{i}}\right)-\sum_{i=1}^{n}\left(\widehat{C_{3,t}^{i}}\right)\frac{U_{i}^{2}e^{-\theta U_{i}}}{\left(1-e^{-\theta U_{i}}\right)^{2}},
\end{aligned}
\]
where $q\in(0,1)$ and $\widehat{C_{j,t}^{i}}\in[0,1]$,
$\forall i,j,t$. We assume that, $\gamma>0$, since average virtual
wait time can not be negative or infinite, and that $\theta>0$, since average customer patience can not be negative or infinite. Additionally, $\sum_{i=1}^{n}\widehat{C_{1,t}^{j}}>0$ for
$j=1,2,3$ and any $t$, since we assume the system has at least one customer from each of the classes defined in Table \ref{tbl:MissingDtaNotation}. Finally, $\left(1-e^{-\theta U_{i}}\right)\in(0,1]$
since $U_{i}>0$ $\forall i$, that is because we expect customers to have some patience, and for the system to offer at least some virtual wait (at least a small $\epsilon$). Therefore, all the second-order partial derivatives are negative. All the second-order mixed derivatives of Eq.\ \eqref{eq:loglikelihoodComplete} are equal to zero, therefore, its Hessian is a $3\mathrm{x3}$ diagonal matrix where the entries are negative, which means that the Hessian is a negative-definite matrix.
\begin{flushright}
$\blacksquare$
\par\end{flushright}

\subsection{Extension: The EM Algorithm with Patience Covariates.}
\label{subsection:EM_covaritates}
Assume that the patience parameter, $\theta$, is influenced by $k$ variables  $\bar{x}=[x_{1},...,x_{k}]$ such that  $\theta | \bar{x}\triangleq e^{\beta_{0}+\beta_{1}x_{1}+...+\beta_{k}x_{k}}=e^{\beta_{0}+\beta^{T} \bar{x}}$. Using similar notations to Section \ref{subsection:EM_algorithm}, the observed data now consists of $D \triangleq \{(U_{i},Y_{i},\triangle_{i},x_{i,1},...,x_{i,k})$, $i=1,...,n\}$. 
We redefine the log-likelihood function as follows:
\begin{equation} \label{eq:log_lik_cov}
\begin{aligned}l(D,\theta,q,\gamma) & =\sum_{i=1}^{n}\left\{ \left(\widehat{C_{1,t}^{i}}\right)\left(-U_{i}e^{\beta_{0}+\beta^T\bar{x}_{i}}\right)\right\} +\sum_{i=1}^{n}\left\{ \log\gamma-\gamma U_{i}\right\} \\
 & +\sum_{i=1}^{n}\left\{ \left(\widehat{C_{2,t}^{i}}\right)\left[\beta_{0}+\beta^T\bar{x}_{i}-U_{i}-\gamma U_{i}e^{\beta_{0}+\beta^T\bar{x}_{i}}+\log(q)-\log\gamma\right]\right\} \\
 & +\sum_{i=1}^{n}\left\{ \left(\widehat{C_{3,t}^{i}}\right)\left[\log(1-q)+\log\left(1-e^{-U_{i}e^{\beta_{0}+\beta^T\bar{x}_{i}}}\right)\right]\right\},
\end{aligned}
\end{equation}
where
 $\beta\triangleq\begin{bmatrix}\beta_{1} & \cdots & \beta_{k}\end{bmatrix}$ and 
$\bar{x}_{i}\triangleq\begin{bmatrix}x_{i,1} & \cdots & x_{i,k}\end{bmatrix}$.

The E-Step of the algorithm does not change. In obtaining $\widehat{C_{1,t}^{i}}$, $\widehat{C_{2,t}^{i}}$, and $\widehat{C_{3,t}^{i}}$,  only $\widehat{C_{3,t}^{i}}$ depends on the current estimation of $\theta$, since $\widehat{C_{3,t}^{i}}= 1_{\{M^{i}=0\}}\left(1-e^{-\widehat{\theta^{(t)}}U_{i}}\right)=1_{\{M^{i}=0\}}\left(1-e^{-{e^{\widehat{\beta_0^{(t)}}+\widehat{\beta^{(t)}}^T\bar{x}_i}}U_{i}}\right)$. The M-Step of the new algorithm  estimates $\gamma$ and $q$, as well as $\beta_{0}$ to $\beta_{k}$. 

We calculate the partial derivative  of \eqref{eq:log_lik_cov} with respect to $\beta_{j}$, as follows:
\[
\begin{aligned}\frac{\partial\ell}{\partial\beta_{j}} & =\sum_{i=1}^{n}\left(\widehat{C_{1,t}^{i}}\right)\left(-x_{i,j}U_{i}e^{\beta_0+\beta^T\bar{x}_i}\right)+\sum_{i=1}^{n}\left(\widehat{C_{2,t}^{i}}\right)x_{i,j}\left(1-U_{i}e^{\beta_0+\beta^T\bar{x}_i}\right)\\
 & +\sum_{i=1}^{n}\left(\widehat{C_{3,t}^{i}}\right)\frac{x_{i,j}U_{i}e^{\beta_0+\beta^T\bar{x}_i}\left(e^{-U_{i}e^{\beta_0+\beta^T\bar{x}_i}}\right)}{1-e^{-\beta_0+\beta^T\bar{x}_i}}\\
 & =\sum_{i=1}^{n}\left(\widehat{C_{1,t}^{i}}\right)\left(-x_{i,j}U_{i}e^{\beta_0+\beta^T\bar{x}_i}\right)+\sum_{i=1}^{n}\left(\widehat{C_{2,t}^{i}}\right)x_{i,j}\left(1-U_{i}e^{\beta_0+\beta^T\bar{x}_i}\right) +\sum_{i=1}^{n}\left(\widehat{C_{3,t}^{i}}\right)\frac{x_{i,j}U_{i}e^{\beta_0+\beta^T\bar{x}_i}}{e^{-U_{i}\left(\beta_0+\beta^T\bar{x}_i\right)}-1}.
\end{aligned}
\]

Equating the partial derivatives above to zero, we get that the estimators for $\beta_{j}^{(t+1)}$ in the $t$th iteration of the algorithm, are given as
the solution to the following $k+1$ estimating equations, where $j=0,...,k$:
\begin{equation}\label{eq:Betas}
\begin{aligned}0= & \sum_{i=1}^{n}\left(\widehat{C_{2,t}^{i}}\right)x_{i,j}\protect\\
+ & \sum_{i=1}^{n}\left\{ \frac{\left[x_{i,j}U_{i}e^{\widehat{\beta_{0}}+{\widehat{\beta^{(t+1)}}}^T\bar{x}_{i}}\right]\left[\widehat{C_{1,t}^{i}}\left(-e^{\widehat{\beta_{0}}+{\widehat{\beta^{(t+1)}}}^T\bar{x}_{i}}\right)+\widehat{C_{2,t}^{i}}\left(-e^{\widehat{\beta_{0}}+{\widehat{\beta^{(t+1)}}}^T\bar{x}_{i}}\right)+1\right]}{e^{-U_{i}\left(\widehat{\beta_{0}}+{\widehat{\beta^{(t+1)}}}^T\bar{x}_{i}\right)}-1}\right\} 
\end{aligned}
\end{equation}
where $\widehat{\beta^{(t+1)}}\triangleq\begin{bmatrix}\widehat{\beta_{1}^{(t+1)}} & \cdots & \widehat{\beta_{k}^{(t+1)}}\end{bmatrix}$.

\vspace{0.3cm}

\begin{algorithm}
\SetAlgoLined
\KwResult{$\widehat{\beta^{(t+1)}_{0}}$,..., $\widehat{\beta^{(t+1)}_{k}}$, $\widehat{q^{(t+1)}}$ and $\widehat{\gamma^{(t+1)}}$.}

Initialization: For every customer $i$, use Eq.~\eqref{eq:ci} to calculate $\widehat{C_{1,0}^{i}}$ and $\widehat{C_{2,0}^{i}}$ and $\widehat{C_{3,0}^{i}}=\hat{\pi}_{i} 1_{\{M^{i}=0\}}$, where $\hat{\pi_{i}}\in\left[0,1\right]$ is chosen randomly. 

To obtain the starting
parameters, $(\widehat{\theta^{(1)}},\widehat{q^{(1)}},\widehat{\gamma^{(1)}})$, solve Equations \eqref{eq:PartialDTheta} and \eqref{eq:PartialDGammaq}, respectively. \\
\vspace{6pt}
 \While{ $ \mid\widehat{\theta^{(t)}}-\widehat{\theta^{(t+1)}}\mid+\mid\widehat{q^{(t)}}-\widehat{q^{(t+1)}}\mid+\mid\widehat{\gamma^{(t)}}-\widehat{\gamma^{(t+1)}}\mid>\epsilon$}{
  E-step: For all  $i \in \{1,...,n\}$ and $j \in \{1,2,3\}$ use Eq.~\eqref{eq:ci} to calculate $\widehat{C_{j,t}^{i}}$ given the observed data
   $D=\{(U_{i},Y_{i},\Delta_{i},x_{i,1},...,x_{i,k})\,,i=1,...,n\}$ and the current estimations of the parameters $(\widehat{\theta^{(t)}},\widehat{q^{(t)}},\widehat{\gamma^{(t)}})$.   \\

\vspace{3pt}
  M-step: Maximize to obtain $(\widehat{\beta^{(t+1)}_{0}},..., \widehat{\beta^{(t+1)}_{k}},\widehat{q^{(t+1)}},\widehat{\gamma^{(t+1)}})$.\
  That is, update the estimations of the parameters using Eqs.~\eqref{eq:Betas} and~\eqref{eq:PartialDGammaq}, respectively; then calculate $\widehat{\theta^{(t+1)}}=e^{\beta_{0}+{\widehat{\beta^{(t+1)}}}^T\bar{x}}$.
 }
 \caption{The EM Algorithm with Covariates}
 \label{EMCV}
\end{algorithm}

\subsection{Patience Estimation in Ticket Queues} 
\label{sec:appendixPatienceTQ}
As mentioned in Section \ref{sec:introduction}, ticketing queues can be viewed as a special case of contact-center operations. 
Here, an arriving customer is given a ticket with their queue number,  then decides whether to join the queue, and then silently abandons. Even if the customer joins the queue, they may silently abandon during the wait. Because the queue is observed and customers that abandon are certainly not served, we do not have the missing-data problem. Instead, like the ED environment of \cite{Yefenof2018}, ticket queue dynamics create both right- and left-censoring. Unlike \cite{Yefenof2018}, ticket queues do not record customer abandonment in real time; hence, we don't have exact patience information on any abandoning customer. Thus, we have only class $C_1$ (served) and $C_3$ (Sab) customers, and no class $C_2$ (Kab) customers.

We formulate a maximum-likelihood estimator (MLE) for customer patience in the ticketing queue as follows:

The likelihood of the recorded data $D=\{(U_{i},Y_{i},\Delta_{i})\,,i=1,...,n\}$
can be written as
\[
\begin{aligned}L(D;\theta,\gamma)= & \prod_{i=1}^{n}\left\{ e^{-\theta U_{i}}\gamma e^{-\gamma U_{i}}\right\} ^{C_{1}^{i}}\left\{ (1-e^{-\theta U_{i}})\gamma e^{-\gamma U_{i}}\right\} ^{C_{3}^{i}}.\end{aligned}
\]

We compute the log of the likelihood
\[
\begin{aligned}l(D,\theta,\gamma) & =\sum_{i=1}^{n}\left\{ \left(C_{1}^{i}\right)\left(\log\gamma-\gamma U_{i}-\theta U_{i}\right)\right\} +\sum_{i=1}^{n}\left\{ \left(C_{3}^{i}\right)\left[\log(1-e^{-\theta U_{i}})+\log\gamma-\gamma U_{i}\right]\right\}. \end{aligned}
\]
The partial derivative with respect to $\gamma$ is
\subsection*{
\[
\frac{\partial\ell}{\partial\gamma}=\frac{n}{\gamma}-\sum_{i=1}^{n}U_{i},
\]
}
which yields
\[
\widehat{\gamma}=n\left\{ \sum_{i=1}^{n}U_{i}\right\} ^{-1}.
\]
The partial derivative with respect to $\theta$ is 
\[
\begin{aligned}\frac{\partial\ell}{\partial\theta} & =\sum_{i=1}^{n}\left(C_{1}^{i}\right)\left(-U_{i}\right)+\sum_{i=1}^{n}\left(C_{3}^{i}\right)\frac{U_{i}e^{-\theta U_{i}}}{1-e^{-\theta U_{i}}}.\end{aligned}
\]
Therefore, $\widehat{\theta}$ is found by setting the partial derivative
with respect to $\theta$ to zero.

\section{EM Algorithm Validation}
\label{app:validation}

\subsection{Accuracy}
\label{subsubsec:EMaccuracy}
As a first examination, we want to evaluate the accuracy of the estimations provided by the EM algorithm and to compare them with the accuracy of previous methods suggested in the literature: \cite{Mandelbaum2013Data} (Method 1) and \cite{Yefenof2018} (Method 2). For this purpose, we simulate data for $\tau$, $W$, and $Y$ with specific parameters $\theta$, $q$, and $\gamma$. We compute $\Delta$ from the realization of $\tau$ and $W$ according to its definition ($\Delta=1_{\{\tau \leq W \}}$). We then estimate  $\widehat{\theta}$, $\widehat{q}$, and $\widehat{\gamma}$ using the EM algorithm to evaluate accuracy. Hence, in this validation strategy, all the assumptions of the EM algorithm hold.

As mentioned, the EM algorithm can cope with missing data, but the other two methods cannot.
In order to use them for this comparison, we need to make certain assumptions on how they handle conversations in the uSab class ($M=0$). To apply  \cite{Yefenof2018}, we have three options of how to classify $M=0$ conversations: classifying them either as served (Sr) customers ($C_{1}=1$) (denoted M2-Sr) or as silent-abandonment customers ($C_{3}=1$) (denoted M2-Sab), or misclassifying them using the same error rate as the SVM model (\S \ref{sec:prob_abnd_WQ}) (denoted M2-SVM).  Here, we simulate the last option by correctly classifying 85\% of the Sab conversations and 76\% of the Sr1 conversations.
To apply the method of \cite{Mandelbaum2013Data}, we have two options of how to classify $M=0$ conversations, either as served customers ($C_{1}=1$) (denoted M1-Sr) or as Kab ($C_{2}=1$) (denoted M1-Ab) , since this method cannot deal with left-censored conversations.

\subsubsection{Accuracy with Regard to the Estimation of $\theta$.}
We generate 200 samples of 2,000 customer conversations using the  parameter combinations stated in Table \ref{tbl:ParametersAcc}. 
Most of the parameters are taken from \cite{Yefenof2018} (Chapter 6), namely, $\theta=4$ and $\gamma=10$ customers per hour (i.e., $E[\tau]=15$ and $E[W]=6$ minutes). We set $q$ to be in the set $\{1,0.9,...,0.1\}$, resulting in a proportion of silent abandonments between 0\% and 26\%. To create higher proportions of Sab customers between 27\% and $44\%$, we need to reduce $\gamma$; we use $\gamma \in \{9,7,5,4\}$ to achieve those abandonment rates. Note that the setting where $\theta<\gamma$ is plausible, since \cite{Brown2005}  found that average customer patience in call centers is greater than average virtual wait time, $E[\tau]>E[W]$. This result has been confirmed to hold in other service environments by several empirical studies, such as \cite{Yefenof2018}, who obtained this result when analyzing data from an ED. All the parameter combinations we choose are designed to keep the simulation within the same $\theta$ less than $\gamma$ setting.

\begin{figure}[htb]
\centering
\subfigure[MSE of $\theta$]{
\includegraphics[width=0.35\textwidth]{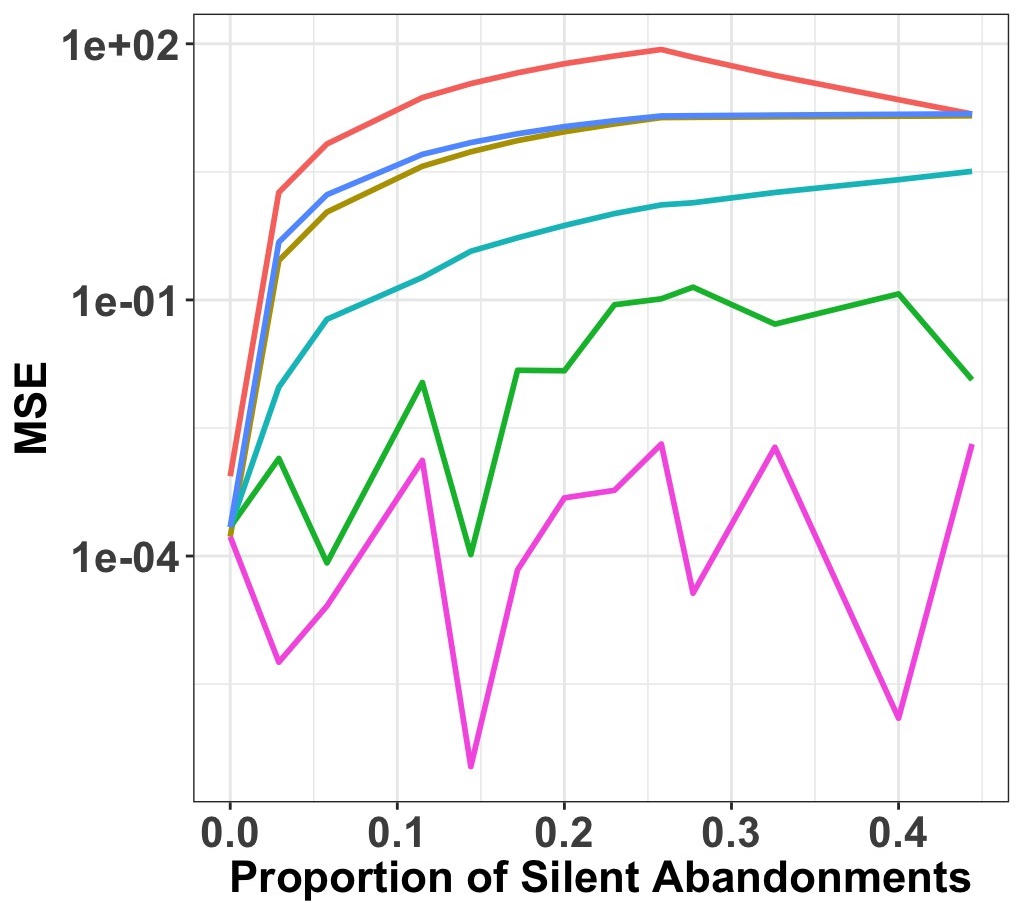} \label{fig:SimulatedData_a}
} 
\subfigure[MSE Ratio of $\theta$ (Baseline of the EM Algorithm)]{
\includegraphics[width=0.56\textwidth]{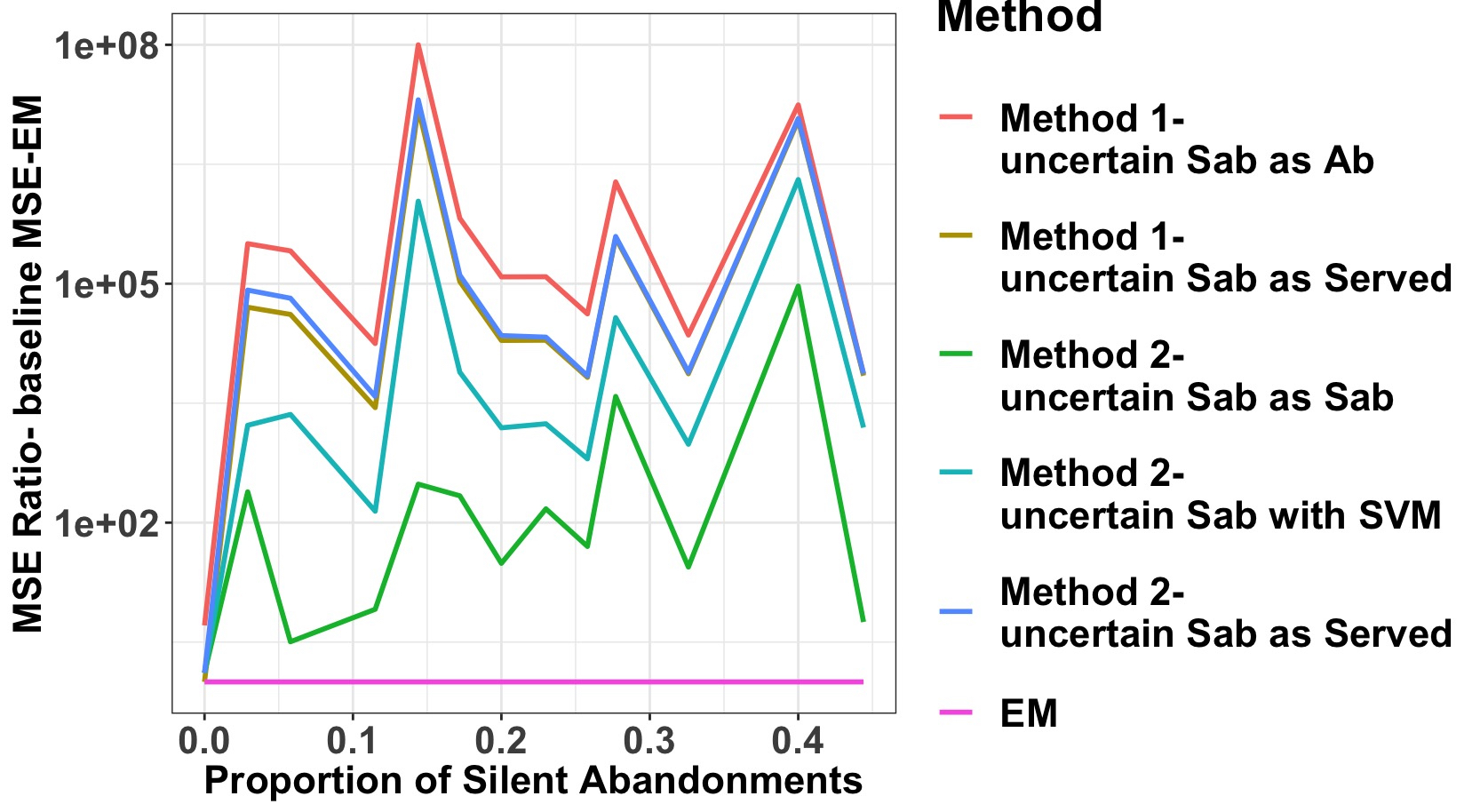}
\label{fig:SimulatedData_b}
}
\caption{Comparison of Accuracy of Customer Patience Parameter, $\theta$, Estimations (Log Scale)}
\label{fig:SimulatedData}
\end{figure}

\begin{figure}[!htb]
\centering
\subfigure[2\% Sab ($\theta=4$, $\gamma=10$, $q=0.9$)]{
\includegraphics[width=0.31\textwidth]{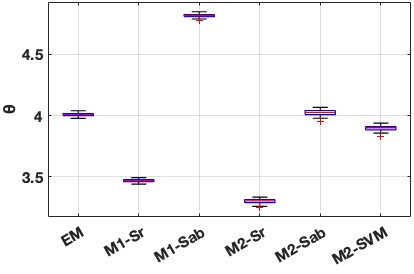} \label{fig:BoxSimulatedD_a}
} 
\subfigure[17\% Sab ($\theta=4$, $\gamma=10$, $q=0.4$)]{
\includegraphics[width=0.31\textwidth]{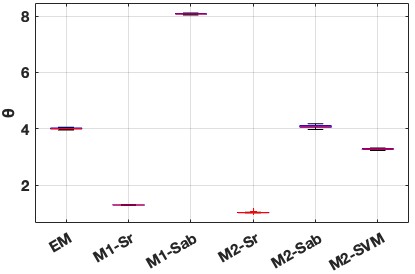} \label{fig:BoxSimulatedD_b}
} 
\subfigure[40\% Sab ($\theta=4$, $\gamma=5$, $q=0.1$)]{
\includegraphics[width=0.31\textwidth]{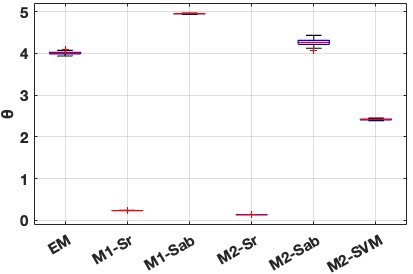} \label{fig:BoxSimulatedD_c}
} 
\caption{Accuracy of Customer-Patience Estimations for Low, Moderate, and High Sab Proportions}
\label{fig:BoxSimulatedD}
\end{figure}

For each sample, we estimate the parameters using the six methods mentioned above. 
We use 100 repetitions of the data sampling and parameter estimation with those methods to calculate the mean and SD of the estimated parameters (see Table \ref{tbl:ParametersAcc}) and create the boxplots (Figure \ref{fig:BoxSimulatedD}). 
Figure \ref{fig:SimulatedData} presents the accuracy results for estimating $\theta$ in a logarithmic scale. Figure EC.\ref{fig:SimulatedData_a} presents the mean squared errors (MSEs) for each model, while Figure EC.\ref{fig:SimulatedData_b} shows the ratio between the MSE of the specific model and the MSE of the EM algorithm (the baseline).
The x-axis in both figures is the proportion of silent abandonments of all arriving customers. Note that we do not report the results of any proportion of silent abandonments that is greater than 45\%, since we would not expect any company to find itself in such a position.

\begin{table}[!htb]
\centering
\caption{Real and Estimated Parameters in EM and Methods 1\&2 Accuracy Tests}
\begin{scriptsize}
\begin{tabular}{llllcrrrcrr} 
\toprule
P(Sab) & \multicolumn{3}{c}{Real} && \multicolumn{3}{c}{Expectation-Maximization} &&  \multicolumn{2}{c}{Method 1 - uSab as Sr} \\ \cline{2-4} \cline{6-8}  \cline{10-11}  
& $\theta$& $\gamma$& $q$ && \multicolumn{1}{c}{$\widehat{\theta}$}& \multicolumn{1}{c}{$\widehat{\gamma}$}& \multicolumn{1}{c}{$\widehat{q}$} && \multicolumn{1}{c}{$\widehat{\theta}$}& \multicolumn{1}{c}{$\widehat{\gamma}$} \tabularnewline
\midrule 
0 & 4& 10& 1.0&&4.01 [3.98, 4.03]&10.02 [9.99, 10.06] & 1.000 [1.000, 1.000] && 4.01 [3.95, 4.03] & 10.03 [9.99, 10.06] \\
0.03 & 4& 10& 0.9&& 4.00 [3.97, 4.02] & 10.00 [9.96, 10.03] & 0.899 [0.898, 0.900] && 3.31 [3.26, 3.33] & 10.00 [9.96, 10.03] \\
0.05 & 4& 10& 0.8&& 3.99 [3.97, 4.01] & 9.99 [9.96, 10.03] & 0.800 [0.798, 0.801] && 2.69 [2.65, 2.71] & 10.00 [9.96, 10.03] \\
0.08 & 4& 10& 0.7&& 4.03 [4.01, 4.05] & 9.98 [9.94, 10.01] & 0.598 [0.596, 0.601] && 1.75 [1.72, 1.76] & 9.98 [9.94, 10.01] \\
0.11 & 4& 10& 0.6&& 4.03 [4.01, 4.05] & 9.98 [9.94, 10.01] & 0.598 [0.596, 0.601] && 1.75 [1.72, 1.76] & 9.98 [9.94, 10.01] \\
0.14 & 4& 10& 0.5&& 4.00 [3.98, 4.01] & 9.99 [9.95, 10.02] & 0.499 [0.497, 0.502] && 1.36 [1.33, 1.37] & 9.99 [9.95, 10.02] \\
0.17 & 4& 10& 0.4&& 3.99 [3.98, 4.00] & 10.02 [9.99, 10.05] & 0.398 [0.396, 0.401] && 1.03 [1.00, 1.03] & 10.02 [9.99, 10.05] \\
0.2 & 4& 10& 0.3&& 4.02 [4.01, 4.03] & 10.00 [9.97, 10.03] & 0.300 [0.297, 0.302] && 0.73 [0.71, 0.73] & 10.00 [9.97, 10.03] \\
0.22 & 4& 10& 0.2&& 4.02 [4.01, 4.03] & 10.00 [9.96, 10.03] & 0.198 [0.196, 0.200] && 0.46 [0.44, 0.46] & 10.00 [9.96, 10.03] \\
0.25 & 4& 10& 0.1&& 4.04 [4.04, 4.05] & 10.00 [9.97, 10.03] & 0.100 [0.098, 0.102] && 0.22 [0.21, 0.22] & 10.00 [9.97, 10.03] \\
0.27 & 4& 9& 0.1&& 3.99 [3.99, 4.00] & 9.00 [8.97, ~9.03] & 0.101 [0.099, 0.102] && 0.21 [0.20, 0.21] & 9.01 [8.97,~ 9.03] \\
0.32 & 4& 7& 0.1&& 3.96 [3.95, 3.96] & 7.01 [6.98, ~7.03] & 0.100 [0.099, 0.102] && 0.18 [0.17, 0.18] & 7.01 [6.98, ~7.03] \\
0.4 & 4& 5& 0.1&& 4.00 [3.99, 4.01] & 4.99 [4.97, ~5.00] & 0.100 [0.098, 0.101] && 0.14 [0.13, 0.14] & 4.99 [4.97, ~5.00] \\
0.44 & 4& 4.1& 0.1&& 3.95 [3.95, 3.96] & 4.10 [4.09, ~4.11] & 0.099 [0.098, 0.100] && 0.11 [0.11, 0.11] & 4.10 [4.09, ~4.11] \\
\bottomrule
\multicolumn{11}{l}{Upper and lower 95\% confidence intervals inside brackets.}
\end{tabular}

\vspace*{5mm}

\begin{tabular}{llllcrrcrr} 
\toprule
P(Sab) & \multicolumn{3}{c}{Real} && \multicolumn{2}{c}{Method 1 - uSab as Ab} &&
\multicolumn{2}{c}{Method 2 - uSab as Sr}\\ \cline{2-4} \cline{6-7}  \cline{9-10}  
& $\theta$& $\gamma$& $q$ && \multicolumn{1}{c}{$\widehat{\theta}$}& \multicolumn{1}{c}{$\widehat{\gamma}$}&& \multicolumn{1}{c}{$\widehat{\theta}$}& \multicolumn{1}{c}{$\widehat{\gamma}$} \tabularnewline
\midrule 
0 & 4& 10& 1.0&& 4.01 [3.95, 4.03] & 10.03 [9.99, 10.06] && 4.01 [3.98, 4.03] & 10.03 [9.99, 10.06] \\
0.03 & 4& 10& 0.9&& 4.04 [3.98, 4.06] & 10.00 [9.96, 10.03] && 3.46 [3.43, 3.48] & 10.00 [9.96, 10.03] \\
0.05 & 4& 10& 0.8&& 3.99 [3.92, 4.01] & 10.00 [9.96, 10.03] && 2.96 [2.94, 2.98] & 10.00 [9.96, 10.03] \\
0.08 & 4& 10& 0.7&& 4.10 [4.00, 4.12] & 9.98 [9.94, 10.01] && 2.08 [2.06, 2.09] & 9.98 [9.94, 10.01] \\
0.11 & 4& 10& 0.6&& 4.10 [4.00, 4.12] & 9.98 [9.94, 10.01] && 2.08 [2.06, 2.09] & 9.98 [9.94, 10.01] \\
0.14 & 4& 10& 0.5&& 4.01 [3.89, 4.02] & 9.99 [9.95, 10.02] && 1.67 [1.65, 1.68] & 9.99 [9.95, 10.02] \\
0.17 & 4& 10& 0.4&& 4.12 [3.96, 4.12] & 10.02 [9.99, 10.05] && 1.29 [1.27, 1.30] & 10.02 [9.99, 10.05] \\
0.2 & 4& 10& 0.3&& 4.12 [3.93, 4.11] & 10.00 [9.97, 10.03] && 0.95 [0.93, 0.95] & 10.00 [9.97, 10.03] \\
0.22 & 4& 10& 0.2&& 4.30 [4.02, 4.25] & 10.00 [9.96, 10.03] && 0.61 [0.59, 0.61] & 10.00 [9.96, 10.03] \\
0.25 & 4& 10& 0.1&& 4.32 [3.85, 4.16] & 10.00 [9.97, 10.03] && 0.30 [0.29, 0.30] & 10.00 [9.97, 10.03] \\
0.27 & 4& 9& 0.1&& 4.38 [3.91, 4.22] & 9.01 [8.97, ~9.03] && 0.29 [0.28, 0.29] & 9.01 [8.97, ~9.03] \\
0.32 & 4& 7& 0.1&& 4.23 [3.83, 4.11] & 7.01 [6.98, ~7.03] && 0.27 [0.26, 0.27] & 7.01 [6.98, ~7.03] \\
0.4 & 4& 5& 0.1&& 4.34 [3.99, 4.26] & 4.99 [4.97, ~5.00] && 0.23 [0.23, 0.24] & 4.99 [4.97, ~5.00] \\
0.44 & 4& 4.1& 0.1&& 4.11 [3.85, 4.07] & 4.10 [4.09, ~4.11] && 0.21 [0.21, 0.21] & 4.10 [4.09, ~4.11] \\
\bottomrule
\multicolumn{9}{l}{Upper and lower 95\% confidence intervals inside brackets.}
\end{tabular}

\vspace*{5mm}

\begin{tabular}{llllcrrcrr} 
\toprule
P(Sab) & \multicolumn{3}{c}{Real} && \multicolumn{2}{c}{Method 2 - uSab as Sab} && \multicolumn{2}{c}{Method 2 - uSab with SVM}\\
\cline{2-4} \cline{6-7}  \cline{9-10} 
&\multicolumn{1}{c}{$\widehat{\theta}$}& \multicolumn{1}{c}{$\widehat{\gamma}$}&
\multicolumn{1}{c}{$\widehat{q}$}&&  \multicolumn{1}{c}{$\widehat{\theta}$}& \multicolumn{1}{c}{$\widehat{\gamma}$} && \multicolumn{1}{c}{$\widehat{\theta}$}& \multicolumn{1}{c}{$\widehat{\gamma}$}\tabularnewline
\midrule 
0 & 4& 10& 1.0&& 4.01 [3.98, 4.03] & 10.03 [9.99, 10.06] && 4.01 [3.95, 4.03] & 10.03 [9.99, 10.06] \\
0.03 & 4& 10& 0.9&& 4.81 [4.78, 4.83] & 8.65 [8.61, 8.68] && 3.90 [3.85, 3.92] & 10.00 [9.96, 10.03] \\
0.05 & 4& 10& 0.8&& 5.56 [5.52, 5.58] & 7.41 [7.37, 7.43] && 3.76 [3.70, 3.77] & 10.00 [9.96, 10.03] \\
0.08 & 4& 10& 0.7&& 6.91 [6.87, 6.93] & 5.16 [5.13, 5.18] && 3.57 [3.51, 3.59] & 9.98 [9.94, 10.01] \\
0.11 & 4& 10& 0.6&& 6.91 [6.87, 6.93] & 5.16 [5.13, 5.18] && 3.57 [3.51, 3.59] & 9.98 [9.94, 10.01] \\
0.14 & 4& 10& 0.5&& 7.51 [7.48, 7.53] & 4.15 [4.12, 4.17] && 3.39 [3.32, 3.41] & 9.99 [9.95, 10.02] \\
0.17 & 4& 10& 0.4&& 8.07 [8.04, 8.10] & 3.24 [3.21, 3.25] && 3.27 [3.21, 3.28] & 10.02 [9.99, 10.05] \\
0.2 & 4& 10& 0.3&& 8.60 [8.57, 8.62] & 2.35 [2.33, 2.36] && 3.14 [3.08, 3.15] & 10.00 [9.97, 10.03] \\
0.22 & 4& 10& 0.2&& 9.10 [9.06, 9.12] & 1.52 [1.50, 1.52] && 2.99 [2.92, 3.00] & 10.00 [9.96, 10.03] \\
0.25 & 4& 10& 0.1&& 9.57 [9.53, 9.59] & 0.74 [0.72, 0.74] && 2.86 [2.79, 2.87] & 10.00 [9.97, 10.03] \\
0.27 & 4& 9& 0.1&& 8.64 [8.61, 8.66] & 0.65 [0.64, 0.66] && 2.83 [2.76, 2.84] & 9.01 [8.97, ~9.03] \\
0.32 & 4& 7& 0.1&& 6.81 [6.78, 6.83] & 0.47 [0.46, 0.47] && 2.65 [2.60, 2.66] & 7.01 [6.98, ~7.03] \\
0.4 & 4& 5& 0.1&& 4.93 [4.91, 4.94] & 0.29 [0.29, 0.29] && 2.40 [2.36, 2.41] & 4.99 [4.97, ~5.00] \\
0.44 & 4& 4.1& 0.1&& 4.09 [4.08, 4.10] & 0.22 [0.22, 0.22] && 2.21 [2.17, 2.22] & 4.10 [4.09, ~4.11] \\
\bottomrule
\multicolumn{9}{l}{Upper and lower 95\% confidence intervals inside brackets.}
\end{tabular}
\end{scriptsize}
\label{tbl:ParametersAcc}
\end{table}

Table \ref{tbl:ParametersAcc} and Figure EC.\ref{fig:SimulatedData_a} show that the errors of the EM algorithm are quite small (less than 0.2\%) in all of the parameter combinations.
Figure EC.\ref{fig:SimulatedData_b} shows that both ways of implementing Method 1 (which accounts only for right-censored data) are very inaccurate. Specifically, estimating customer patience while ignoring the silent-abandonment phenomenon altogether results in an error rate that is $O(10^{8})$ higher than that of the EM baseline. A similar picture emerges when implementing Method 2, which assumes that all the uncertain conversations are served. Here too, the error rate is $O(10^{8})$ higher than that of the EM algorithm. 
If we take silent-abandonment conversations into account to the extent that we regard them as left-censored conversations but ignore the missing data, we obtain a (relatively) lower error rate. This is apparent when we look at the other two ways of implementing Method 2: either by considering all missing data to be Sab conversations or by completing the data with an SVM model. The problem with the latter approach is that the classification is considered correct, whereas a classification model is not completely accurate but has certain  sensitivity and specificity proportions. However, both of the abovementioned options yield less accurate results than the EM algorithm does: the respective error rates are $O(10^{5})$ and $O(10^{7})$ greater than that of the EM algorithm. To conclude, our algorithm outperforms all other methods for estimating customer patience. Note that when there is no silent abandonment in the system (0\% in Figure \ref{fig:SimulatedData}), all methods achieve the same performance level; this suggests that the EM algorithm can be used also in cases where the company does not have Sab customers or is unsure whether they exist. 


In order to analyze whether the estimations are biased or just have larger variance, we present the boxplots in Figure \ref{fig:BoxSimulatedD}. We include boxplots for only three of the parameter combinations we simulated. The parameters were chosen to enable comparison of estimations of parameters that result in low (2\%), moderate (17\%), and high (40\%) levels of silent abandonment in subfigures (a), (b), and (c), respectively. We see that regardless of the level of silent abandonment, the EM algorithm produces the most accurate estimation of customer patience, followed by Method 2 taking uSab as Sab (M2-Sab), which overestimates $\theta$ (underestimates average customer patience). 


\subsubsection{Accuracy with Regard to the Estimation of $q$ and $\gamma$.}
\label{sec:ApendixEMaccuracy}

\begin{figure}[htb]
\centering
\subfigure[MSE of $q$]{
\includegraphics[width=0.355\textwidth]{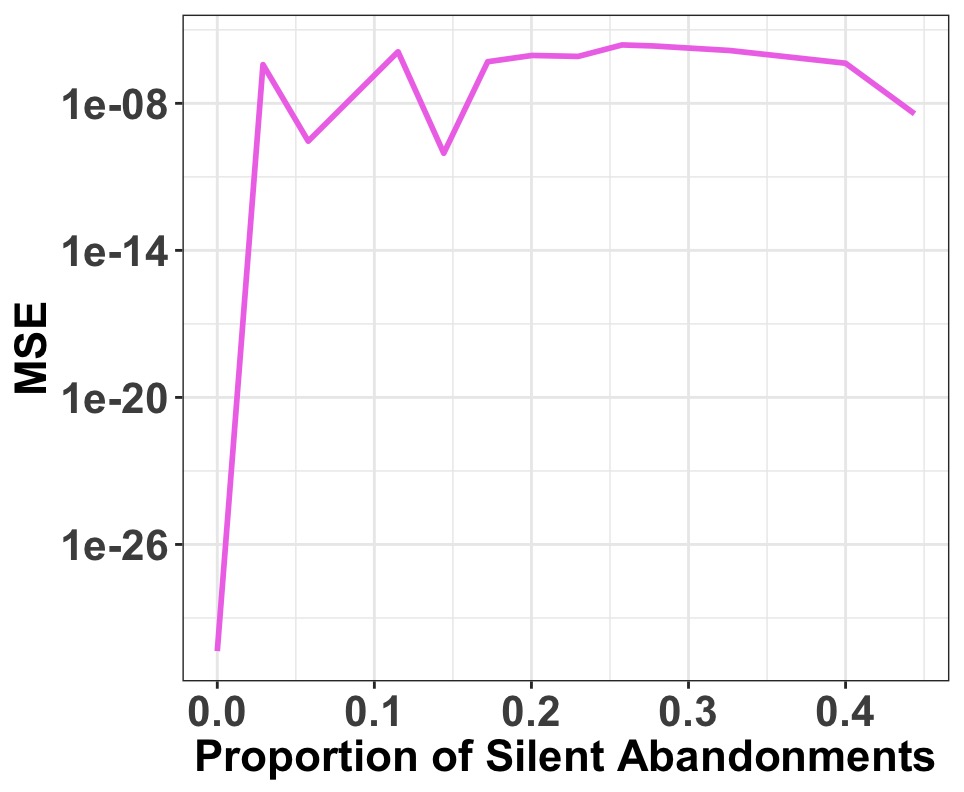} \label{fig:SimulatedDataq_a}
} 
\subfigure[MSE of $\gamma$]{
\includegraphics[width=0.555\textwidth]{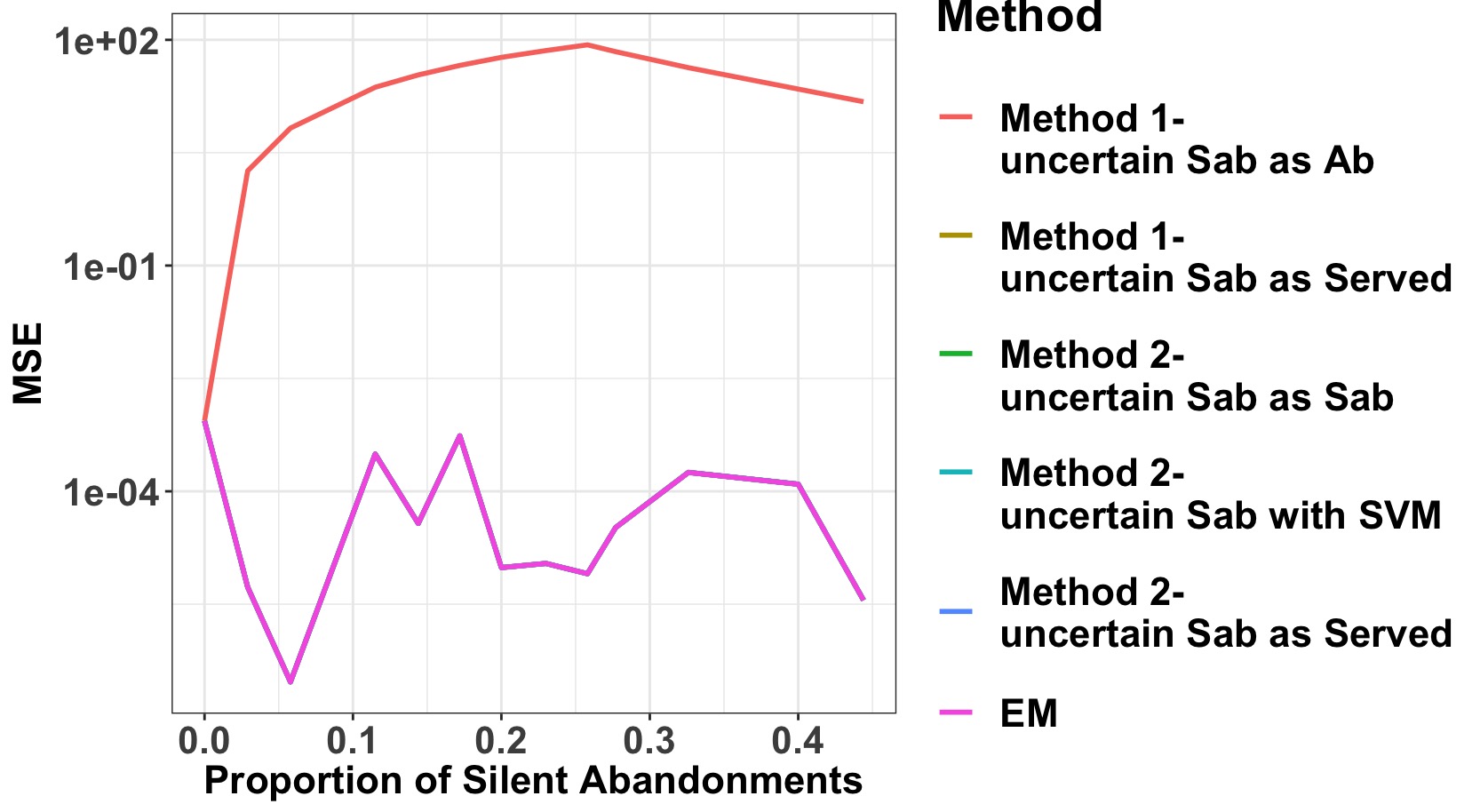}
\label{fig:SimulatedDataGamma_b}
}
\caption{Comparison of Accuracy of Probability of Indicating Abandonment ($q$) and Virtual Wait Time Parameter ($\gamma$), Estimations (Log Scale). Note That the $\gamma$ Estimation of All the Methods (Except One) Is Exactly the Same as for the EM Algorithm}
\label{fig:SimulatedDataGamma}
\end{figure}

Here, we provide the MSEs of the estimations of $q$ and $\gamma$ using the simulated data described in Appendix \ref{subsubsec:EMaccuracy}. Note that here we cannot compare estimation of $q$ to other methods since only the EM algorithm estimates this parameter. Figure EC.\ref{fig:SimulatedDataq_a} presents the MSE of $q$ as a function of the proportion of silent-abandonment customers (out of all the arriving customers) in logarithmic scale. We show here that the error rate is very small; therefore, the estimation is very accurate. 


Figure EC.\ref{fig:SimulatedDataGamma_b} presents the MSE results for estimating $\gamma$ in a logarithmic scale. 
The x-axis is the proportion of silent-abandonment customers (out of all the arriving customers). We note that the estimation of most of the methods is exactly the same, except for the method of \cite{Mandelbaum2013Data}, where we take the uncertain conversations ($M=0$) to be $C_{2}$. For this reason, most of the lines in Figure EC.\ref{fig:SimulatedDataGamma_b} are exactly the same as in the estimation of the EM algorithm. We conclude that the error rate of the EM algorithm as well as for most of the other methods  is quite small. 

\subsection{Sensitivity Analysis} \label{subsubsec:Sensitivity}
The next tests are designed to investigate the sensitivity of the EM algorithm under the initial conditions. In addition, we would like to know whether starting the algorithm under some sophisticated initial conditions, for example, by using a classification model, such as the one we developed in Section \ref{sec:prob_abnd_WQ}, helps the model converge to a more accurate estimation. Accordingly, we first investigate the sensitivity of the EM algorithm to $\hat{\pi}_{i}$. Note that by Algorithm \ref{EM}, $\hat{\pi}_{i}$  affects $\widehat{C_{3,0}^{i}}$ and $\widehat{C_{1,0}^{i}}$ only for the class of uSab customers.  

We generated 200 samples of 2,000 customer conversations using the following parameters: $\theta=4, \gamma=10$
, and $q=0.5$. For each sample, we estimate the parameters ($\widehat{\theta}, \widehat{q}$, $\widehat{\gamma}$) using the EM algorithm (with 100 repetitions) and consider the average of those parameters as the final estimator for that sample.
We present here four variants of the starting weights for all $M^{i}=0$ conversations.
\begin{lyxlist}{00.00.0000}
\item[\emph{All Sab:}] Setting all $M^{i}=0$ conversations to be silent-abandonment conversations with probability 1. Formally,  $\widehat{C_{3,0}^{i}}=1$ and $\widehat{C_{1,0}^{i}}=0$ for all conversations with $M^{i}=0$.   
\item[\emph{All Sr:}] Setting all $M^{i}=0$ conversations to be served-in-one-exchange conversations with probability 1. Formally,  $\widehat{C_{3,0}^{i}}=0$ and $\widehat{C_{1,0}^{i}}=1$ for all conversations with $M^{i}=0$.
\item[\emph{50:50:}] Setting 50\% of the conversations to be short-service conversations and 50\% to be Sab conversations; that is, for 50\% of the conversations with $M^{i}=0$, we set $\widehat{C_{3,0}^{i}}=1$, and for the rest of the $M^{i}=0$ conversations, we set  $\widehat{C_{1,0}^{i}}=1$. We choose this option because about 50\% of the uSab conversations are Sab and about 50\% are Sr1 within our data (see~\S\ref{sec:prob_abnd}).
\item [\emph{Best classifier:}] For conversations with $M^{i}=0$, we simulate a classification with sensitivity and specificity proportions according to our best classification model from Section \ref{sec:prob_abnd}; therefore, 85\% of the Sab conversations are classified correctly, as are 76\% of the Sr1 conversations. That is, 85\% of the actual $C_{3}=1$ conversations and 76\% of the actual $C_{1}=1$ conversations are identified as such; therefore, we set the correct values to $\widehat{C_{3,0}^{i}}$ and $\widehat{C_{1,0}^{i}}$ for those. 
For the remainder of the conversations, we set wrong values on  $\widehat{C_{3,0}^{i}}=1$ and $\widehat{C_{1,0}^{i}}=1$; for example,
for an actual $C_{3}=1$: $\widehat{C_{3,0}^{i}}=0$, $\widehat{C_{1,0}^{i}}=1$. 

\end{lyxlist}

\begin{figure}[!htb]
\centering
\includegraphics[width=0.3\textwidth]{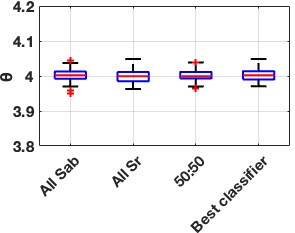}
\caption{Sensibility Analysis (Setting: $\theta=4$)}
\label{fig:Sensitivity}
\end{figure}

Figure \ref{fig:Sensitivity} shows that the estimations of customer patience are stable and do not change when different initial values are inserted in the EM algorithm. This suggests that one may not need to use the output of the classification model we developed in Section \ref{sec:prob_abnd}  (or any model with similar sensitivity and specificity proportions) as starting probabilities in the EM algorithm. 



\begin{figure}[!htb]
\centering
\subfigure[$q=0.5$]{
\includegraphics[width=0.3\textwidth]{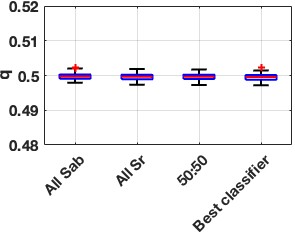} \label{fig:Sensitivityq}
} 
\subfigure[$\gamma=10$]{
\includegraphics[width=0.3\textwidth]{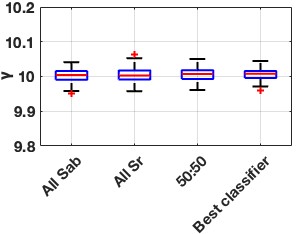} \label{fig:SensitivityGamma}
} 
\caption{Sensitivity Analysis}
\label{fig:More_Sensitivity}
\end{figure}

Next, we provide a sensitivity estimation for $q$ and $\gamma$ (see Figures EC.\ref{fig:Sensitivityq} and EC.\ref{fig:SensitivityGamma}, respectively),  using the same simulated data. The results are consistent, showing that the EM algorithm is not sensitive to the initial values.

\subsection{Robustness Check of Parameter Estimation of Real Data}
\label{sec:ApendixEMRobustness}

A potential problem with EM algorithms is that they might converge to a saddle point (Chapter 8 of \citealt{little2002statistical}). 
To verify that this does not happen here, we started our EM algorithm with different weights. Specifically, we estimated the parameters by using the EM algorithm and taking the starting weights $\widehat{C_{3,0}^{i}}$ for the $M^{i}=0$ conversations to be $1$, $0$, $0.5$, or $\hat{\pi_{i}}$  from the SVM model, 
presented in Section \ref{sec:prob_abnd}. 
In every case, the obtained parameters ($\widehat{\theta}, \widehat{q}$, $\widehat{\gamma}$) were consistent, verifying that the algorithm did  not converge to a saddle point when applied to the real  data. 

\begin{figure}[!htb]
\centering
\includegraphics[width=0.4\textwidth]{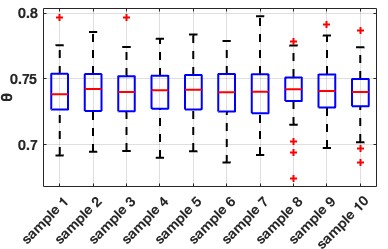}
\caption{Estimations of the Parameter of Customer Patience ($\theta$) Using the EM Algorithm in Subsamples of the Write-in-queue Dataset (May 2017). Horizontal Line Indicates Estimation Based on the Entire Dataset $\hat{\theta}=0.739$}
\label{fig:tetaSamples}
\end{figure}

\begin{figure}[!htb]
\centering
\subfigure[Probability of Indicating Abandonment ($q$)]{
\includegraphics[width=0.4\textwidth]{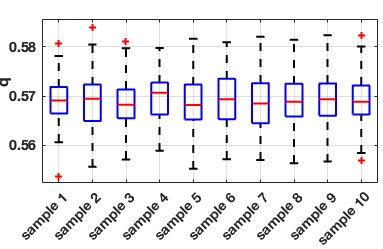} \label{fig:qSamples}
} 
\subfigure[Virtual Wait Time Parameter  ($\gamma$)]{
\includegraphics[width=0.4\textwidth]{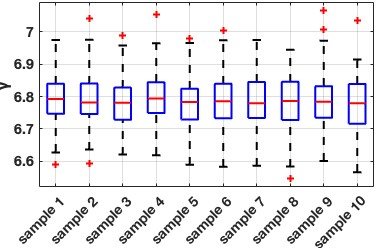} \label{fig:virtualWaitSamples}
} 
\caption{Estimated $q$ and $\gamma$ Using the EM Algorithm in Samples of the Write-in-queue Data, Horizontal Lines Indicate Estimation Based on the Entire Dataset $\hat{q}=0.58$ and $\hat{\gamma}=6.78$}
\label{fig:More_Samples}
\end{figure}

Finally, we performed several robustness checks by dividing the data set into 10--15 samples and estimating patience in each one using the EM algorithm 100 times. We performed these tests to make sure that the results that we obtained from the monthly data ($\hat{\theta}=0.739$, $\hat{q}=0.58$, and $\hat{\gamma}=6.78$) are robust. We find that the estimations of $\theta$, $q$, and $\gamma$ from subsamples of the dataset are consistent with those of the whole dataset corpus. See Figures \ref{fig:tetaSamples}, EC.\ref{fig:qSamples}, and EC.\ref{fig:virtualWaitSamples}. 
%
Note that the estimation of $q$ using a smaller sample results in a small bias. 

\subsection{Accuracy of Estimations Using the Queueing Model}
\label{sec:ApendixEMqueueing}

The accuracy analysis presented in Appendixes  \ref{subsubsec:EMaccuracy} and \ref{subsubsec:Sensitivity} assumed that customer patience and virtual wait time are independent and exponentially distributed, just like the EM algorithm assumption. However, the queueing model developed in Section \ref{sec:managerial} assumed a more realistic setting in which the data may incorporate some dependencies arising from the system dynamics. Specifically, the queueing simulation does not impose the exponential assumption on the virtual wait time; instead, this variable is determined by the system load (that depends on arrival and service processes of the queueing system). Using this model, we can test the EM algorithm's robustness to the assumptions we made in Assumption \ref{assumption:1}. To do so, we check here the EM algorithm's performance using simulated data obtained by that queueing model in two settings: (a) with the parameters of \cite{Yefenof2018} (\S\ref{subsec:queue_yefenof}) and (b) with the parameters of our write-in-queue dataset (\S\ref{subsec:queue_data}).

\subsubsection{Queueing Simulation with the Parameters of \cite{Yefenof2018}.}
\label{subsec:queue_yefenof}

First, we place ourselves in the setting found in \cite{Yefenof2018}, where the parameters used in the simulation are $\theta=4$ and $q=0.5$. Since the queueing simulation does not impose a specific virtual wait time (i.e., a specific $\gamma$), we need to set this variable using the queueing dynamics. To do so, we calibrate $\mu_{Sr}=9.5$ and $\mu_{Sab}=30$, so that the virtual wait time parameter is $\gamma=10$ (as in \citealt{Yefenof2018}). 
Figure EC.\ref{fig:PQue} shows the accuracy of estimating $\theta$ using the five methods we compared in Appendix \ref{subsubsec:EMaccuracy}. The results are consistent with our previous results, suggesting that only the EM algorithm is able to estimate $\theta$ accurately. Figures EC.\ref{fig:qQue} and EC.\ref{fig:VWQue} show the accuracy of estimating $q$ and $\gamma$, respectively. It can be seen that the EM algorithm accurately estimates $q$, whereas for $\gamma$, it estimates as accurately as most of the other methods.
\begin{figure}[!htb]
\centering
\subfigure[Patience Parameter ($\theta$)]{
\includegraphics[width=0.31\textwidth]{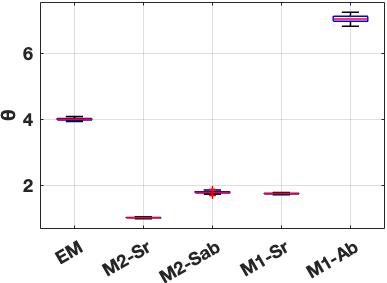} \label{fig:PQue}
} 
\subfigure[Probability of Indicating Abandonment ($q$)]{
\includegraphics[width=0.31\textwidth]{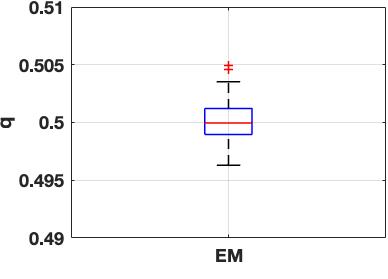} \label{fig:qQue}
} 
\subfigure[Virtual Wait Time Parameter ($\gamma$)]{
\includegraphics[width=0.31\textwidth]{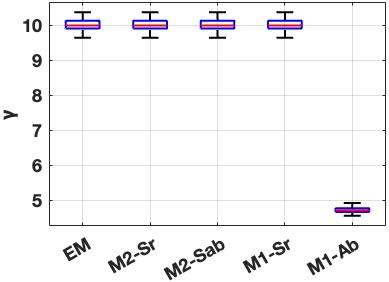} \label{fig:VWQue}
} 
\caption{Queueing Model Simulation ($\theta=4, q=0.5$, $\gamma=10$ as Determined by System Dynamics)}
\label{fig:More_Queue}
\end{figure}

\subsubsection{Queueing Simulation with the Parameters of the Write-in-queue Dataset.}
\label{subsec:queue_data}

In this simulation, we used parameters that were evaluated from the real messaging data at a highly loaded time of the day (weekdays at 12:00). Specifically, the arrival rate is $\lambda=753$ customers per hour and the service rate is $\mu_{Regular}=1.22$, calculated using the net time a conversation stayed open in the system (from agent assignment until the last message written by either the customer or the agent). 
For simplicity, we first assume that $\mu_{Sr}=\mu_{Sab}$.  
The number of full-time equivalent ``agents" is $n=452$, which is taken to be the average number of online agents ($113$) per hour (either idling or serving customers) times 4, a typical average concurrency level in the system. Thus, $n$ represents the number of slots available for service, and 
$q$ is taken to be 0.332, a result that follows from the following calculation: 

\begin{equation} \label{eq:qrelation}
q=\frac{Pr\left\{ C_2=1\right\} }{Pr\{C_2=1\}+Pr\{C_3=1\}} =\frac{Pr\left\{ C_2=1\right\} }{Pr\left\{ C_3=1\mid M=0\right\} Pr\left\{ M=0\right\} +Pr\left\{ C_2=1\right\} },
\end{equation}
where $Pr\left\{ C_2=1\right\} =0.0716$ and $Pr\left\{ M=0\right\} =0.2616$ are
calculated using data on conversations without uncertainty (with complete data). $Pr\left\{ C_3=1\mid M=0\right\} =0.55$ is obtained from the classification model of Section \ref{sec:prob_abnd_WQ}.
The patience parameter $\theta=60/81.2=0.739$ (rate per hour) is the result we obtained from the estimation of the EM algorithm in Section \ref{subsubsec:EMValidation_data}. 
We simulate  our queueing model for a period of 1 month (with 100 repetitions), excluding a warm-up period of 2 hours till the system achieves steady state.



\begin{figure}[!htb]
\centering
\subfigure[Patience Parameter ($\hat{\theta}=0.739$ EM)]{
\includegraphics[width=0.31\textwidth]{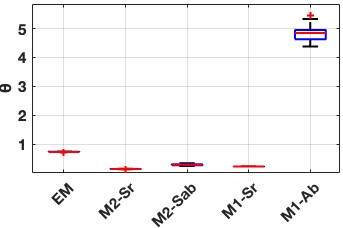} \label{fig:PQueMessag}
} 
\subfigure[Probability of Indicating Abandonment ($\hat{q}=0.332$ Data)]{
\includegraphics[width=0.31\textwidth]{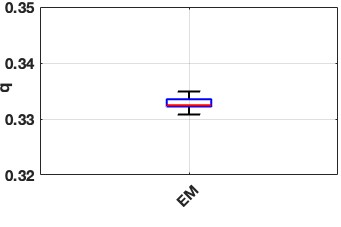} \label{fig:qQueMesg}
} 
\subfigure[Virtual Wait Time Parameter ($\hat{\gamma}=6.782$ EM)]{
\includegraphics[width=0.31\textwidth]{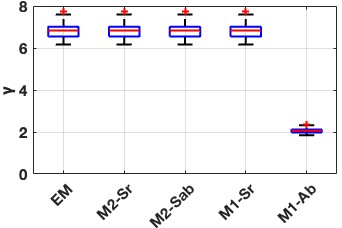} \label{fig:GammaQueMesg}
} 
\caption{Queueing Model Simulation, with Write-in-queue System Parameters}
\label{fig:More_Queue_messaging}
\end{figure}

We estimate $\theta$, $q$, and $\gamma$ from the simulated data using the EM algorithm as well as the methods described in the main text \citep{Mandelbaum2013Data,Yefenof2018}. The results are presented in Figure EC.\ref{fig:PQueMessag}. Again,  the EM algorithm is the only method that is able to accurately estimate the customer patience rate, $\theta$, and it has no close competitor. The closest estimator to the EM algorithm is \citet{Yefenof2018} with $M=0$ considered as silent abandonment. However, even this model is far from accurate, estimating the average customer patience rate per hour to be $\hat{\theta}=0.113$, implying that the average customer patience is $E[Patience]=\frac{60}{\hat{\theta}}=\frac{60}{0.113}=531.45$ minutes. This number is six times larger than the parameter we entered to this simulation. 
Accuracy results for the estimation of $q$ and $\gamma$ in this simulation are presented in Figures EC.\ref{fig:qQueMesg} and EC.\ref{fig:GammaQueMesg}, respectively.   

\section{Estimated Performance Measures  and Parameter Estimation of No-write-in-queue System Dataset}
\label{sec:apexMOPS}

Table \ref{tbl:MOPS} presents the average performance measures of the no-write-in-queue dataset and performance estimations using the  simulation models presented in  Section \ref{sec:model_fitting}. 
Table \ref{tbl:ParametersHour} presents the arrival rate, service rate, and number of servers in each hour of the day estimated using the no-write-in-queue dataset and used for the simulations in Section \ref{sec:model_fitting}.

\begin{table}[!htb]
 \centering
  \caption{Estimated Performance Measures No-write-in-queue System Using Simulation}
  \begin{scriptsize}
  \begin{tabular}{lcccccc} \toprule
  Performance & \multicolumn{1}{c}{Weekdays, Feb 2017}   & (1) Ignoring & (2) Sab as left-censored, &(3) Sab as left-censored, &(4) Considering Sab \\
  Measure & Mean (SD) &Sab & nonparametric & parametric & as time-consuming\\ 
  \midrule
  $P\{\text{Wait}>0\}$ &0.59 (0.03) & 0.54 (0.27) & 0.38 (0.21) & 0.31 (0.17) & 0.54 (0.17)\tabularnewline
  $P\{\text{Ab}\}$ &0.23 (0.04) & 0.09 (0.10) & 0.12 (0.10) & 0.14 (0.10) & 0.18 (0.12)\tabularnewline
  E[Queue] & 1.72 (0.46) & 3.13 (3.02) & 0.49 (1.71) & 0.38 (0.31) & 1.11 (0.39) \tabularnewline
  E[Wait] & 2.41 (0.72) & 2.90 (2.83) & 0.94 (1.46) & 0.54 (0.43) & 2.17 (0.81) \tabularnewline
  E[Wait$|$Served] & 2.04 (0.58) &3.01 (3.37) & 0.92 (2.06) & 0.20 (0.17) &1.79 (0.39)\tabularnewline
  \bottomrule
 \multicolumn{4}{l}{Standard deviations (SD) are provided in parentheses.}
  \end{tabular}
  \end{scriptsize}
 \label{tbl:MOPS}
\end{table}

\begin{table}[!htb]
 \centering
  \caption{Hourly Parameters No-write-in-queue Dataset (February 2017)}
  \begin{scriptsize}
  \begin{tabular}{lccccc} \toprule
  Hour of the day & $\lambda_t$ & $\mu_{Sr}$ & $\mu_{Sab}$& $\mu_{All}$ & $n_t$  \\
  \midrule
  8:00 & 47.30 & 4.49 & 12.63 &4.12& 8 \tabularnewline
  9:00 & 51.15 & 4.86 & 11.01 &4.81& 8 \tabularnewline
  10:00 & 49.10 & 5.06 & 16.51 &4.85& 8 \tabularnewline
  11:00 & 51.10 & 4.95 & 10.95 &4.88& 8 \tabularnewline
  12:00 & 51.40 & 5.18 & 10.06 &5.07& 8   \tabularnewline
  13:00 & 52.80 & 4.89 & 9.79 &4.82& 8 \tabularnewline
  14:00 & 64.60 & 4.95 & 10.95 &4.97& 11 \tabularnewline
  15:00 & 66.10 & 5.23 & 14.29 &5.16& 11 \tabularnewline
  16:00 & 51.00 & 4.92 & 13.37 &4.90& 8 \tabularnewline
  17:00 & 48.30 & 4.93 & 11.59 &4.87& 8   \tabularnewline
  18:00 & 44.45 & 4.86 & 9.51 &4.65& 8 \tabularnewline
  19:00 & 49.65 & 5.12 & 12.57 &5.03& 7 \tabularnewline
  20:00 & 41.95 & 4.56 & 13.55 &4.45& 7 \tabularnewline
  21:00 & 37.50 & 5.03 & 12.43 &5.04& 7 \tabularnewline
  \bottomrule
  \end{tabular}
  \end{scriptsize}
 \label{tbl:ParametersHour}
\end{table}

\section{Mathematical Formulations for Method 1 \citep{Mandelbaum2013Data} and Method 2 \citep{Yefenof2018}}
\label{app:mandelbaum_yefenot_formulas}
For your convenience, we provide here details on Method 1 and 2, used in Section \ref{sec:patience_estimate} and Appendix \ref{app:validation}.

\textbf{Method 1---\cite{Mandelbaum2013Data}:}
In this method, only two classes of customers exist: served and known abandonments. Let $Y_{i}$ be an indicator of whether the customer abandoned, and $U_{i}$ the observed time of the customer in the queue. \cite{Mandelbaum2013Data} suggest the following estimates:
\[
\hat{\theta}=\frac{\sum_{i=1}^{n}Y_{i}\triangle_{i}}{\sum_{i=1}^{n}U_{i}},\;\hat{\gamma}=\frac{\sum_{i=1}^{n}\left(1-\triangle_{i}\right)}{\sum_{i=1}^{n}U_{i}}.
\]
When applying this method in Section \ref{sec:patience_estimate} and Appendix \ref{app:validation}, we can assume that the uSab customers are either known abandonment (Method 1---Uncertain silent abandonment is abandonment) or served (Method 1---Uncertain silent abandonment is service). For the first option, $Y_i=1$ for Kab and uSab customers and 0 otherwise, and in the second option, $Y_i=1$ only for Kab customers.

\textbf{Method 2---\cite{Yefenof2018}:}
In this method, three classes of customers exist: served, known abandonments, and silent abandonments. Let $Y_{i}$ be an indicator of whether the customer abandoned, $U_{i}$ the observed time of the customer in the queue, and $\triangle_{i}$ an indicator of whether the customer was served. \cite{Mandelbaum2013Data} suggest the following estimates:
\[
\theta=\frac{\sum_{i=1}^{n}\left(1-\triangle_{i}\right)}{\sum_{i=1}^{n}U_{i}\left(1-\triangle_{i}\right)}-\frac{n-\sum_{i=1}^{n}Y_{i}\triangle_{i}}{\sum_{i=1}^{n}U_{i}},\;\gamma=\frac{n-\sum_{i=1}^{n}Y_{i}\triangle_{i}}{\sum_{i=1}^{n}U_{i}}.
\]

When applying this method in Section \ref{sec:patience_estimate} and Appendix \ref{app:validation}, we can assume that the uSab customers are either silent abandonment (Method 2—Uncertain silent abandonment is silent abandonment) or served (Method 2---Uncertain silent abandonment is service). For the first option, $Y_i=1$ for Kab and uSab customers and 0 otherwise, and in the second option, $Y_i=1$ only for Kab customers. Moreover, for the first option, $\triangle_{i}=1$ for served customers only, and for the second option, $\triangle_{i}=1$ for served and uSab customers.

\end{document}